\documentclass[aps,prd,twocolumn,twoside,superscriptaddress,floatfix, nofootinbib]{revtex4-1}
\pdfoutput=1
\usepackage{graphicx}
\usepackage{color}
\usepackage{hyperref}
\usepackage{float}
\usepackage{aas_macros}
\usepackage{ifthen,amsmath,amssymb, bm}

\newcommand{\beq} {\begin{equation}}
\newcommand{\eeq} {\end{equation}}
\newcommand{\bal} {\begin{aligned}}
\newcommand{\eal} {\end{aligned}}

\begin{document}

\title{The Atacama Cosmology Telescope: Combined kinematic and thermal Sunyaev-Zel'dovich measurements from BOSS CMASS and LOWZ halos}


\author{Emmanuel Schaan}
\email{eschaan@lbl.gov}
\affiliation{Lawrence Berkeley National Laboratory, One Cyclotron Road, Berkeley, CA 94720, USA}
\affiliation{Berkeley Center for Cosmological Physics, UC Berkeley, CA 94720, USA}
\author{Simone Ferraro}
\affiliation{Lawrence Berkeley National Laboratory, One Cyclotron Road, Berkeley, CA 94720, USA}
\affiliation{Berkeley Center for Cosmological Physics, UC Berkeley, CA 94720, USA}
\author{Stefania Amodeo}
\affiliation{Department of Astronomy, Cornell University, Ithaca, NY 14853, USA}
\author{Nicholas Battaglia}
\affiliation{Department of Astronomy, Cornell University, Ithaca, NY 14853, USA} 

\author{Simone Aiola}
\affiliation{Center for Computational Astrophysics, Flatiron Institute, New York, NY, USA 10010}
\author{Jason E. Austermann}
\affiliation{Quantum Sensors Group, NIST, 325 Broadway, Boulder, CO 80305}
\author{James A. Beall}
\affiliation{Quantum Sensors Group, NIST, 325 Broadway, Boulder, CO 80305}
\author{Rachel Bean}
\affiliation{Department of Astronomy, Cornell University, Ithaca, NY 14853, USA}
\author{Daniel T. Becker}
\affiliation{Quantum Sensors Group, NIST, 325 Broadway, Boulder, CO 80305}
\author{Richard J. Bond}
\affiliation{Canadian Institute for Theoretical Astrophysics, 60 St. George Street, University of Toronto, Toronto, ON, M5S 3H8, Canada}
\author{Erminia Calabrese} 
\affiliation{School of Physics and Astronomy, Cardiff University, The Parade, Cardiff, Wales, UK CF24 3AA} 
\author{Victoria Calafut}
\affiliation{Department of Astronomy, Cornell University, Ithaca, NY 14853, USA}
\author{Steve~K.~Choi}
\affiliation{Department of Physics, Cornell University, Ithaca, NY, USA 14853}
\affiliation{Department of Astronomy, Cornell University, Ithaca, NY 14853, USA}
\author{Edward V. Denison}
\affiliation{Quantum Sensors Group, NIST, 325 Broadway, Boulder, CO 80305}
\author{Mark J. Devlin}
\affiliation{Department of Physics and Astronomy, University of Pennsylvania, 209 South 33rd Street, Philadelphia, PA, USA 19104}
\author{Shannon M. Duff}
\affiliation{Quantum Sensors Group, NIST, 325 Broadway, Boulder, CO 80305}
\author{Adriaan J. Duivenvoorden}
\affiliation{Joseph Henry Laboratories of Physics, Jadwin Hall, Princeton University, Princeton, NJ, USA 08544}
\author{Jo~Dunkley}
\affiliation{Joseph Henry Laboratories of Physics, Jadwin Hall,
Princeton University, Princeton, NJ, USA 08544}
\affiliation{Department of Astrophysical Sciences, Peyton Hall, 
Princeton University, Princeton, NJ USA 08544}
\author{Rolando D\"unner}
\affiliation{Instituto de Astrof\'isica and Centro de Astro-Ingenier\'ia, Facultad de F\'isica, Pontificia Universidad Cat\'olica de Chile, Av. Vicu\~na Mackenna 4860, 7820436, Macul, Santiago, Chile}
\author{Patricio A. Gallardo}
\affiliation{Department of Physics, Cornell University, Ithaca, NY, USA 14853}
\author{Yilun Guan}
\affiliation{Department of Physics and Astronomy, University of Pittsburgh, Pittsburgh, PA, USA 15260}
\author{Dongwon Han}
\affiliation{Physics and Astronomy Department, Stony Brook University, Stony Brook, NY 11794}
\author{J.~Colin Hill}
\affiliation{Department of Physics, Columbia University, New York, NY, USA 10027}
\affiliation{Center for Computational Astrophysics, Flatiron Institute, New York, NY, USA 10010}
\author{Gene C. Hilton}
\affiliation{Quantum Sensors Group, NIST, 325 Broadway, Boulder, CO 80305}
\author{Matt Hilton}
\affiliation{Astrophysics Research Centre, University of KwaZulu-Natal, Westville Campus, Durban 4041, South Africa}
\affiliation{School of Mathematics, Statistics \& Computer Science, University of KwaZulu-Natal, Westville Campus, Durban 4041, South Africa}
\author{Ren\'ee Hlo\v{z}ek}
\affiliation{David A. Dunlap Department of Astronomy and Astrophysics, University of Toronto, 50 St. George Street, Toronto ON M5S3H4}
\affiliation{Dunlap Institute for Astronomy and Astrophysics, University of Toronto, 50 St. George Street, Toronto ON M5S3H4}
\author{Johannes Hubmayr}
\affiliation{Quantum Sensors Group, NIST, 325 Broadway, Boulder, CO 80305}
\author{Kevin M. Huffenberger}
\affiliation{Department of Physics, Florida State University, Tallahassee, FL 32306, USA}
\author{John P. Hughes}
\affiliation{Department of Physics and Astronomy, Rutgers, the State
University of New Jersey, 136 Frelinghuysen Road, Piscataway, NJ
08854-8019, USA}
\author{Brian J. Koopman}
\affiliation{Department of Physics, Yale University, New Haven, CT 06520}
\author{Amanda MacInnis}
\affiliation{Physics and Astronomy Department, Stony Brook University, Stony Brook, NY 11794}
\author{Jeff McMahon}
\affiliation{Kavli Institute for Cosmological Physics, University of Chicago, Chicago, IL 60637, USA}
\affiliation{Department of Astronomy and Astrophysics, University of Chicago, Chicago, IL 60637, USA}
\affiliation{Department of Physics, University of Chicago, Chicago, IL 60637, USA}
\affiliation{Enrico Fermi Institute, University of Chicago, Chicago, IL 60637, USA}
\author{Mathew~S.~Madhavacheril}
\affiliation{Centre for the Universe, Perimeter Institute, Waterloo, ON N2L 2Y5, Canada}
\author{Kavilan Moodley}
\affiliation{Astrophysics Research Centre, University of KwaZulu-Natal, Westville Campus, Durban 4041, South Africa}
\affiliation{School of Mathematics, Statistics \& Computer Science, University of KwaZulu-Natal, Westville Campus, Durban 4041, South Africa}
\author{Tony Mroczkowski}
\affiliation{European Southern Observatory (ESO), Karl-Schwarzschild-Strasse 2, Garching, 85748, Germany}
\author{Sigurd Naess}
\affiliation{Center for Computational Astrophysics, Flatiron Institute, New York, NY, USA 10010}
\author{Federico Nati}
\affiliation{Department of Physics, University of Milano-Bicocca, Piazza della Scienza 3, 20126 Milano, Italy}
\author{Laura B. Newburgh}
\affiliation{Department of Physics, Yale University, 217 Prospect St, New Haven, CT 06511}
\author{Michael D. Niemack}
\affiliation{Department of Physics, Cornell University, Ithaca, NY, USA 14853}
\affiliation{Department of Astronomy, Cornell University, Ithaca, NY 14853, USA} 
\author{Lyman A. Page}
\affiliation{Joseph Henry Laboratories of Physics, Jadwin Hall,
Princeton University, Princeton, NJ 08544, USA}
\author{Bruce Partridge}
\affiliation{Department of Physics and Astronomy, Haverford College, 370 Lancaster Ave, Haverford, PA 19041, USA}
\author{Maria Salatino} 
\affiliation{Physics Department, Stanford University, 382 via Pueblo, Stanford, CA 94305, USA}
\affiliation{Kavli Institute for Particle Astrophysics and Cosmology, 452 Lomita Mall, Stanford, CA 94305-4085, USA}
\author{Neelima Sehgal}
\affiliation{Physics and Astronomy Department, Stony Brook University, Stony Brook, NY 11794}
\author{Alessandro Schillaci}
\affiliation{Department of Physics, California Institute of Technology, Pasadena, CA91125, USA}
\author{Crist\'obal Sif\'on}
\affiliation{Instituto de F\'isica, Pontificia Universidad Cat\'olica de Valpara\'iso, Casilla 4059, Valpara\'iso, Chile}
\author{Kendrick M. Smith}
\affiliation{Perimeter Institute for Theoretical Physics, Waterloo, ON N2L 2Y5, Canada}
\author{David N. Spergel}
\affiliation{Center for Computational Astrophysics, Flatiron Institute, New York, NY, USA 10010}
\affiliation{Department of Astrophysical Sciences, Peyton Hall, Princeton University, Princeton, NJ, USA 08544}
\author{Suzanne Staggs}
\affiliation{Joseph Henry Laboratories of Physics, Jadwin Hall,
Princeton University, Princeton, NJ, USA 08544}
\author{Emilie R. Storer}
\affiliation{Joseph Henry Laboratories of Physics, Jadwin Hall, Princeton University, Princeton, NJ 08544, USA}
\author{Hy Trac}
\affiliation{McWilliams Center for Cosmology, Department of Physics, Carnegie Mellon University, Pittsburgh, PA 15213, USA}
\author{Joel N. Ullom}
\affiliation{Quantum Sensors Group, NIST, 325 Broadway, Boulder, CO 80305}
\author{Jeff Van Lanen}
\affiliation{Quantum Sensors Group, NIST, 325 Broadway, Boulder, CO 80305}
\author{Leila R. Vale}
\affiliation{Quantum Sensors Group, NIST, 325 Broadway, Boulder, CO 80305}
\author{Alexander van Engelen}
\affiliation {School of Earth and Space Exploration, Arizona State University, Tempe AZ, 85287, USA}
\author{Mariana Vargas Maga\~na}
\affiliation{Instituto de F\'isica, Universidad Nacional Aut\'onoma de M\'exico, Apdo. Postal 20-364, Ciudad de M\'exico, M\'exico}
\author{Eve~M. Vavagiakis}
\affiliation{Department of Physics, Cornell University, Ithaca, NY, USA 14853}
\author{Edward J. Wollack}
\affiliation{NASA / Goddard Space Flight Center, Greenbelt, MD 20771, USA}
\author{Zhilei Xu}
\affiliation{Department of Physics and Astronomy, University of Pennsylvania, 209 South 33rd Street, Philadelphia, PA, USA 19104}

\begin{abstract}
The scattering of cosmic microwave background (CMB) photons off the free-electron gas in galaxies and clusters leaves detectable imprints on high resolution CMB maps:
the thermal and kinematic Sunyaev-Zel'dovich effects (tSZ and kSZ respectively).  
We use combined microwave maps from the Atacama Cosmology Telescope (ACT) DR5 and \textit{Planck} in combination
with the CMASS (mean redshift $\langle z\rangle = 0.55$ and host halo mass $\langle M_\text{vir} \rangle = 3\times 10^{13} M_\odot$) and LOWZ ($\langle z\rangle = 0.31$, $\langle M_\text{vir} \rangle = 5\times 10^{13} M_\odot$) galaxy catalogs from the Baryon Oscillation Spectroscopic Survey (BOSS DR10 and DR12), to study the gas associated with these galaxy groups.
Using individual reconstructed velocities, we perform a stacking analysis and reject the no-kSZ hypothesis at 6.5~$\sigma$, the highest significance to date.
This directly translates into a measurement of the electron number density profile, and thus of the gas density profile.
Despite the limited signal to noise, the measurement shows at high significance that the gas density profile is more extended than the dark matter density profile, for any reasonable baryon abundance (formally $>90$\,$\sigma$ for the cosmic baryon abundance).
We simultaneously measure the tSZ signal, i.e. the electron thermal pressure profile of the same CMASS objects, and reject the no-tSZ hypothesis at 10~$\sigma$.
We combine tSZ and kSZ measurements to estimate the electron temperature to 20\% precision in several aperture bins, and find it comparable to the virial temperature.
In a companion paper,
we analyze these measurements to constrain the gas thermodynamics and the properties of feedback inside galaxy groups.
We present the corresponding LOWZ measurements in this paper,
ruling out a null kSZ (tSZ) signal at 2.9 (13.9)\,$\sigma$,
and leave their interpretation to future work.
This paper and the companion paper demonstrate that current CMB experiments can detect and resolve gas profiles in low mass halos and at high redshifts, which are the most sensitive to feedback in galaxy formation and the most difficult to measure any other way.
They will be a crucial input to cosmological hydrodynamical simulations, thus improving our understanding of galaxy formation.
These precise gas profiles are already sufficient to reduce the main limiting theoretical systematic in galaxy-galaxy lensing: baryonic uncertainties.
Future such measurements will thus unleash the statistical power of weak lensing from the Rubin, Euclid and Roman observatories.
Our stacking software \texttt{ThumbStack}\footnote{\url{https://github.com/EmmanuelSchaan/ThumbStack}} 
is publicly available and directly applicable to future Simons Observatory and CMB-S4 data.

\end{abstract}

\maketitle

\section{Introduction}

Observations of present day galaxies account for only 10\% of the cosmological abundance of baryons \cite{Fukugita:2004ee}. 
The majority of the baryons is thought to reside outside of the virial radius of galaxies in an ionized, diffuse and colder gas known as the warm-hot intergalactic medium (WHIM) \cite{Fukugita:2004ee, Cen:2006by}. 

Localizing these ``missing baryons'' will improve our understanding of the rich physical processes involved in galaxy formation and evolution. 
Moreover, since baryons account for more than 15\% of the total matter in the Universe, knowing their distribution is required for the interpretation of future percent-precision large-scale structure surveys carried out by the Vera Rubin Observatory \cite{2009arXiv0912.0201L}, Euclid \cite{Amendola:2016saw} and the Nancy Grace Roman Space Telescope \cite{2013arXiv1305.5425S}.

Quasar absorption lines \cite{Kovacs:2018uap, Nicastro:2018eam, 2018MNRAS.479.2547C, 2019MNRAS.484.2257Z}, X-ray observations \cite{2002ARA&A..40..539R, 2009ApJ...693.1142S, 1998ApJ...499...82T, 1998ApJ...495...80B, 1999Natur.397..135P, 2010arXiv1007.1980R} and dispersion measure variations in Fast Radio Bursts (FRBs, \cite{Macquart:2020lln, Munoz:2018mll, Madhavacheril:2019buy}) for a few specific systems have helped find the missing baryons.
Previous Sunyaev-Zel'dovich measurements have also made progress towards the characterization of the WHIM \cite{Hand:2012ui, 2013ApJ...778...52S, Ade:2015lza, Schaan:2015uaa, 2016ApJ...820..101S, Hill:2016dta, Soergel:2016mce, Soergel:2017ahb, 2019ApJ...880...45S, 2019A&A...624A..48D, 2019MNRAS.483..223T, 2020arXiv200702952T}.
However, X-ray observations \cite{Bregman:2007ac} require modeling the clumping and temperature of the gas, and are limited to relatively high mass and nearby objects, while the use of absorption lines requires modeling the metallicity profile, which is subject to considerable uncertainty. 

Through Compton scattering, ionized gas around galaxies and clusters leaves several distinct imprints on the CMB.
The two main effects are the Doppler shifts of CMB photons due to the bulk motion of the gas, the kinematic Sunyaev-Zel'dovich (kSZ) effect, and due to the velocity dispersion of the gas, i.e. the thermal Sunyaev-Zel'dovich effect (tSZ) \cite{SZ80, 1972CoASP...4..173S}.
Being independent of redshift, these SZ effects are uniquely well-suited for studying high redshift galaxies and clusters.
Since the kSZ signal is linearly proportional to the electron number density, the integrated kSZ scales linearly with halo mass
and is well-suited to probe the low density and low temperature outskirts of lower mass galaxies and groups.
Furthermore, its interpretation is particularly straightforward, as the kSZ effect simply counts the number of free electrons, independent of electron temperature or clumping.
On the other hand, the tSZ signal is proportional to the integrated pressure ($P_e \propto n_e T_e$).
Because the electron temperature is higher in more massive halos, the tSZ signal effectively scales as a higher power of halo mass ($\propto M^{5/3}$), and therefore receives most of its contribution from the most massive objects in the sample.  
The tSZ and kSZ thus provide complementary information on the electron density and temperature in galaxies and clusters.
In principle, by combining the kSZ, tSZ and lensing mass measurements from the same galaxies or clusters, we can fully determine the thermodynamic properties of the sample, including the amount of energy injected by feedback or the fraction of non-thermal pressure support \cite{Battaglia:2017neq, 2019arXiv190704473A, paper2}.
In the absence of kSZ measurements, this approach would be limited by the modeling of the gas temperature (for tSZ) and clumping (X-rays) \cite{2019PhRvD.100f3519P, 2020MNRAS.491..235S}.
This joint tSZ and kSZ measurement also informs the ``lensing is low'' tension, where the galaxy-galaxy lensing signal of BOSS galaxies is found to be anomalously low, compared to the expected signal based on their clustering \cite{2016MNRAS.460.1457S, 2017MNRAS.467.3024L, 2019MNRAS.488.5771L}.
This paper and companion paper \cite{paper2} are a first step in constraining the gas thermodynamics in galaxy groups and directly measuring the baryonic effects in weak lensing.
Refs.~\cite{CalafutInPrep, VavagiakisInPrep} present complementary kSZ and tSZ measurements using the same microwave temperature maps, but consider different galaxy samples, 
and instead focus on the luminosity dependence of the signals and the velocity correlation function, which contains information on neutrino masses \cite{Mueller:2014dba}, dark energy and modifications to General Relativity \cite{Mueller:2014nsa} and primordial non-Gaussianity \cite{2019PhRvD.100h3508M}.
To do so, they use a pairwise difference estimator instead of the velocity reconstruction from the density field used here.
The results of both studies are thus complementary \cite{2018arXiv181013423S}, and the relationship between the two estimators has been investigated in \cite{2018arXiv181013423S}.

\begin{figure}[h!]
\centering
\includegraphics[width=0.95\columnwidth]{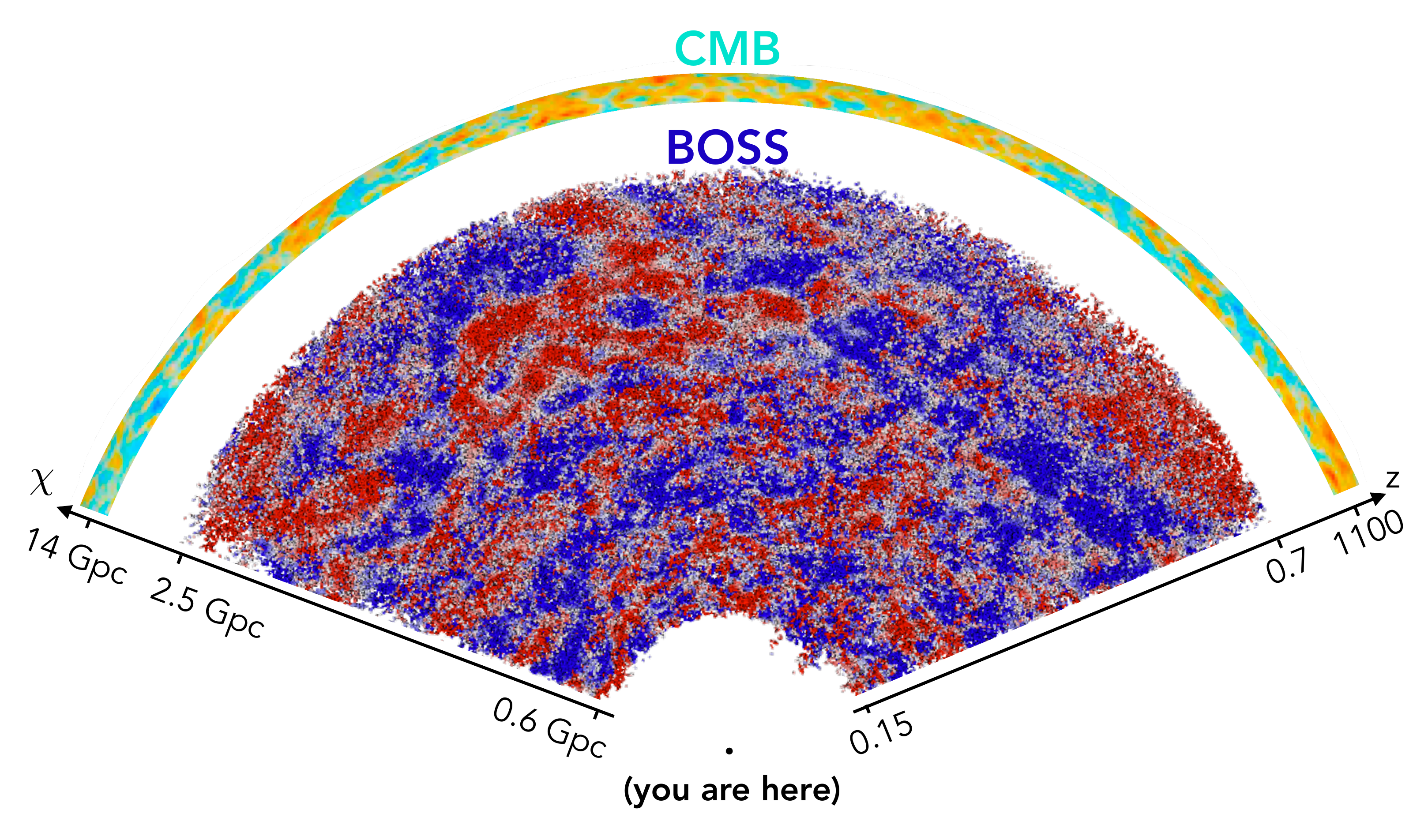}
\caption{
In this visualization, the BOSS galaxies are color coded based on their LOS peculiar velocity (blue towards us, red away from us), estimated from their 3D number density.
As the CMB photons travel towards us, they are Compton scattered by the free electrons associated with the BOSS galaxies, producing the tSZ and kSZ signals.
We detect the tSZ by stacking the CMB map at the position of the BOSS galaxies. For the kSZ, we weight the stack by the estimated LOS velocity.
In this fixed declination slice, the radial direction corresponds to the galaxy comoving distances, estimated from their redshifts, and the angular position corresponds to right ascension. 
Axes indicate redshift (right) and comoving radial distance (left). 
The CMB position and image are not to scale.
The large-scale velocity structure apparent in this visualization is signal-dominated.
}
\label{fig:boss_cmb_schematic}
\end{figure}

In this paper, we combine data from the Baryon Oscillation Spectroscopic Survey (BOSS \cite{2014ApJS..211...17A, 2017MNRAS.470.2617A, 2017MNRAS.464.1493S}), the Atacama Cosmology Telescope (ACT \cite{2007ApOpt..46.3444F, 2011ApJS..194...41S, 2016ApJS..227...21T, 2016JLTP..184..772H})
and \textit{Planck} \cite{2018arXiv180706205P}.
We use spectroscopic galaxy catalogs from BOSS and stack the CMB temperature maps from ACT at the positions of these galaxies
as illustrated in Fig.~\ref{fig:boss_cmb_schematic}.
The tSZ signal is detected by its characteristic spectral signature in our multifrequency CMB data, in which it yields a temperature decrement (increment) at frequencies below (above) 217 GHz.
Thermal emission from dust inside the galaxy groups produces a smaller and more concentrated temperature excess, which we also measure and correct for in several ways \cite{paper2}.
This tSZ stacking procedure nulls the kSZ signal, which changes sign depending on the galaxy group's bulk velocity, and thus cancels on average.
To measure the kSZ signal, we perform a weighted stack, where each galaxy group's temperature signal is multiplied by an estimate of the group's line-of-sight (LOS) velocity \cite{2006NewAR..50..918S, 2009arXiv0903.2845H, 2011MNRAS.413..628S, 2014MNRAS.443.2311L, Schaan:2015uaa}. 
The estimated LOS velocity is obtained through ``linear reconstruction from the density field'' \cite{2012MNRAS.427.2132P, 2015arXiv150906384V}:
using the galaxy redshifts, the spectroscopic galaxy catalog can be placed on a 3D grid, yielding an estimate of the 3D density field, which is then converted to velocities via the Zel'dovich approximation \cite{1970A&A.....5...84Z}.
This velocity-weighted stacking has the added benefit of suppressing the tSZ and dust contamination to kSZ, as well as any other foreground uncorrelated with the galaxy velocities
\cite{Schaan:2015uaa, 2018arXiv181013423S}.

The remainder of this paper is organized as follows. 
In Section~\ref{sec:theory}, we review the origin of the kSZ and tSZ effects.
Section~\ref{sec:data} presents our microwave temperature maps and galaxy catalogs, and Section~\ref{sec:analysis} describes the analysis techniques to extract both tSZ and kSZ. 
The results are in Section~\ref{subsec:results}, followed by a discussion of systematics and null tests. Finally, our conclusions are found in Section~\ref{sec:conclusions}. 
The interpretation of the measurements is presented in detail in \cite{paper2}.

\section{Theory: kSZ and tSZ effects}
\label{sec:theory}

The kSZ effect is the Doppler shift of CMB photons due to the bulk motion of the ionized gas in and around galaxies and clusters.
It preserves the blackbody frequency spectrum of the CMB
and shifts its thermodynamic temperature as \cite{SZ80}:
\begin{equation}
\frac{\delta T _{\rm kSZ} (\hat{\bm{n}})}{T_{\rm CMB}} 
= 
-
\int \frac{d \chi}{1+z} 
n_e(\chi\hat{\bm{n}}, z)
\sigma_T \
e^{-\tau(z)}  \ 
\frac{\bm{v}_e \cdot \bm{\hat{n}} }{c},
\label{eq:kSZdef}
\end{equation}
where $\sigma_T$ is the Thomson cross-section,
$\tau(z)$ is the optical depth to Thomson scattering between the observer and redshift $z$, along the line of sight considered:
\beq
\tau(z)
\equiv
\int \frac{d \chi}{1+z} 
n_e(\chi\hat{\bm{n}}, z)
\sigma_T 
,
\label{eq:tau_def}
\eeq
$\chi$ is the comoving distance to redshift $z$,
$n_e$ is the \textit{free}-electron physical (not comoving) number density and $\bm{v}_e$ the peculiar velocity, $c$ the speed of light and $\bm{\hat{n}}$ is the line-of-sight (LOS) direction, defined to point away from the observer.
For the redshift range $z=0.4\text{--}0.7$ of interest in this measurement, 
the mean optical depth $\bar{\tau}(z)$ is well below percent-level (e.g., Fig~16 in \cite{2016A&A...596A.108P}).
Furthermore, the galaxy groups in this analysis are optically thin.
We can therefore take $e^{-\tau(z)} \approx 1$ in the integral to a percent level accuracy.
Finally, our stacking analysis selectively extracts the kSZ signal correlated with the galaxy group of interest.
The kSZ signal thus simplifies to
\beq
\frac{\delta T _{\rm kSZ} (\hat{\bm{n}})}{T_{\rm CMB}}
=
- \tau_\text{gal} \; \left(\frac{v_{e, r}}{c} \right),
\eeq
where $v_{e,r}$ is the free electron bulk LOS velocity and $\tau_\text{gal}$ refers to the optical depth to Thomson scattering of the galaxy group considered, i.e. the contribution from the galaxy group to Eq.~\eqref{eq:tau_def}.

The tSZ effect also comes from relativistic Doppler shifts, but it is due to the thermal motion of the electrons in the gas.
Each electron, moving at its own speed and in its own direction, Doppler-boosts some of the CMB photons to a blackbody spectrum with a different temperature. 
Averaging all these different blackbody spectra together leads to a $y$-type spectral distortion (see \cite{2012MNRAS.426..510C} for a more rigorous derivation), proportional to the square of the electron thermal velocity $v_\text{th}$, and thus to the electron temperature $T_e$:
\begin{equation}
\frac{\delta T _{\rm tSZ} (\hat{\bm{n}})}{T_{\rm CMB}} 
=
f_\text{tSZ}(\nu) y(\bm{\hat{n}}),
\label{eq:tSZdef}
\end{equation}
where the frequency dependence is
$f_\text{tSZ}(\nu) = x \coth{(x/2)} -4$ 
with 
$x=h\nu / k_B T_{\rm CMB}$,
and the amplitude is given by the Compton $y$ parameter:
\begin{equation}
    y(\hat{\bm{n}})  = \frac{k_B \sigma_T}{m_e c^2} \int \frac{d \chi}{1+z} n_e(\chi\hat{\bm{n}}, z) T_e(\chi\hat{\bm{n}}).
\end{equation}
In the expression above, $k_B$ is the Boltzmann constant and $m_e$ the electron mass.

The fractional temperature changes due to kSZ and tSZ
can be written intuitively as 
$\tau_\text{gal} v_e/c$
and
$\tau_\text{gal} (k_bT_e/m_ec^2) \sim \tau_\text{gal} (v_\text{th}/c)^2$ respectively,
with $\tau_\text{gal} = \int a d\chi n_e \sigma_T$ as above ($a$ is the scale factor).
From this, we can infer the order of magnitude of the kSZ and tSZ signals \cite{Hu:2001bc}.
Considering a circular aperture with radius $\sim 1'$, similar to the beam widths of the maps used in this analysis ($\text{FWHM}=1.3\text{--}2.4'$), the mean optical depth is typically $\tau_\text{gal}\sim 10^{-5}$ for our galaxy groups ($M_{200} = 3\text{--}5\times 10^{13}M_\odot$).
We can assume $v_\text{bulk}/c \sim 10^{-3}$ for the electron bulk motion (the cosmological RMS) and $v_\text{th}/c\sim 0.1$ for their thermal motion ($T_e\sim 10^{7}$K).
For $T_\text{CMB}\sim 3$~K, the mean kSZ and tSZ signals within the aperture are thus of order 0.1~$\mu$K, compared to the 100~$\mu$K primary CMB fluctuations.
As we explain below, this large-scale noise from the CMB can be reduced with high-pass filtering (aperture photometry filters in this analysis) and by averaging over many galaxies.

\section{Data sets}
\label{sec:data}

\subsection{BOSS galaxy sample}

\begin{figure}[h!]
\centering
\includegraphics[width=0.95\columnwidth]{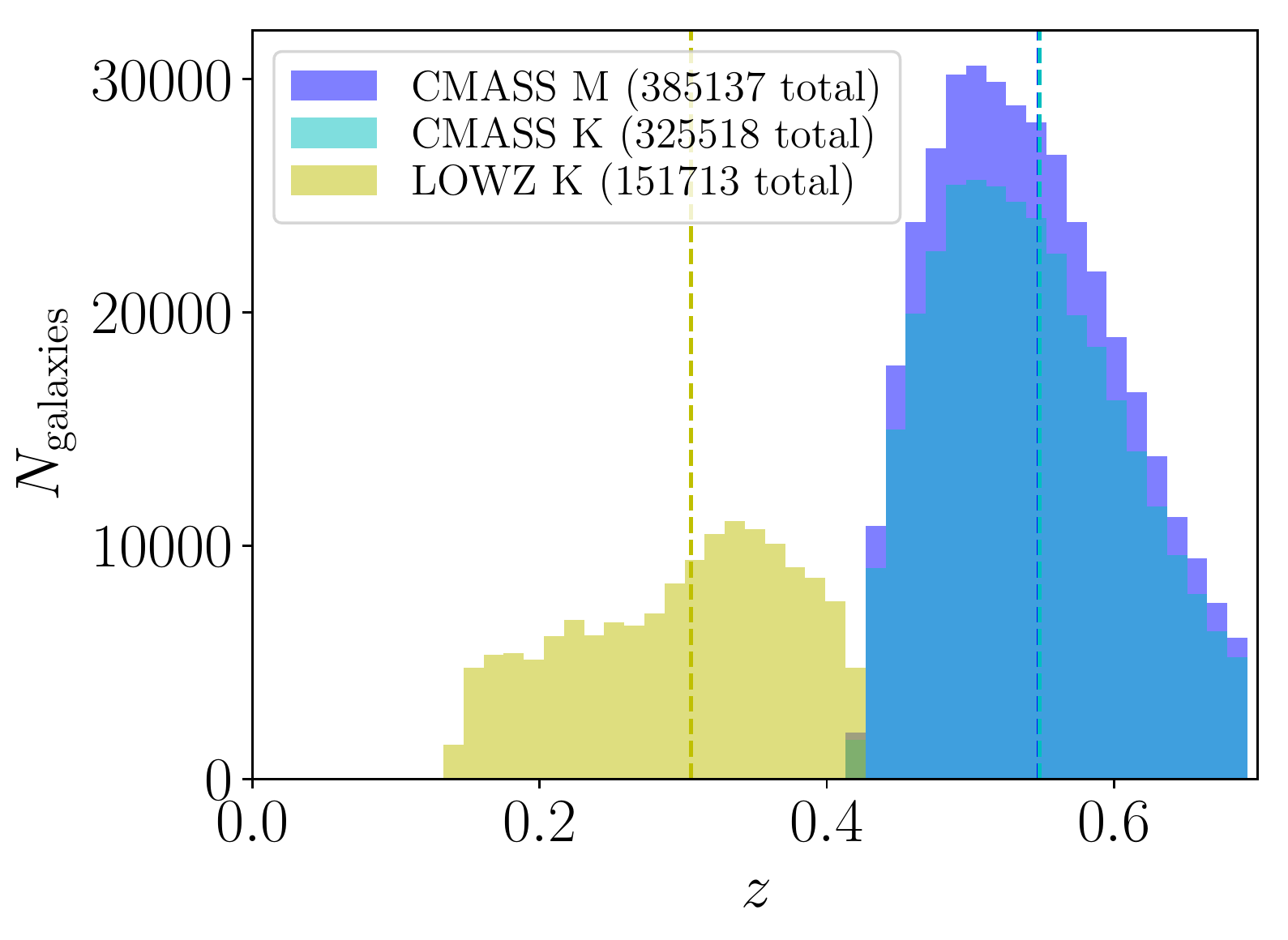}
\caption{
Redshift distribution of the LOWZ K (DR10), CMASS K (DR10) and CMASS M (DR12)
spectroscopic galaxies whose positions on the sky overlap with the ACT DR5 microwave maps.
The mean redshifts are 0.31 for LOWZ K and 0.54 for CMASS K and CMASS M. They are indicated by the vertical dashed lines.
}
\label{fig:dndz}
\end{figure}

\begin{figure}[h!]
\centering
\includegraphics[width=0.95\columnwidth]{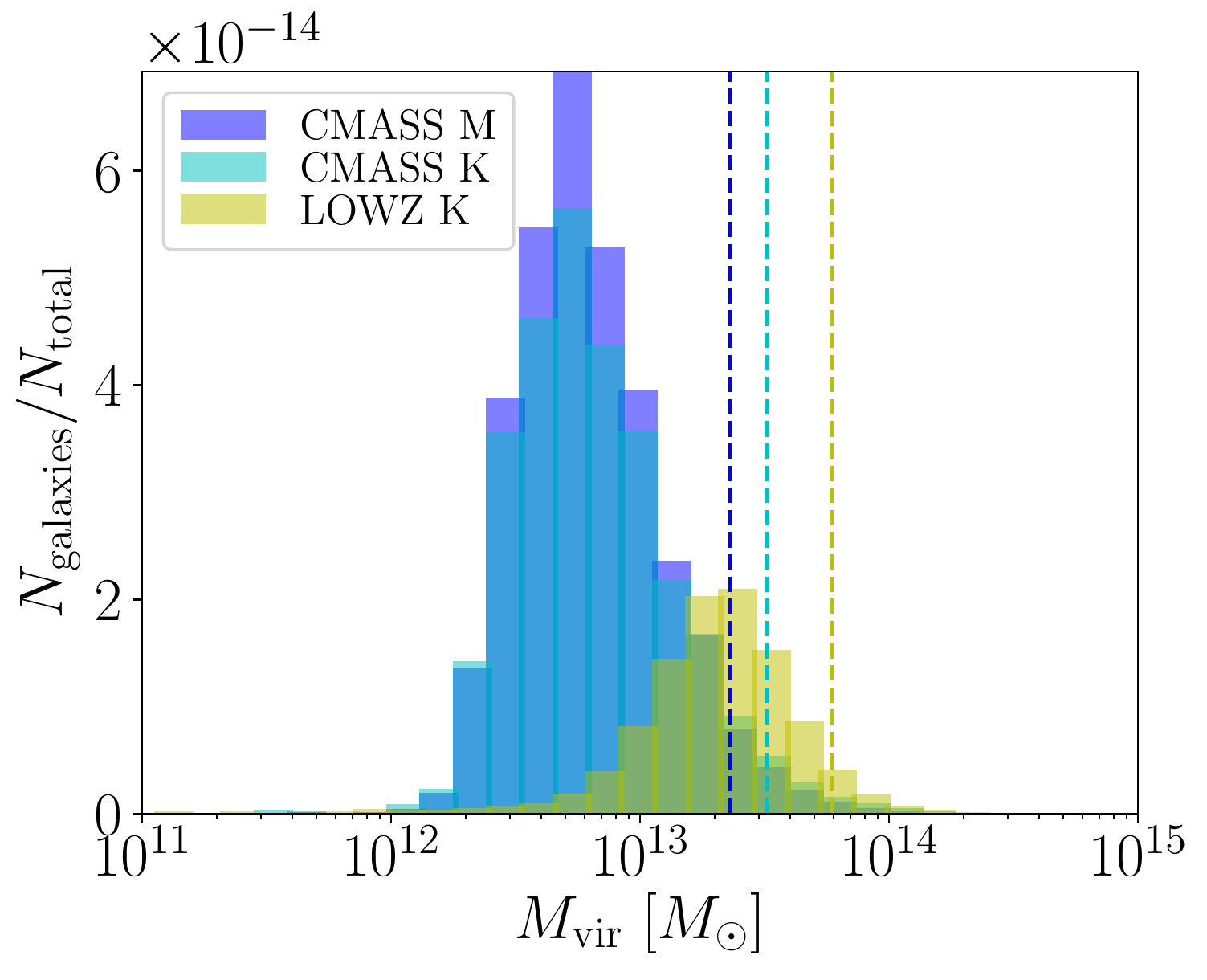}
\caption{
Host halo virial masses of the LOWZ K (DR10), CMASS K (DR10) and CMASS M (DR12) galaxies, as inferred from their stellar masses in Appendix~\ref{app:mass_distribution}.
The dashed lines indicate the mean halo masses for each sample,
$\langle M_\text{vir} \rangle = 3\times 10^{13} M_\odot$ for CMASS K and 
$\langle M_\text{vir} \rangle = 5\times 10^{13} M_\odot$ for LOWZ K.
These do not coincide with the modes of the mass distributions, due to the high mass tails (the x-axis is logarithmic).
In this analysis, we further discard the objects with $M_\text{vir} > 10^{14} M_\odot$ to avoid tSZ contamination to the kSZ signal, as explained in Sec.~\ref{subsec:systematics}.
}
\label{fig:halo_masses}
\end{figure}

In the fiducial analysis, we use the CMASS (``constant mass'') and LOWZ (``low redshift'') galaxy catalogs from the Baryon Oscillation Spectroscopic Survey (BOSS) 
DR10 \cite{2014ApJS..211...17A}, for which we have reconstructed velocities
(see next subsection) and which we refer to as CMASS K and LOWZ K.
In the Appendix, as a null test, we also compare the results to a different velocity reconstruction algorithm for CMASS \cite{2015arXiv150906384V}, which is based on the DR12 catalog \cite{2017MNRAS.470.2617A, 2017MNRAS.464.1493S} and which we refer to as CMASS M.
The redshift distributions of the LOWZ and CMASS samples are shown in Fig.~\ref{fig:dndz}, and their host halo masses are shown in Fig.~\ref{fig:halo_masses}. 
The latter are inferred from the stellar mass estimates of \cite{2013MNRAS.435.2764M} and the Wisconsin PCA method\footnote{\url{https://data.sdss.org/sas/dr12/boss/spectro/redux/galaxy/v1_1/}} \cite{2012MNRAS.421..314C} using the stellar population model of \cite{2011MNRAS.418.2785M}.
The stellar masses are then converted to halo masses using
 the stellar-to-halo mass relation of \cite{2018AstL...44....8K}.
The resulting mean halo mass obtained for CMASS ($\langle M_\text{vir} \rangle = 3\times 10^{13} M_\odot$) is in agreement with galaxy lensing measurements \cite{2015ApJ...806....1M}.

\begin{figure}[h]
\centering
\includegraphics[width=0.95\columnwidth]{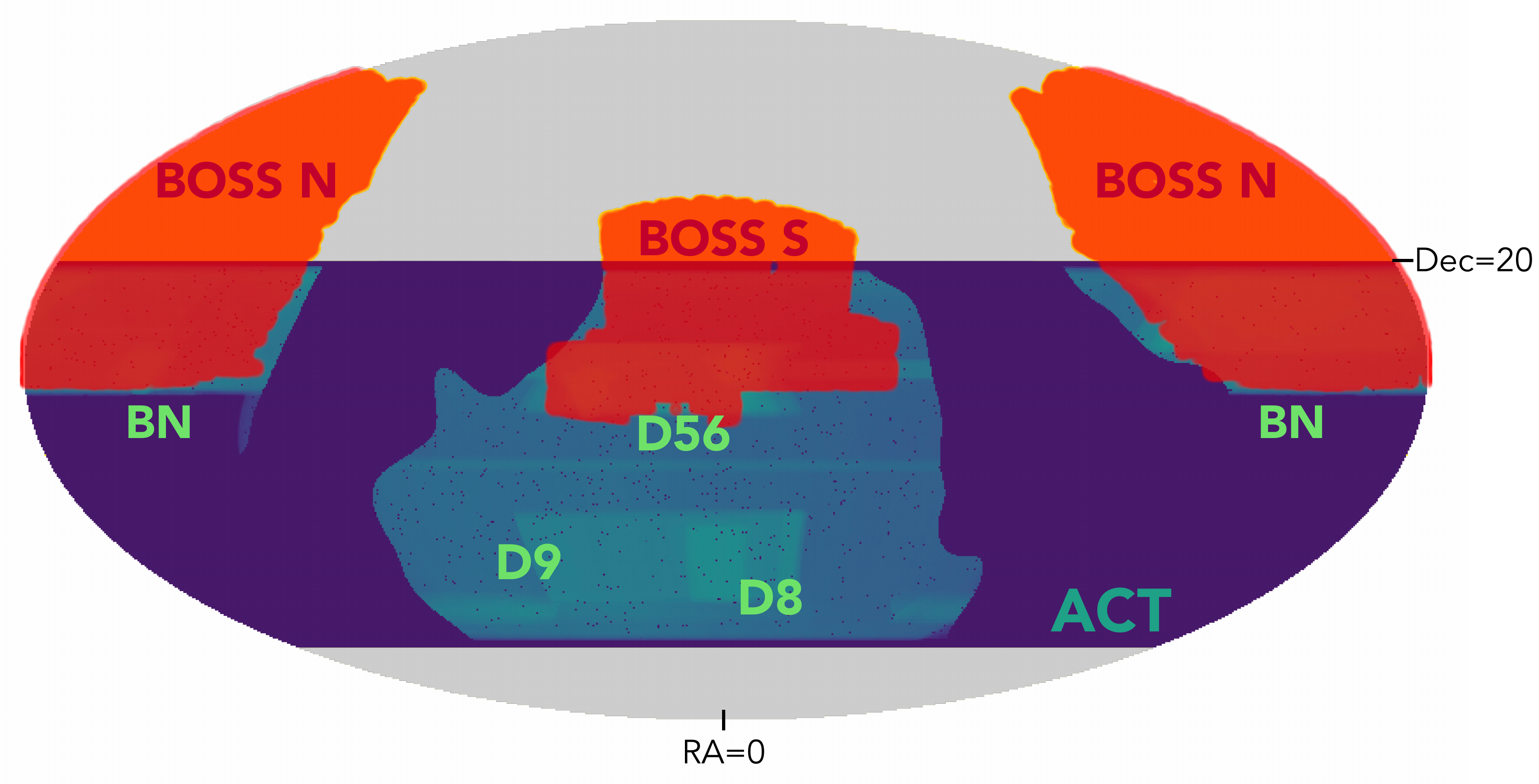}
\caption{
50--70\% of the BOSS galaxies (red) overlap with the ACT footprint (blue-green).
With about 350,000 CMASS galaxies and 150,000 LOWZ galaxies, this is a large increase over our previous analysis \cite{Schaan:2015uaa}.
The green map shows the ACT inverse noise variance map, masked for point sources and with the \textit{Planck} $f_\text{sky} = 60\%$ Galactic mask, in equatorial coordinates.
The Northern Galactic cap of BOSS overlaps with the ``BN'' (for BOSS North) ACT field, and the Southern Galactic cap of BOSS overlaps with the D56 (for Deep 5-6) ACT field.
}
\label{fig:overlap}
\end{figure}

The overlap of the BOSS catalogs with the ACT temperature maps is shown in Fig.~\ref{fig:overlap}.
It includes 
325,518 CMASS K galaxies (out of 501,844),
385,137 CMASS M galaxies (out of 777,202)
and 151,713 LOWZ K galaxies (out of 218,905).
After masking for point sources and for the Milky Way (see Sec.~\ref{subsec:temperature_maps}), 
312,708 CMASS K,
368,701 CMASS M
and 145,714 LOWZ K galaxies are left.
Finally, discarding the objects with $M_\text{vir}>10^{14}M_\odot$ (see Sec.~\ref{subsec:systematics}) leaves
311,309 CMASS K,
360,084 CMASS M
and 134,702 LOWZ K galaxies for the tSZ and kSZ analyses.

\subsection{Velocity reconstruction}

The kSZ signal changes sign depending on whether the galaxy group is moving towards us or away from us.
To avoid cancellation when stacking, we use an estimate of the peculiar velocity of each galaxy, 
reconstructed from the 3D galaxy number density.
Similarly to the Baryon Acoustic Oscillations (BAO) reconstruction method, an estimate of the peculiar velocity field along the line of sight $v_{\rm rec}$ can be obtained by solving the linearized continuity equation in redshift-space \cite{2012MNRAS.427.2132P, 2015arXiv150906384V}:
\begin{equation}
\bm{\nabla}\cdot \bm{v}
+ f  \bm{\nabla}\cdot \left[ \left( \bm{v}\cdot \bm{\hat{n}} \right) \bm{\hat{n}} \right]
= -a H f \; \frac{\delta_g}{b}
\label{eq:vrec}
\end{equation}
Here $\delta_g$ is the galaxy overdensity,
$f = d \ln \delta / d \ln a$ is the logarithmic linear growth rate and $b$ is the linear bias. 
Importantly, Eq.~\eqref{eq:vrec} takes into account the linear redshift-space distortion (Kaiser effect). Since the kSZ effect is only sensitive to the \textit{radial} component of the velocity field, the scalar $v$ will always refer to the radial velocity in the remainder of the paper.
The velocity reconstruction is not perfect, due to shot noise, non-linearities and the finite volume observed.
This reduces the kSZ signal-to-noise (SNR), multiplying it by a factor equal to the real-space correlation coefficient between true and reconstructed galaxy velocities
\begin{equation}
    r_v = \frac{\langle v_{\rm true} v_{\rm rec} \rangle}{v_{\rm rms}^{\rm true} \ v_{\rm rms}^{\rm rec}},
\end{equation}
where $v_{\rm rms}^{\rm true}$ and $v_{\rm rms}^{\rm rec}$ are the standard deviations of the true and reconstructed galaxy radial velocities, respectively. 
In our fiducial analysis, we use the velocity reconstruction from a Wiener Filter analysis of CMASS and LOWZ DR10 (described in Ref.~\cite{kendrickvelrec} in preparation), already used in \cite{Schaan:2015uaa}.
The correlation coefficient $r_v$ is estimated by comparing the true and reconstructed galaxy velocities in realistic BOSS mock catalogs \cite{2013MNRAS.428.1036M, 2015MNRAS.447..437M}.
This yields $r_v=0.7$, which we use throughout this paper.
In Appendix~\ref{app:null_tests}, we compare our results to the reconstructed velocities for CMASS DR12 from \cite{2015arXiv150906384V}.
This reconstruction uses a fixed smoothing scale, instead of the optimal Wiener filtering, and achieves $r_v \approx 0.5$ on mock catalogs.

Below, we use the velocity correlation coefficient $r_v=0.7$ to correct the kSZ estimator, making it unbiased with respect to imperfections of the velocity reconstruction.
However, the kSZ SNR is still reduced by a factor $r_v$:
a perfect velocity reconstruction ($r_v=1$ instead of $r_v=0.7$) would improve our kSZ SNR by $40\%$.

The uncertainty on the value of $r_v$ is less than a few percent (as described in \cite{kendrickvelrec} in preparation), making it a negligible contribution to our overall kSZ noise budget.
However, upcoming measurements with higher kSZ SNR will need to quantify this uncertainty carefully \cite{2020arXiv200713721N}.

\subsection{Microwave Temperature maps}
\label{subsec:temperature_maps}

Our measurement relies crucially on high resolution and high sensitivity microwave temperature maps from the Atacama Cosmology Telescope (ACT)
\cite{2007ApOpt..46.3444F, 2016ApJS..227...21T, 2016JLTP..184..772H}.
This experiment, located in northern Chile, produces arcminute-resolution maps of the microwave sky, both in temperature and polarization.

Since the kSZ measurement is a velocity-weighted stack, most foregrounds automatically cancel because they are uncorrelated with the velocity field.
As a result, we use the temperature maps without performing foreground cleaning as it is not needed to measure the kSZ effect. If foreground cleaning were to be applied, the optimal temperature map to measure kSZ would be the one with unit response to the CMB blackbody spectrum, e.g. the result of the standard internal linear combination (ILC) foreground cleaning method \cite{2003PhRvD..68l3523T}.
Here, we use two coadded CMB temperature maps at 98 GHz (called f90 hereafter for consistency with \cite{naessetal20}) and 150 GHz (called f150) produced by combining data from ACT \cite{2007ApOpt..46.3444F, 2011ApJS..194...41S, 2016ApJS..227...21T, 2016JLTP..184..772H} and \textit{Planck} \cite{2018arXiv180706205P}.
We use the ACT DR5 2008--2018 day \& night maps, which combine
data from the first generation ACT receiver MBAC (the Millimeter Bolometric Array
Camera) \cite{2011ApJS..194...41S}, the second generation polarization-sensitive receiver ACTPol (Atacama Cosmology Telescope Polarimeter) \cite{2016ApJS..227...21T} and the AdvACT receiver (Advanced ACTPol) \cite{2016JLTP..184..772H}.

\begin{figure}[h]
\centering
\includegraphics[width=0.95\columnwidth]{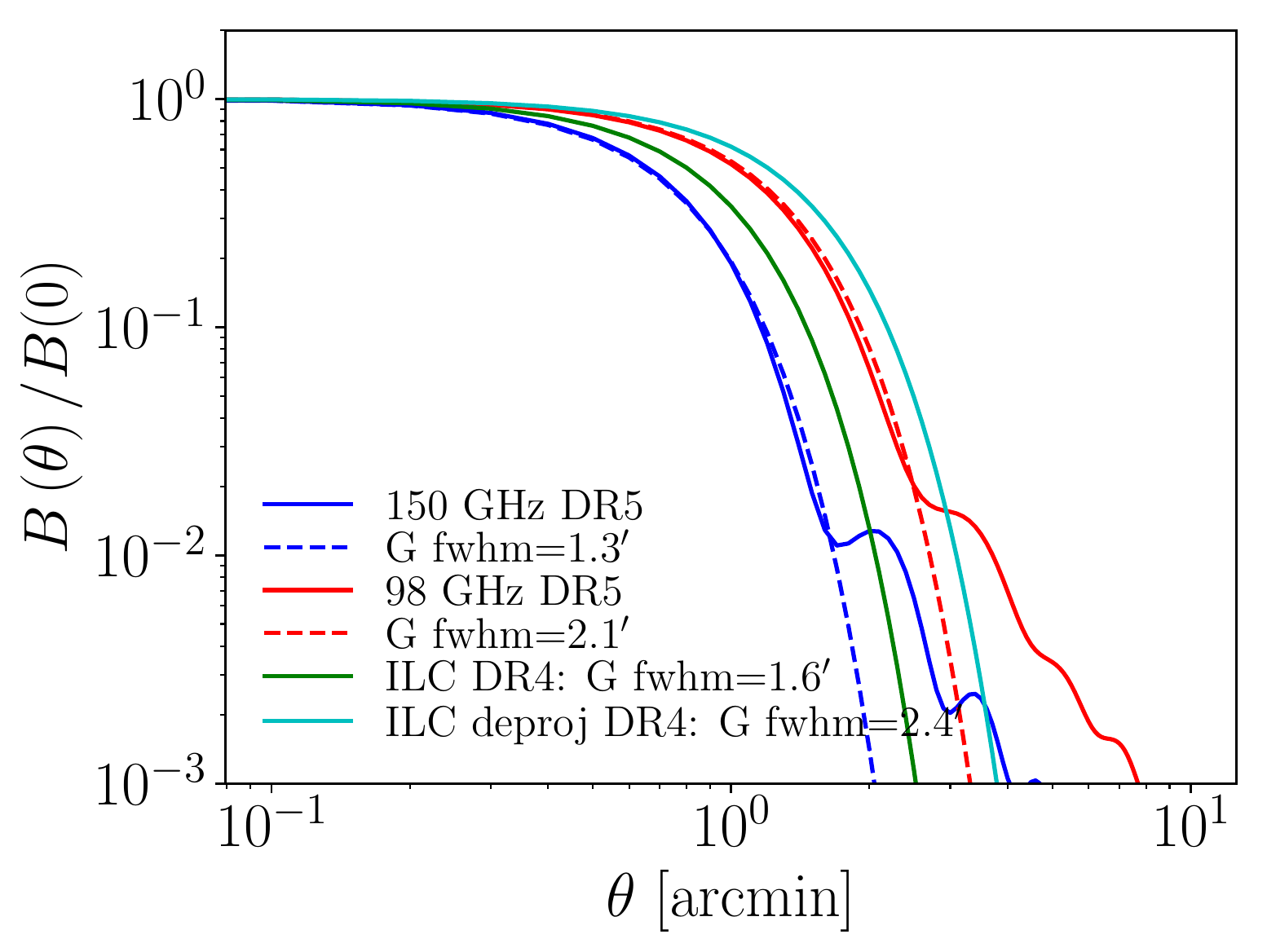}
\caption{
The effective beam profiles for the coadded f90 and f150 DR5 maps from \cite{naessetal20} are shown in solid blue and red, and compared to Gaussian beams with the same FWHM. Percent-level sidelobes are visible at $2\text{--}4'$. These are included in the modeling of the signal in \cite{paper2}.
The beams for the ILC maps with and without deprojection from \cite{2019arXiv191105717M} are shown in green and cyan. These are Gaussian by construction.
}
\label{fig:beams}
\end{figure}

Most of the kSZ SNR comes from multipoles of a few thousand, where ACT dominates the coadd over \textit{Planck} due to its resolution and sensitivity.
The addition of \textit{Planck} data is helpful on larger scales, as it is not affected by atmospheric noise.
These DR5 maps are described in detail in \cite{naessetal20}.
Their beams are shown in Fig.~\ref{fig:beams}, and are close to Gaussian with $\text{FWHM}= 2.1, 1.3$ arcmin for f90, f150 respectively.
By construction, the beams are uniform over the whole map area. 
However, as described in \cite{naessetal20}, the coadds contain 2017--2018 and daytime ACT data, where the beam characterization is more preliminary, and the beam size could vary by as much as $10\%$ from patch to patch.
The resulting beam uncertainty after averaging over the wide area encompassing all of the galaxies is substantially reduced.
The 2017--2018 and daytime ACT data also have a percent-level gain calibration uncertainty, resulting in a percent-level uncertainty on the measured kSZ signal.
A more detailed characterization of ACT beams and calibration for post-2016 and daytime data is in progress.

We measure the kSZ profiles separately on the f90 and f150 maps, including their covariance.
The microwave maps are deepest in the so-called Deep56 region (``D56'', 8--12 $\mu$K$\cdot$arcmin in f150 and 12--18 $\mu$K$\cdot$arcmin in f90) and the BOSS North region (``BN'', 8--10 $\mu$K$\cdot$arcmin in f150 and 8--12 $\mu$K$\cdot$arcmin at 98 GHz) and shallower in the wide area in between (up to $\simeq 30 \mu$K$\cdot$arcmin in f90 and f150).

To measure the stacked tSZ profiles, we used two distinct sets of maps.
First, we use the temperature coadds f90 and f150 described above.
As shown in Fig.~18 in \cite{naessetal20}, because the coadds combine maps with different bandpasses and noise levels, the response of these maps to tSZ is scale-dependent. 
We include this scale dependence in the interpretation of the measured profiles in \cite{paper2}.
Furthermore, because the maps combined in the coadds have different spatial noise variations, the tSZ response is also position-dependent.
However, the tSZ response only varies at the percent level (Fig.~19 in \cite{naessetal20}) across the map, so its average over the positions of the BOSS galaxies should be accurate to better than a percent, and therefore any spatial variation is negligible.
Finally, differences due to the inverse-variance weighting (instead of uniform weighting) in the stack are an even smaller effect.

Unlike for kSZ, foreground contamination is a major concern for tSZ, especially the thermal dust emission from the BOSS galaxies and other galaxies correlated with them.
We handle this in two independent ways.
With the first method, a thermal dust emission profile from BOSS galaxies is obtained by stacking on Herschel data. For this purpose we use three fields of Herschel/H-ATLAS data \cite{eales+10} in the three bands centered at 600, 857 and 1200 GHz, which overlap with both ACT and about 9000 CMASS halos.
This measurement and the corresponding modeling is presented in details in the companion paper  \cite{paper2}.
A second method, which constitutes our fiducial analysis, involves using the 
internal linear combination (ILC) component-separated maps of \cite{2019arXiv191105717M}.
Specifically, we use the Compton-$y$ map with deprojected CIB, 
which nulls any thermal dust emission with a fixed frequency dependence (see Eq.~\eqref{eq:tilec_dust_sed}).
However, this map has higher noise, in part because it does not include the latest post-2016 ACT data included in the f90 and f150 DR5 coadds, and in part because of the foreground deprojection.
It has a Gaussian beam with $\text{FWHM}=2.4'$.

Finally, we perform a number of null tests, comparing the stacks on the f90 and f150 coadds, and several of the ILC component separated maps, with and without deprojection.
These null tests are shown in Appendix~\ref{app:null_tests}.
In all cases, we mask the Milky Way using the \textit{Planck} $60\%$ galactic mask
\footnote{\href{https://irsa.ipac.caltech.edu/data/Planck/release_2/ancillary-data}{Planck release 2 website}}
and the point sources detected at $> 5\sigma$ in the maps, corresponding to roughly $15$ mJy (variable with the map position).
This leaves 312,708 CMASS K galaxies,
368,701 CMASS M galaxies
and 145,714 LOWZ K galaxies.

In summary, we use the following maps:
\begin{itemize}
\item{\textbf{ACT DR5 + \textit{Planck} coadds f90 and f150}} to measure the kSZ signal and the tSZ + dust signal;
\item{\textbf{ACT DR4 + \textit{Planck} ILC Compton-$y$ map with deprojected CIB}} to measure the tSZ signal without CIB contamination;
\item{\textbf{Various ACT DR4 + \textit{Planck} ILC maps with or without deprojection}} for the null tests.
\end{itemize}
The map beams are summarized in Fig.~\ref{fig:beams}, shown in configuration space (see the Fourier beams for the DR5 coadds in Fig.~4 in \cite{naessetal20}). 

\section{Analysis}
\label{sec:analysis}

\subsection{Filtering}

\begin{figure}[h]
\centering
\includegraphics[width=0.95\columnwidth]{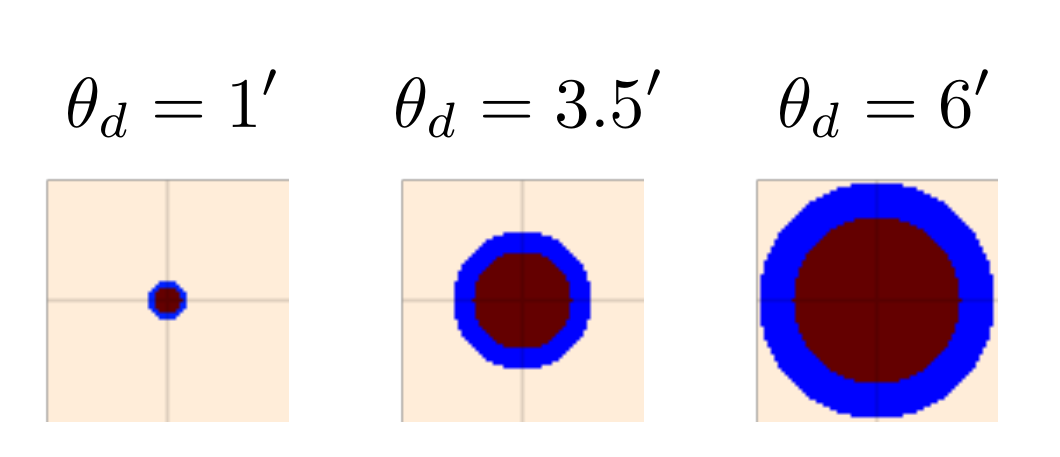}
\caption{
Cutout pixelation and CAP filters (smallest, intermediate and largest).
Each cutout is $17.25'$ and 69 pixels on the side, with $0.25'$ pixels.
Given the large number of galaxies in our catalogs, many of the CAP filters from different objects overlap, which affects the covariance matrix (see Appendix.~\ref{app:mock_cov}). 
As the disk radius $\theta_d$ increases, the CAP filter output behaves similarly to a cumulative (integrated) profile.
}
\label{fig:ap_filters}
\end{figure}

For both kSZ and tSZ, we use compensated aperture photometry (CAP) filters with varying aperture radius $\theta_d$, centered around each galaxy. 
The output of the CAP filter on a temperature map $\delta T$ is defined by: 
\begin{equation}
\label{eq:ap}
\mathcal{T}(\theta_{d})=
\int d^2\theta \, \delta T(\theta) \, W_{\theta_{d}}(\theta) \,.
\end{equation}
where the filter $W_{\theta_{d}}$ is chosen as:
\begin{equation}
W_{\theta_{d}}(\theta) =
\left\{
\begin{aligned}
1& &  &\text{for} \, \theta < \theta_d \,, \\
-1& &  &\text{for} \, \theta_d \leq \theta \leq \sqrt{2}\theta_d \,, \\
0& & &\text{otherwise}. \\
\end{aligned}
\right.
\end{equation}
This corresponds to measuring the integrated temperature fluctuation in a disk with radius $\theta_d$ and subtracting the same signal measured in a concentric ring of the same area around the disk,
as illustrated in Fig.~\ref{fig:ap_filters}.
Since our CMB maps have units of $\mu$K, the CAP output units are $\mu {\rm K}\cdot$arcmin$^2$.

As the disk radius $\theta_d$ is increased,
the CAP filter output behaves similarly to a cumulative (integrated) profile: for small disk radii, the output vanishes; for large radii, where all the gas profile is included inside the disk, the output is equal to the integrated gas profile.
Intuitively, the CAP filter profiles shown in this paper can thus almost be thought of as cumulative gas density/temperature profiles.

Since the CAP filter is compensated (i.e. $W$ integrates over area to zero),
it has the desirable property that fluctuations with wavelength longer than the filter size will cancel in the subtraction.
This significantly reduces the noise from degree-scale CMB fluctuations, and the correlation between the various CAP filter sizes.
This basically corresponds to band-pass filtering the temperature map before stacking.
However, it allows us to use a different band-pass filter with each CAP filter radius.

If the tSZ and kSZ profiles were known, a matched filter would be the minimum variance unbiased linear estimator of the profile's amplitude.
However, the profile is not known, and measuring it is the goal of our study. 
For this reason, we adopt the simple CAP filter, and vary its size $\theta_d$ between 1 and 6 arcmin.
This corresponds to approximately $0.5-4$ virial radii,
which are the physical scales relevant to study feedback and baryonic effects.
Beyond 6 arcmin, the kSZ CAP filter measurements become very highly correlated, due to the common degree-scale CMB fluctuations acting as the dominant noise. 
As a result, the kSZ SNR saturates at these large aperture values.

\subsection{Stacking}

For a given CAP filter radius $\theta_d$, we wish to combine the measured temperatures $\mathcal{T}_i(\theta_d)$ around each galaxy $i$.
Let us first assume that the CAP filter noise is independent from one galaxy to the other.
For tSZ, the minimum-variance unbiased linear estimator of the signal is simply the inverse-variance weighted mean:
\begin{equation}
    \hat{T}_{\rm tSZ}(\theta_d) = 
    \frac{\sum_i \mathcal{T}_i (\theta_d) / \sigma_i^2}{\sum_i 1 / \sigma_i^2}.
    \label{eq:tSZ_est},
\end{equation}
where $\sigma_i$ is the noise standard deviation for the CAP filter on galaxy $i$.
Because the detector and atmospheric noise in our maps is inhomogeneous,
the noise $\sigma_i$ on the CAP filter for galaxy $i$ depends on the galaxy $i$. We describe how we estimate it below.
For kSZ, the minimum-variance unbiased linear estimator is the velocity weighted, inverse-variance weighted mean:
\begin{equation}
    \hat{T}_{\rm kSZ}(\theta_d) = -
    \frac{1}{r_v}
    \frac{ v_{\rm rms}^{\rm rec}}{  c}
    \frac{\sum_i \mathcal{T}_i(\theta_d) (v_{{\rm rec}, i}/c) / \sigma_i^2}{\sum_i (v_{{\rm rec}, i}/c)^2 / \sigma_i^2}
    \label{eq:kSZ_est}
\end{equation}
where again $v_{\rm rms}^{\rm rec}$ refers to the rms of the radial component of the reconstructed velocity (computed from the catalog of reconstructed velocities), and the factor $r_v^{-1}$ ensures that the estimator is not biased by the imperfections in the velocity reconstruction.
The velocity weighting is crucial: without it, the kSZ signal would cancel in the numerator, since it is linear in the galaxy LOS velocities ($\mathcal{T}_i(\theta_d) \propto v$), which are equally likely to be pointing away or towards us.
With the velocity weighting, both numerator and denominator now scale as the mean squared velocity, which avoids the cancellation and selectively extracts the kSZ signal.

Interestingly, Eq.~\eqref{eq:kSZ_est} implies that the kSZ estimator is insensitive to any overall multiplicative rescaling of the velocities.
In practice, the noise on the CAP filters around two nearby galaxies are not necessarily uncorrelated, 
especially for the large apertures where the CAP filters can overlap.
This makes the stack estimators above slightly suboptimal but does not bias them.
Indeed, they are unbiased for any choice of the weights $\sigma_i$.
However, this has an impact on the noise covariance, which we discuss in detail in Sec~\ref{sec:covariance}.

The noise $\sigma_i$ receives contributions from the detector and atmospheric noise, but also the primary CMB and all other foregrounds.
Maps of the inverse (detector plus atmospheric) noise variance ``$ivar$'' per pixel are available for our coadded f90 and f150.
Since we also want to include the CMB and other foregrounds in $\sigma_i$, we do not simply use $\sigma_i^2 = 1/ivar$, but instead $\sigma_i^2 = f_{\theta_d}\left(1/ivar\right)$, where the function $f_{\theta_d}$ is determined empirically for each CAP filter radius $\theta_d$, by measuring the CAP filter variance at the galaxy positions in bins of $ivar$ and interpolating it.
Using the same measured CAP filters in the stack and to determine the weights $\sigma_i$ is not a problem here, since the tSZ and kSZ from galaxy $i$ only contributes $\sim 0.01\%$ of the variance of the CAP filter $\mathcal{T}_i$.
We repeat the same analysis separately on the f90 and f150 maps.

This approach is formally equivalent to measuring the cross-power spectrum of the temperature map with a template map, built by adding the velocities of all galaxies falling in a given map pixel \cite{2018arXiv181013423S}. 
The cross-power spectrum approach in Fourier space has the advantage of having more independent $\ell$ bins, but both approaches have the same SNR.
Furthermore, our goal is to learn about the configuration-space gas density and pressure profiles, so the configuration-space estimator is better suited to our purposes.

Individual mass estimates for the CMASS and LOWZ galaxies are publicly available 
\cite{2012MNRAS.421..314C, 2013MNRAS.435.2764M}
\footnote{\url{https://data.sdss.org/sas/dr12/boss/spectro/redux/galaxy/v1_1/}}
.
In principle, one could include an additional weight in the stack from the dependence of the kSZ and tSZ signals with mass.
We tried converting the stellar masses into halo masses and performing halo mass weighting (see Appendix~\ref{app:mass_distribution}), assuming that the gas mass scales linearly with halo mass.
However, we did not see the improvement in SNR expected from the mass distribution, and therefore do not adopt this approach here.
This is likely due to the scatter in any one or all of 
the stellar mass estimates, the stellar-to-halo mass relation and the halo to gas mass relation.
 
Our publicly available pipeline \texttt{ThumbStack}
\footnote{\url{https://github.com/EmmanuelSchaan/ThumbStack}}
implements the CAP filters, estimates the optimal weights $\sigma_i$, performs the stacking and estimates the covariance matrix.

\subsection{Covariance matrix}
\label{sec:covariance}

In order to interpret the measured kSZ and tSZ profiles, knowing the covariance of measurements at different apertures and on different maps is required.
For a map with uniform sensitivity, the covariance of different CAP filters can be computed analytically from the power spectrum of the map. 
However, the depth in our maps is non-uniform, making this more difficult. 
Furthermore, different maps (temperature maps f90 and f150 and component-separated maps) have some components in common (CMB, foregrounds) and some uncorrelated components (foreground decorrelation, detector and atmospheric noise), making the analytical calculation more complicated.
Another approach consists of running the stacking analyses on many realistic mock temperature maps. However, this requires mock maps with the correct correlation across maps and the correct noise non-uniformity within each map.

For these reasons, our fiducial covariance matrices are estimated by bootstrap resampling the individual galaxies.
Specifically, we draw with repetition from the galaxy catalog to generate a resampled galaxy catalog, with the same number of objects. From this resampled galaxy catalog, we measure the stacked tSZ and kSZ CAP profiles. We then repeat this process with a large number (10,000) of resampled galaxy catalogs, and infer the covariance matrices from the scatter across the corresponding resampled tSZ and kSZ stacked profiles.
This produces an unbiased estimate of the covariance, in the limit of independent noise realizations from galaxy to galaxy. 
The assumption of independent noise from one galaxy to another can fail if the projected galaxy number density is high enough that the CAP filters overlap.
We thus expect this issue to be worse for the larger apertures.
To check this, we use Gaussian mocks (to quantify the effect of aperture overlap, having the correct noise non-uniformity is not crucial).
We show in Appendix~\ref{app:mock_cov} that the bootstrap covariance is accurate to 10\%, which is sufficient for this analysis.
In Figs.~\ref{fig:kSZ_summary_cmass} and \ref{fig:tSZ_tilec_cmass}, we show the measured kSZ and tSZ stacked profiles along with their covariance matrices.
We have checked that the correlation matrix depends only on the map power spectrum.
It is thus identical for LOWZ and CMASS, and for the tSZ and kSZ estimators run on the same map.
Measurements at small apertures are dominated by the detector noise in the temperature maps.
Since this noise is mostly white and uncorrelated across frequencies, the various low aperture measurements are mostly uncorrelated within each map and across maps.
On the other hand, measurements at large apertures receive a larger contribution from the large-scale CMB fluctuations, which are shared across apertures and frequency maps.
As the aperture increases, the measurements become more and more correlated within each map and across maps, thus contributing less and less to the overall SNR.
This motivates our maximum aperture choice of $6'$ radius.

\subsection{Dust contamination to tSZ and kSZ}

Thermal emission from dust in our galaxy sample or galaxies spatially correlated with it can bias the inferred tSZ signal. 
Dust emission may contribute a positive signal to both the f150 and f90 maps, partially canceling the tSZ signal and thus biasing our inference about the circumgalactic gas.

To avoid this, the fiducial tSZ results in this paper are obtained from the CIB-deprojected ILC \cite{2019arXiv191105717M} $y$ map which deprojects any signal with the following frequency dependence:
\beq
f_\text{dust}(\nu)
=
\frac{\nu^{3+\beta}}{e^{h\nu/ (k_B T_\text{dust})}-1}
\left( \frac{dB}{dT}(\nu, T_\text{CMB}) \right)^{-1}.
\label{eq:tilec_dust_sed}
\eeq
This is a modified blackbody spectrum with temperature $T_\text{dust} = 24$ K and power law index $\beta=1.2$, converted from specific intensity to temperature units using the Planck function $B$. 
These parameter values were selected in \cite{2019arXiv191105717M} by fitting the SED of the mean CIB intensity, as predicted by the halo model in \cite{2014A&A...571A..30P}.
These parameters may appear to differ from \cite{2014A&A...571A..30P} for two reasons.
First, $T_d$ is redshift-dependent in \cite{2014A&A...571A..30P}, and the mean intensity is thus some average of it.
Second, $\beta$ and $T_d$ are somewhat degenerate in the fit of the mean SED, such that a change in the effective $T_d$ also changes the best fit $beta$.
Dust residuals in the CIB-deprojected ILC map may have either sign, and may be non-negligible if the assumed dust SED differs from the correct one.

We also pursue an independent approach, by measuring the stacked tSZ + dust signals in the f90 and f150 maps, and jointly modeling them with dust-dominated measurements from Herschel data.
In the companion paper \cite{paper2}, we use 161 deg$^2$ of overlapping Herschel data from the H-ATLAS \cite{eales+10} survey at 250 $\mu$m, 350 $\mu$m, and 500 $\mu$m. To simplify the modeling, we use the same CAP filters used for measuring the tSZ to measure the dust emission, and we refer the reader to \cite{paper2} for details of this analysis. Dust residuals in this method may also have either sign.

As shown in \cite{paper2}, the two independent methods to subtract the dust contamination recover the same tSZ profile. 
This is reassuring and suggests that the dust SED assumed in the CIB-deprojected ILC map is sufficient for our purposes.
A caveat is that both methods assume that the dust SED for all CMASS galaxies can be described by one single smooth SED, parameterized as a modified blackbody.
Testing this assumption would be a valuable project.

We do not expect the thermal dust emission to significantly contaminate the kSZ measurement.
It is zero on average when weighted with the galaxy velocities which have both positive and negative sign with equal probability, so the only residual signal could come from imperfect cancellation because of the finite number of galaxies in our sample. Dust is a small correction to the already small residual tSZ (see below), and for these reasons, we do not consider dust contamination to kSZ further. We note however that the Doppler boosting of the dust emission could in principle bias our measurements. An estimate for the size of the effect is in Section \ref{subsec:systematics} and we find it to be completely subdominant to the other sources of error. 
Galactic dust is not expected to be a major source of contamination since it is uncorrelated with our galaxy sample, and further suppressed by the galactic mask used.

\subsection{KSZ systematics and null tests}
\label{subsec:systematics}

In this section, we discuss in detail the various systematic effects affecting the kSZ estimates.

The filtering pipeline and estimators are thoroughly tested on simulated maps with known profiles and velocities to ensure a correct measurement. This includes testing the effects of pixelation, interpolation, reprojection and subpixel weighting, as well as the estimators themselves. 
These tests are discussed in detail in Appendices~\ref{app:pipeline_description} and~\ref{app:pipeline_tests_2halo}.
They show that the present pipeline is accurate to sub-percent level, and is therefore appropriate for this measurement as well as upcoming ones from Simons Observatory \cite{2019JCAP...02..056A} and CMB-S4 \cite{2019JCAP...02..056A}.

One important concern is the potential leakage from tSZ to the kSZ estimator.  
Since the tSZ signal is independent of the galaxy's peculiar velocity, it vanishes {\it on average} when weighting galaxies by their velocities as in the kSZ estimator (Eq. \ref{eq:kSZ_est}). 
However, because the tSZ signal scales steeply with mass ($\propto M^{5/3}$ in the self-similar regime), a few massive clusters can dominate it. Since they are rare, the cancellation due to the velocity weighting is only approximate, potentially causing a large residual tSZ contamination to kSZ.
In \cite{Schaan:2015uaa} for example, we masked the 1,000--3,000 most massive galaxies (as inferred by their measured stellar masses), in order to keep the tSZ contamination to less than 10\% of the kSZ signal.
In Appendix~\ref{app:foreground_outliers}, we estimate this tSZ leakage to kSZ, and find it to be smaller than 10\% of the signal and 10\% of the noise for both CMASS and LOWZ. 
To be prudent, and facilitate the interpretation of the signal, we keep the maximum halo mass cut of $10^{14} M_\odot$, similar to \cite{Schaan:2015uaa}.
In practice, we  perform this cut by rejecting any galaxy with stellar mass larger than $5.5\times 10^{11}M_\odot$, which corresponds to a halo mass of $10^{14} M_\odot$ in the mean stellar-to-halo mass relation \cite{2018AstL...44....8K}.
This discards 1,399 CMASS K galaxies (out of 312,708), 8,617 CMASS M galaxies (out of 368,701) and 11,013 LOWZ K galaxies (out of 145,714).

Another caveat is that any emission from our tracers, including thermal dust emission, is also Doppler boosted by the peculiar motion of the galaxies. 
To lowest order, this is proportional to the LOS galaxy velocity, i.e. $\delta T_\text{Doppler dust} = \delta T_\text{dust} v/c$, just like the kSZ signal.
The Doppler-boosted dust emission $\delta T_\text{Doppler dust}$ would then bias the kSZ estimator, just like the usual dust emission $\delta T_\text{dust}$ biases the tSZ estimator.
However, since $v/c \sim 10^{-3}$, we know that the Doppler-boosted dust emission $\delta T_\text{Doppler dust}$ is smaller than the usual dust emission $\delta T_\text{dust}$ by three orders of magnitude.
Furthermore, our statistical error bars on tSZ and kSZ are very similar (e.g., in $\mu$K$\cdot$arcmin$^2$).
Therefore, for the Doppler-boosted dust emission to be a $1\sigma$ bias to kSZ, 
the usual (non-Doppler boosted) dust emission would have to be a $1000\sigma$ bias to tSZ.
If so, it would completely overwhelm the measured tSZ signal, turning the observed temperature decrements into very large increments, which are not seen.
For this reason, we know that the Doppler-boosted dust is several orders of magnitude subdominant to kSZ.
We thus neglect this effect here, and simply note that it may be an interesting signal \textit{per se} at higher frequency.

Figure~\ref{fig:boss_cmb_schematic} highlights the large correlation length of the velocity fields ($\sim 100$ Mpc/$h$) and the fact that the BOSS survey contains a finite and relatively small number of independent velocity regions. 
However, because the kSZ estimator Eq.~\eqref{eq:kSZ_est} is a ratio of velocities, the cosmic variance of the velocity field does not affect the measurement.
However, the small number of independent velocity regions implies that the cancellation of foregrounds in the kSZ estimator is imperfect. 
We show that it is sufficient for our purposes in Appendix~\ref{app:foreground_outliers}.

Furthermore, correctly interpreting the measured kSZ and tSZ profiles requires a detailed knowledge of the halo occupation distribution (HOD) of our galaxy sample.
For instance, a large offset of the CMASS galaxies from the center of the gas profiles would artificially extend the size of the observed gas profiles.
If, for example, a significant fraction of CMASS galaxies were satellites in more massive halos, the observed gas profiles would also be affected.
We discuss these issues in Appendix~\ref{app:foreground_outliers} and in greater detail in \cite{paper2}.

Finally, the measured tSZ and kSZ signals also contain a ``2-halo'' term, from other gas correlated with our tracer sample.
An estimate of this effect is given in Appendix~\ref{app:pipeline_tests_2halo}, and a first-principle calculation is included in the modeling in \cite{paper2}.

\subsection{Results: kSZ \& tSZ profiles}
\label{subsec:results}

In this section, we present the measured kSZ, tSZ and tSZ+dust CAP profiles for the CMASS and LOWZ galaxies, along with the relevant covariance matrices. 
We take CMASS K and LOWZ K as our fiducial sample and we compare the results to the CMASS M sample in Appendix~\ref{app:null_tests}.

To assess the significance of the measurements, we use the $\chi^2$ statistic, defined as:
\beq
\chi^2_\text{model} = (\text{data}-\text{model})^t\, \text{Cov}^{-1}\, (\text{data}-\text{model}),
\label{eq:def_chi_squared}
\eeq
where ``model'' stands for either the null hypothesis (producing $\chi^2_\text{null}$), a baryon profile following the dark matter ($\chi^2_\text{dark matter}$), or the best-fit profile ($\chi^2_\text{best fit}$) \cite{paper2}.
The various models used for the best fit curves and the fitting method are described in detail in \cite{paper2}.
To compute the significance of the rejection of the null hypothesis, we convert the measured $\chi^2_\text{null}$ into a probability to exceed (PTE) such a high chi squared value, given the number of data points. We then express this PTE in terms of equivalent Gaussian standard deviations $\sigma$.
To compute the significance of the preference of the best fit model over the null hypothesis, we simply compute
\beq
\text{SNR}_\text{model}
\equiv
\sqrt{\Delta \chi^2_\text{null - best fit}}=
\sqrt{\chi^2_\text{null} - \chi^2_\text{best fit}}.
\label{eq:snr}
\eeq
This quantity corresponds to the SNR on the amplitude of a free amplitude multiplying the best fit profile.
It therefore corresponds to the detection significance of the best fit profile.
The SNR values for the various maps and catalogs are summarized in Table~\ref{tab:snr}.
For the tSZ+dust stacks, 
the best fit models from \cite{paper2} exclude the smallest aperture where the dust contamination is a large fraction of the signal, as described in Sec.~III of \cite{paper2}. 
This smallest aperture is therefore not included in the $\chi^2_\text{null}$ and $\text{SNR}_\text{model}$ for tSZ+dust.
It is however included for kSZ, where dust contamination is negligible.
\begin{table}[H]
\centering
\begin{tabular*}{0.95\columnwidth}{@{\extracolsep{\fill}}| l | c c c | c |}
\hline
Stack & $\chi^2_\text{null}$ & dof & PTE\,$\sigma$ & SNR$_\text{model}$ \\
\hline
\hline
CMASS kSZ & 86.2 & 18 dof & \bf{6.5\,$\sigma$} & \bf{7.9\,$\sigma$} \\
CMASS tSZ & 131.8 & 9 dof & \bf{10.1\,$\sigma$} & \bf{11.0\,$\sigma$} \\
CMASS tSZ+dust & 421.6 & 16 dof & \bf{18.9\,$\sigma$} & \bf{19.7\,$\sigma$} \\
\hline
\hline
LOWZ kSZ & 38.3 & 18 dof & \bf{2.9\,$\sigma$} & - \\
LOWZ tSZ & 229.7 & 9 dof & \bf{13.9\,$\sigma$} & -  \\
LOWZ tSZ+dust & 330.3 & 16 dof & \bf{16.4\,$\sigma$} & - \\
\hline
\end{tabular*}
\caption{
For each stacked measurement, we quote $\chi^2_\text{null}$ (Eq.~\eqref{eq:def_chi_squared}), which quantitifes the rejection of the null hypothesis. Given the number of data points (dof), we convert $\chi^2_\text{null}$ into a probability to exceed (PTE), expressed in terms of number of Gaussian sigmas.
This quantifies the significance at which the ``no-signal'' hypothesis is rejected.
We further quote the model SNR from Eq.~\eqref{eq:snr}. This quantifies the significance of the preference of the best fit model to the no-signal hypothesis.
The kSZ and tSZ+dust stacks correspond to the joint fits to the DR5 f90 and f150 fits (the first data point weas excluded in the fitting process for the tSZ+dust stacks, see \cite{paper2}); the tSZ stacks correspond to the DR4 ILC $y$ map with deprojected CIB.
This explains the lower significance of the tSZ measurements.
We only model the profiles for the simpler CMASS sample \cite{paper2}, so the model SNR is only available for these stacks.
}
\label{tab:snr}
\end{table}

\subsubsection{CMASS}

\begin{figure}[h!!]
\includegraphics[width=0.9\columnwidth]{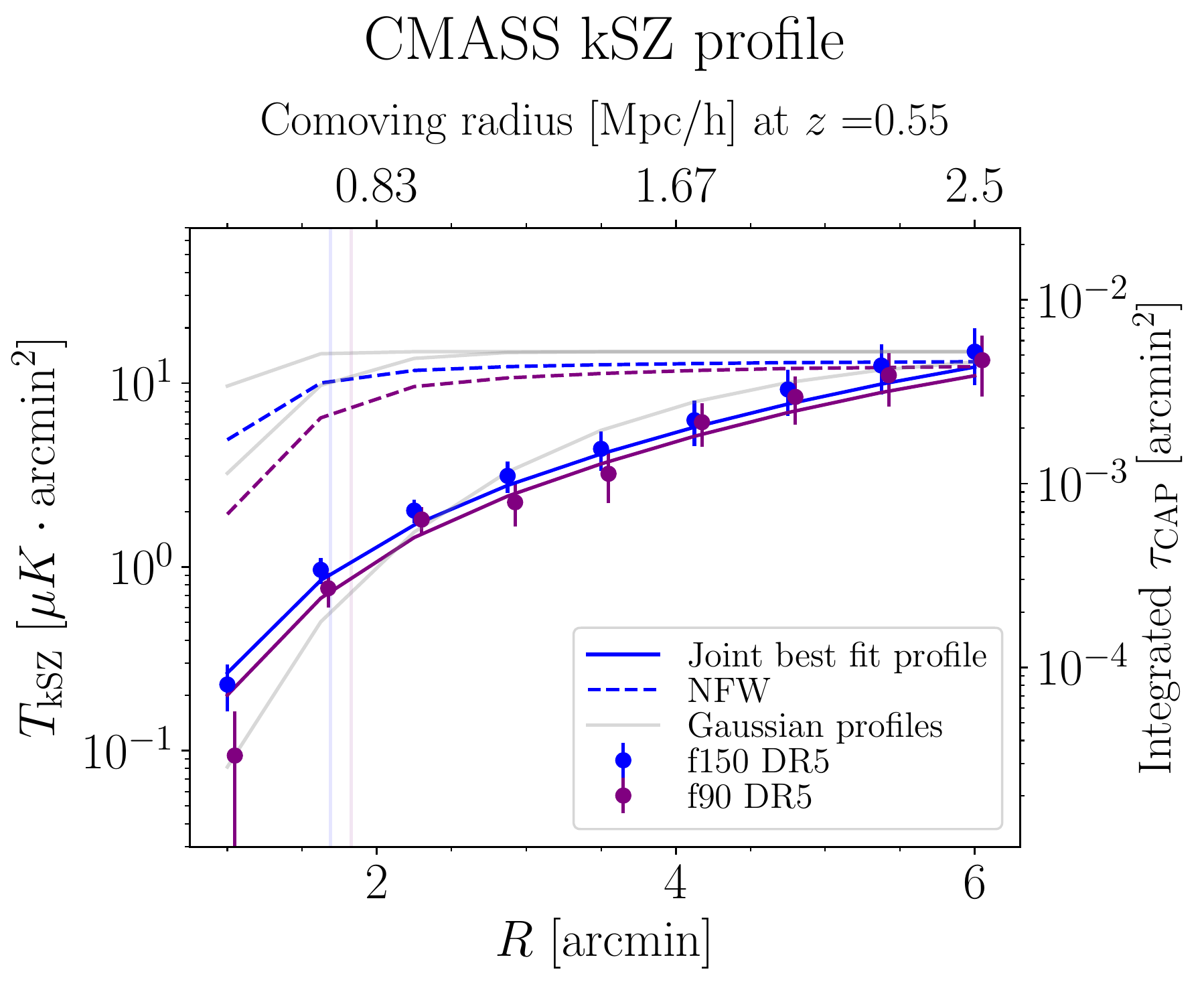}
\includegraphics[width=0.9\columnwidth]{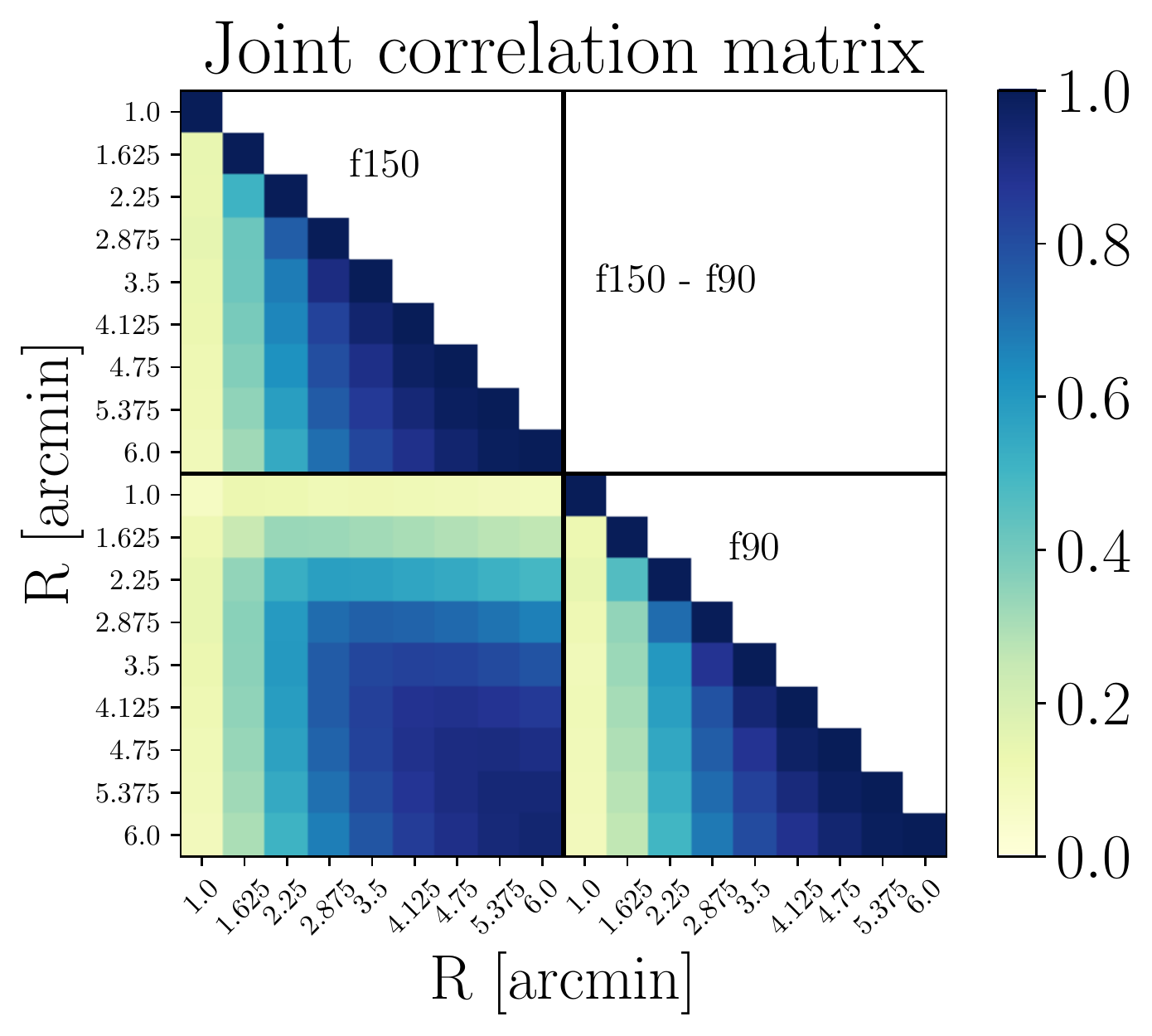}
\caption{
\textbf{Top:} The mean CMASS kSZ signal in each compensated aperture photometry filter with radius $R$ (see Eq.~\eqref{eq:kSZ_est}),
obtained by stacking the single-frequency temperature maps f90 and f150.
The joint best fit kSZ profile from \cite{paper2}, convolved with the beams of f90 and f150, is shown in solid lines.
The kSZ signal is detected at 7.9 $\sigma$ 
(i.e. $\text{SNR}_\text{model} = \sqrt{\Delta \chi^2} = 7.9$).
The dashed lines show the expected kSZ signal if the gas followed the dark matter (NFW) profile (convolved with the beams and CAP filters).
The data show that the electron profile is more extended than the dark matter profile at very high significance ($\sqrt{\chi^2_\text{NFW} - \chi^2_\text{best fit}} = 96$).
The vertical lines show the halo virial radius ($1.6'$ at $z=0.55$) added in quadrature with the beam standard deviations ($\sigma=\text{FWHM}/\sqrt{8\ln 2}= 0.55'$ in f150 and 0.89' in f90).
To guide the eye, the gray solid lines correspond to Gaussian profiles with $\text{FWHM}=1.3'$ (f150 beam), $\text{FWHM}=2.1'$ (f90 beam) and $\text{FWHM}=6'$ (similar to the measured profile) from left to right. They are normalized to match the largest aperture in f150.
The y-axis on the right converts the measured kSZ signal into the CAP optical depth to Thomson scattering, which counts the number of free electrons within the CAP filter.
Null tests are shown in Figs.~\ref{fig:pipe_null_tests_cmass} and \ref{fig:fg_null_tests_ksz_cmass}.
\textbf{Bottom panel:} correlation matrix between the different CAP filters and frequencies.
}
\label{fig:kSZ_summary_cmass}
\end{figure}

Focusing first on CMASS, the stacked kSZ profiles from the f90 and f150 temperature maps are shown in Fig.~\ref{fig:kSZ_summary_cmass}.
The kSZ signal is detected at 7.9 $\sigma$.
The best fit theory profiles derived in \cite{paper2} match the measurements in f90 and f150, taking into account the differing beams of these two maps.
This fit of the theory profile takes into account the range of CMASS host halo masses.
The 2-halo term is also included in the theory curves, although its contribution is not significantly detected \cite{paper2}.
The best-fit model is a good fit to the data, 
with $\chi^2_\text{best fit} = 23.6$ for 17 degrees of freedom, i.e. $\text{PTE} = 0.13$.

For comparison, the dashed lines in Fig.~\ref{fig:kSZ_summary_cmass} show the expected kSZ signal if the gas followed the dark matter.
Specifically, these curves are computed by assuming a Navarro-Frenk-White (NFW) profile \cite{1996ApJ...462..563N} for the dark matter in each of the CMASS halos.
To guide the eye, Fig.~\ref{fig:kSZ_summary_cmass} also shows the CAP profiles for three Gaussian profiles (grey lines).
The first two are point sources convolved with Gaussians with $\text{FWHM}=1.3', 2.1'$, corresponding to the beams in f150 and f90, respectively.
The last one is a Gaussian profile with $\text{FWHM}=6'$, chosen because it resembles the measured CAP profile.
This shows that the dark matter profiles would be barely resolved, being close to point sources.
In contrast, the measured profile is much more similar to the Gaussian profile with $\text{FWHM}=6'$, showing that the actual gas profile is well-resolved, and much more extended.

The kSZ CAP profile, in the case where the gas would follow the dark matter, is computed as follows.
For each CMASS halo, we use the individual halo mass estimate and redshift to infer the corresponding NFW profile, using the mass-concentration relation from \cite{2008MNRAS.390L..64D}.
The 3D NFW profile is truncated at one virial radius, such that the total mass enclosed is exactly one virial mass.
The NFW matter density profile is then converted to number density of free electrons (assuming cosmological baryon abundance, and a fully ionized gas with primordial helium abundance), then convolved with the beams in f90 and f150, and propagated through the CAP filters.
The assumption of a fully ionized gas ignores the $5\text{--}10\%$ of the baryons in the form of stars or other neutral gas \cite{1998ApJ...503..518F}.
The resulting CAP profiles are finally averaged over all the individual halo mass estimates in the catalog.
In summary, the dark matter dashed lines are not a fit to the data, but rather a prediction based on the individual host halo masses and redshifts of the CMASS galaxies.
In particular, they do not correspond to a single NFW profile, but to the average of many NFW profiles.
The measured electron density CAP profile, from the kSZ measurement, lies well below the predicted NFW lines at the smaller apertures.
Because these are CAP filters, this result indicates that the NFW profile is much steeper than the gas profile, i.e. the gas profile is much more extended than the dark matter profile.
Indeed, the dark matter profile is highly discrepant with the data, with an extremely high $\chi_\text{dark matter}^2 = 9344$, compared to the expected $\sim 18$ for 18 degrees of freedom. 
This indicates a very poor fit, i.e. a very strong rejection of the hypothesis that the gas follows the dark matter.
Similarly, the hypothesis that the gas follows the best fit profile is preferred over the hypothesis that the gas follows the dark matter at 97\,$\sigma$, in the sense that $\sqrt{\chi^2_\text{dark matter} - \chi^2_\text{best fit}} = 96.54$.
In fact, because the dark matter prediction is so high, even the hypothesis of no kSZ signal is preferred over the hypothesis that the gas follows the dark matter at 96\,$\sigma$, i.e.
$\sqrt{\chi^2_\text{dark matter} - \chi^2_\text{no gas}} = 96.45$.
This is still completely compatible with the best fit profile being preferred over the null at 7.9\,$\sigma$.

The rejection of the hypothesis that the gas follows the dark matter is robust when relaxing a number of analysis assumptions.
First, we truncated the NFW profiles at one virial radius. 
However, undoing this truncation would increase the predicted dark matter profile at large apertures ($> 2.5'$), making it even more inconsistent with the data.
Second, miscentering between the positions of the CMASS galaxies and the halo centers could smooth the measured profile, making it look artificially more extended.
However, this miscentering was estimated to be $\sim 0.2'$ \cite{2012ApJ...757....2G}, much too small to reconcile the dark matter profile with the data.
Finally, this analysis does not account for the $\sim 15\%$ of CMASS galaxies which could be satellites in more massive halos. This could alter the predicted dark matter profile, but would likely enhance it.

We also assumed that the total baryon mass in CMASS halos matches the cosmic mean.
One may therefore wonder if the discrepancy with the measured kSZ profile can be alleviated by lowering the baryon mass in CMASS halos.
To answer this question, we define a free parameter $f_\text{c}$ corresponding to the ratio of free electrons in CMASS halos to the expectation based on the cosmic mean.
We then simply rescale the predicted profile by $f_\text{c}$, and evaluate the $\chi^2$ statistics.
We find that 
$\sqrt{\chi^2_\text{dark matter} - \chi^2_\text{best fit}} = 97, 45, 25$
for
$f_\text{c} = 1, 0.5, 0.3$.
A baryon mass as low as one $0.3$ times the cosmic mean would thus still be rejected at very high significance.
In fact, the best fit amplitude is $f_\text{c} = 0.065$, requiring a baryon mass more than 15 times smaller than the cosmic mean.
This model is still rejected at 4~$\sigma$ 
(i.e. $\sqrt{\chi^2_\text{dark matter} - \chi^2_\text{best fit}} = 4.2$).
Given that the mass of CMASS host halos is known to 4\% from galaxy lensing \cite{2015ApJ...806....1M}, this would require a ratio of baryon to dark matter mass more than 15 times smaller than the cosmic mean, highly unlikely.

One could in principle reduce the discrepancy between the data and the NFW profile by allowing the NFW normalization or concentration to vary.
In practice, our result that the gas does not follow the dark matter is robust to this.
Indeed, reducing the normalization of the dark matter to match the smaller apertures would amount to dividing the halo mass by more than ten, which is excluded from the lensing mass estimates and our individual halo mass estimates.
Even then, the larger apertures would still be discrepant.
One would need to change the mass-concentration relation by a large factor, on top of the unphysical total halo mass.

In summary, while our measurement (Fig.~\ref{fig:kSZ_summary_cmass}) is only a 7.9\,$\sigma$ detection of the kSZ effect, it is sufficient to reject the hypothesis that the electrons follow the dark matter at much higher significance, $>90$\,$\sigma$
\footnote{This can be understood intuitively by considering the smaller apertures. There, our data constrains the kSZ signal with a precision of $\sim 10\%$ (for a $\sim$10 sigma detection). Because the NFW profile overpredicts the kSZ signal by a factor $\sim$10, it is rejected by the data with a significance of $\sim 10\times 10=100$\,$\sigma$.}.
This can be understood since the dark matter profile would produce a much higher kSZ signal at low apertures, which is not seen.
This is a key result of this paper. 
It shows that even a modest significance kSZ measurement contains high significance information about the gas profile.

We convert the kSZ temperatures into integrated optical depth to Thomson scattering in the CAP filter via
$T_\text{kSZ} = \tau_\text{CAP} T_\text{CMB} (v_\text{rms}^\text{true} / c)$,
with $T_\text{CMB} = 2.726 K$ and $v_\text{rms}^\text{true} = 313$ km/s at $z=0.55$, according to linear theory.
The resulting values are shown on the y axis of Fig.~\ref{fig:kSZ_summary_cmass}.
In Appendix~\ref{app:consistency_schaan+16}, we confirm the consistency of this kSZ measurement with our previous one from \cite{Schaan:2015uaa}, where we used the same galaxy sample and a smaller map with higher noise. The increase in SNR shown in Fig.~\ref{fig:consistency_schaan+16} is striking.

\begin{figure}[h]
\centering
\includegraphics[width=0.9\columnwidth]{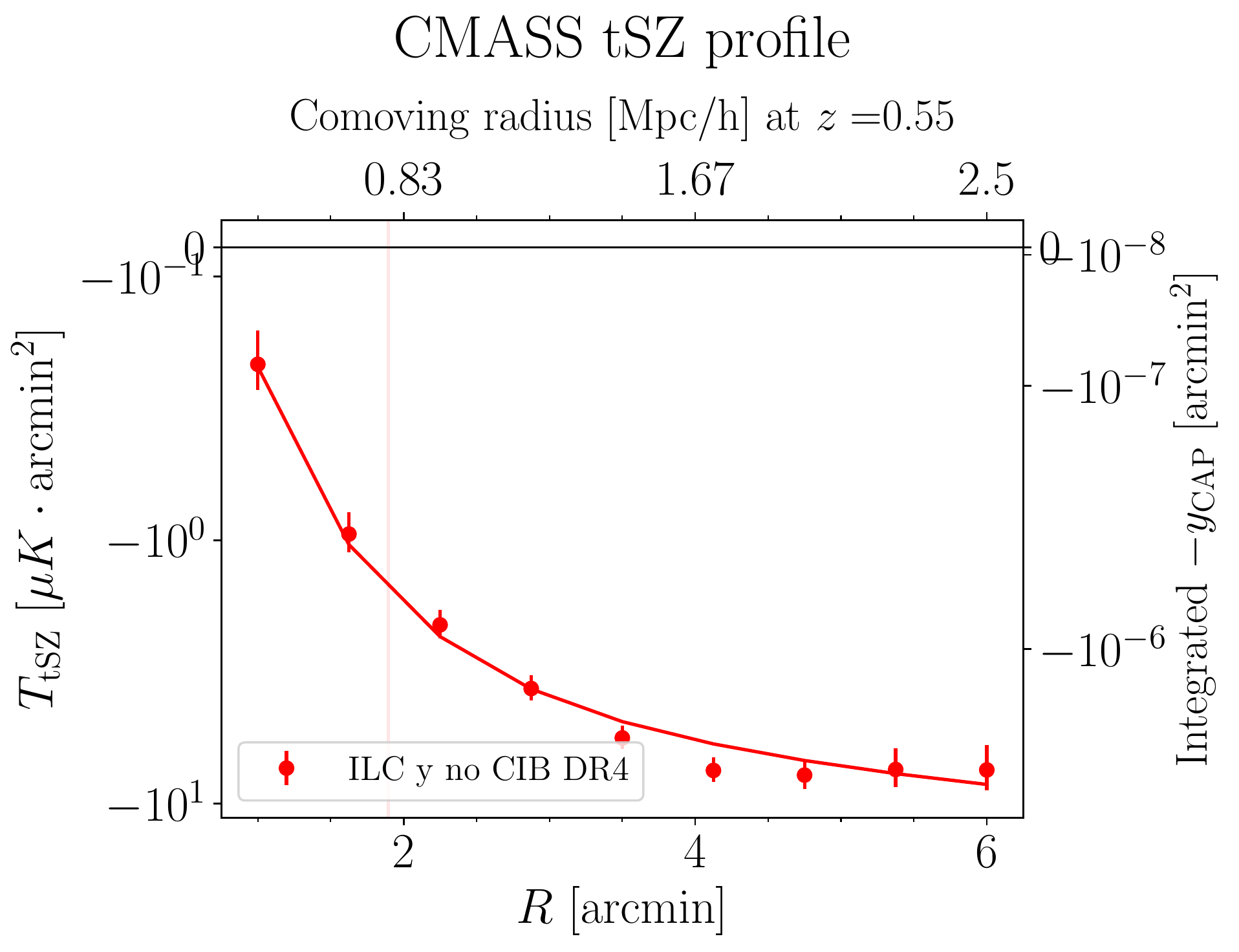}
\includegraphics[width=0.9\columnwidth]{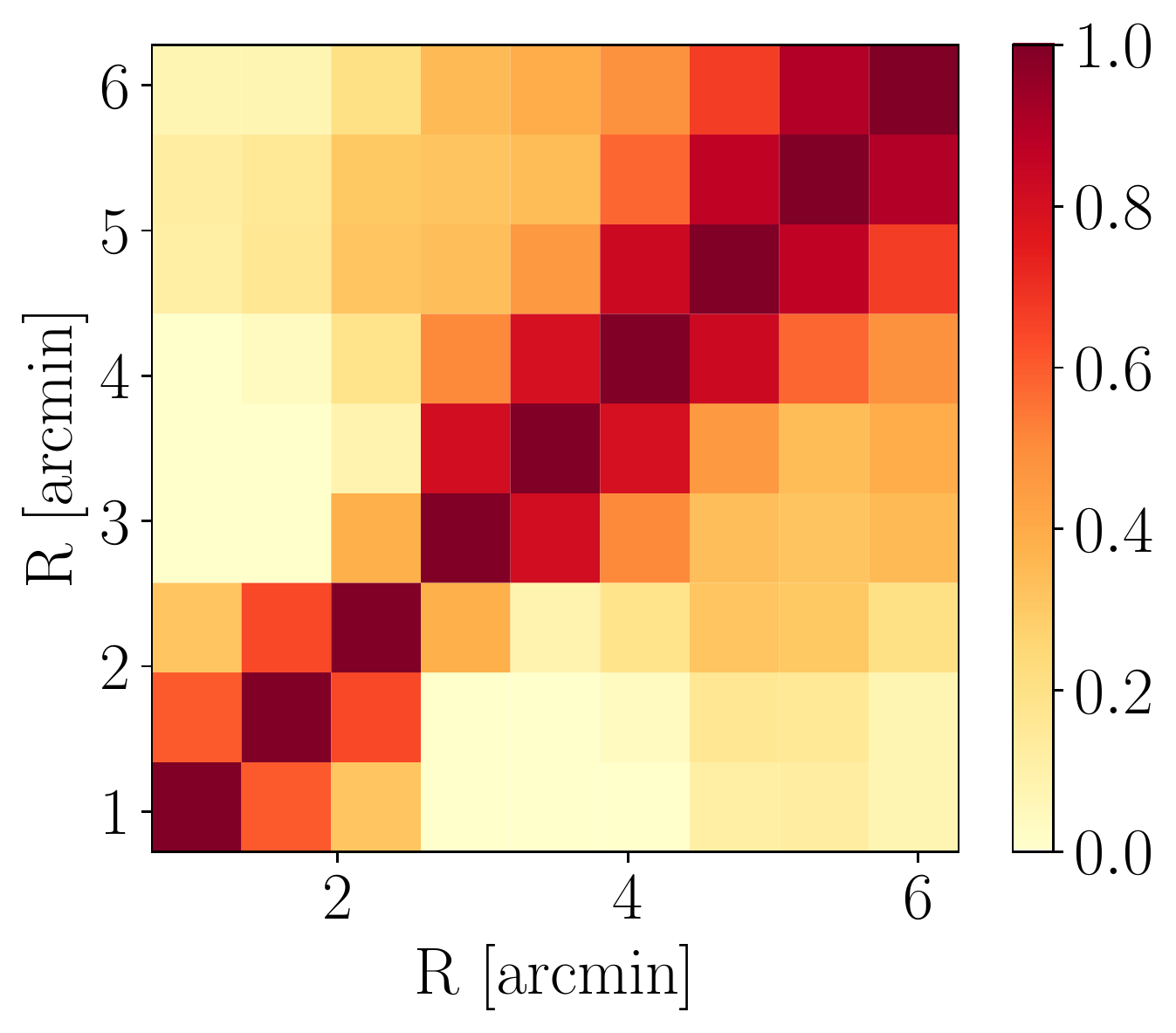}
\caption{\textbf{Top panel:}
The mean CMASS tSZ signal in all compensated aperture photometry filters, measured from the ILC $y$ map with deprojected dust, as defined in Equation \ref{eq:tSZ_est}.
The solid line shows the best fit tSZ profile, from \cite{paper2}.
The $y$ profiles were converted to $\mu$K at 150 GHz with $f_\text{tSZ}(\nu=150 \text{GHz}) T_\text{CMB} = -2.59 \times 10^{6} \mu$K, to allow the reader to compare the tSZ and kSZ signal amplitudes.
The tSZ signal is detected at 11 $\sigma$ 
(i.e. $\text{SNR}_\text{model} = \sqrt{\Delta \chi^2} = 11$).
The vertical line shows the halo virial radius ($1.6'$ at $z=0.55$) added in quadrature with the beam standard deviation ($\sigma=\text{FWHM}/\sqrt{8\ln 2}= 1.5'$ in f150 and 0.89' in f90). \textbf{Bottom panel:} correlation matrix between the different CAP filters.
}
\label{fig:tSZ_tilec_cmass}
\end{figure}

Our fiducial tSZ profile is obtained by stacking on the ILC Compton-$y$ map with deprojected dust, as explained above.
Figure~\ref{fig:tSZ_tilec_cmass} shows that it is detected at 11 $\sigma$.
The best-fit tSZ model, presented in \cite{paper2},
is a good fit to the data:
$\chi^2_\text{best fit} = 9.8$ for 8 degrees of freedom, i.e. $\text{PTE} = 0.28$.
In Fig.~\ref{fig:tSZ_tilec_cmass}, we show the tSZ signal both in units of Compton $y$ and temperature decrement at 150 GHz, to allow the reader to compare the amplitudes of the kSZ, tSZ and tSZ+dust signals in the same unit.

\begin{figure}[h!!]
\centering
\includegraphics[width=0.9\columnwidth]{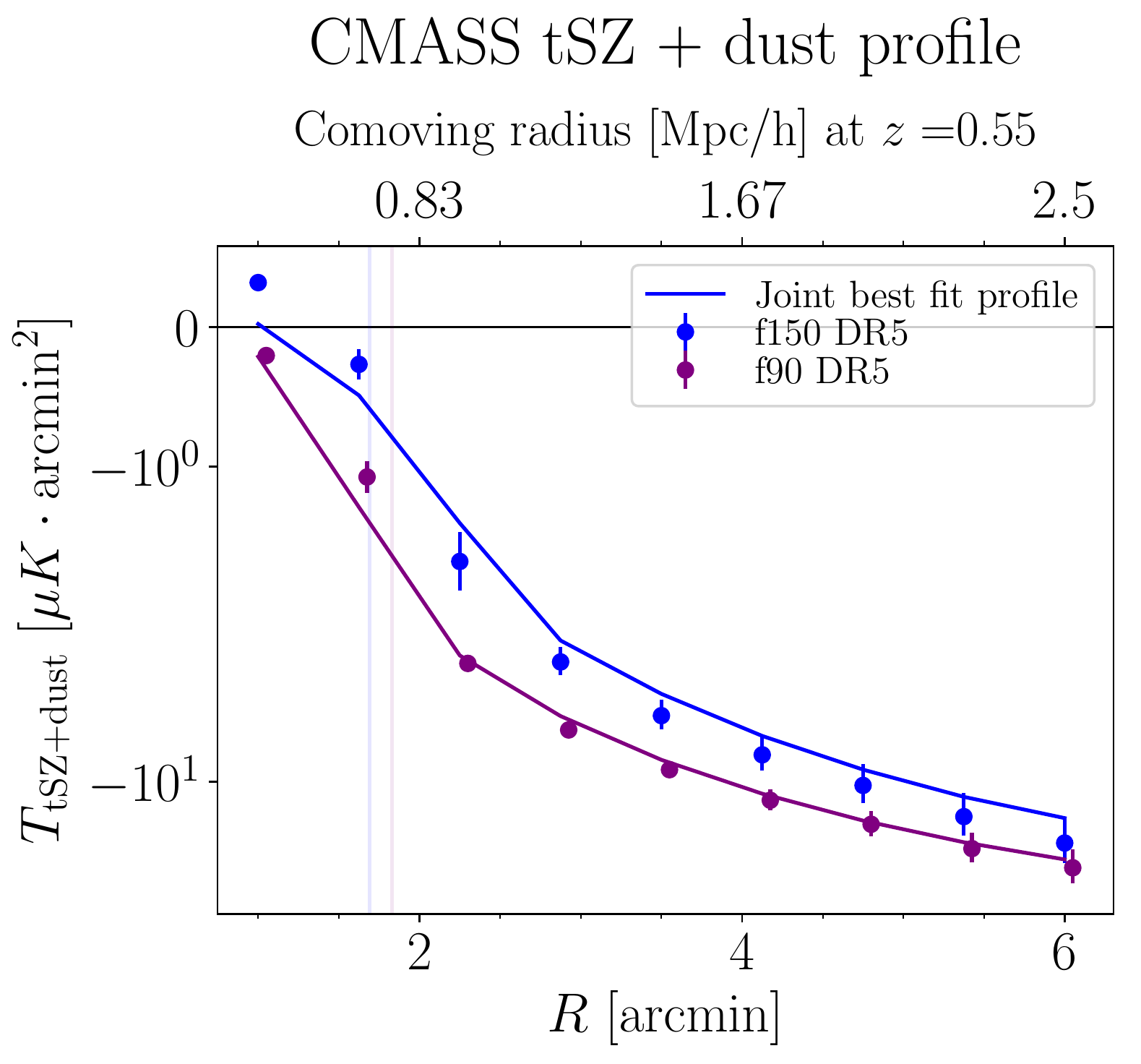}
\caption{
Mean tSZ + dust signal in all compensated aperture photometry filters, as defined in Equation \ref{eq:tSZ_est}.
These were obtained by stacking on the single-frequency temperature maps f90 and f150.
The best joint fit tSZ+dust profile to the f90, f150 and Herschel data from \cite{paper2} is shown at these frequencies in solid lines.
The no-signal hypothesis is rejected at 18.9~$\sigma$ (see Table~\ref{tab:snr}).
The impact of dust emission is seen in the difference between these profiles and Fig.~\ref{fig:tSZ_tilec_cmass}, not at the large apertures where the noise is different, but at the smallest apertures where the dust signal fills in the tSZ decrement (causing even a ``negative tSZ decrement'' at 150 GHz).
The vertical lines show the halo virial radius ($1.6'$ at $z=0.55$) added in quadrature with the beam standard deviations ($\sigma=\text{FWHM}/\sqrt{8\ln 2}= 0.55'$ in f150 and 0.89' in f90).
The correlation matrix for the different CAP filters and frequencies is identical to Fig.~\ref{fig:kSZ_summary_cmass}.
}
\label{fig:tSZ_coadd_cmass}
\end{figure}

Using the single frequency temperature maps f90 and f150, we measure the tSZ + dust profiles, shown in Fig.~\ref{fig:tSZ_coadd_cmass}.
In \cite{paper2}, these are used in combination with Herschel measurements to jointly fit for the tSZ and dust signals.
Once corrected for the dust emission, they are found to be consistent with our tSZ-only measurement (see Fig.~4 in \cite{paper2}).
In particular, as described in Sec.~III.C of \cite{paper2},
the best-fit tSZ + dust model is a good fit to the f90, f150 and Herschel data, with $\text{PTE}=0.45$.

\begin{figure}[h]
\centering
\includegraphics[width=0.9\columnwidth]{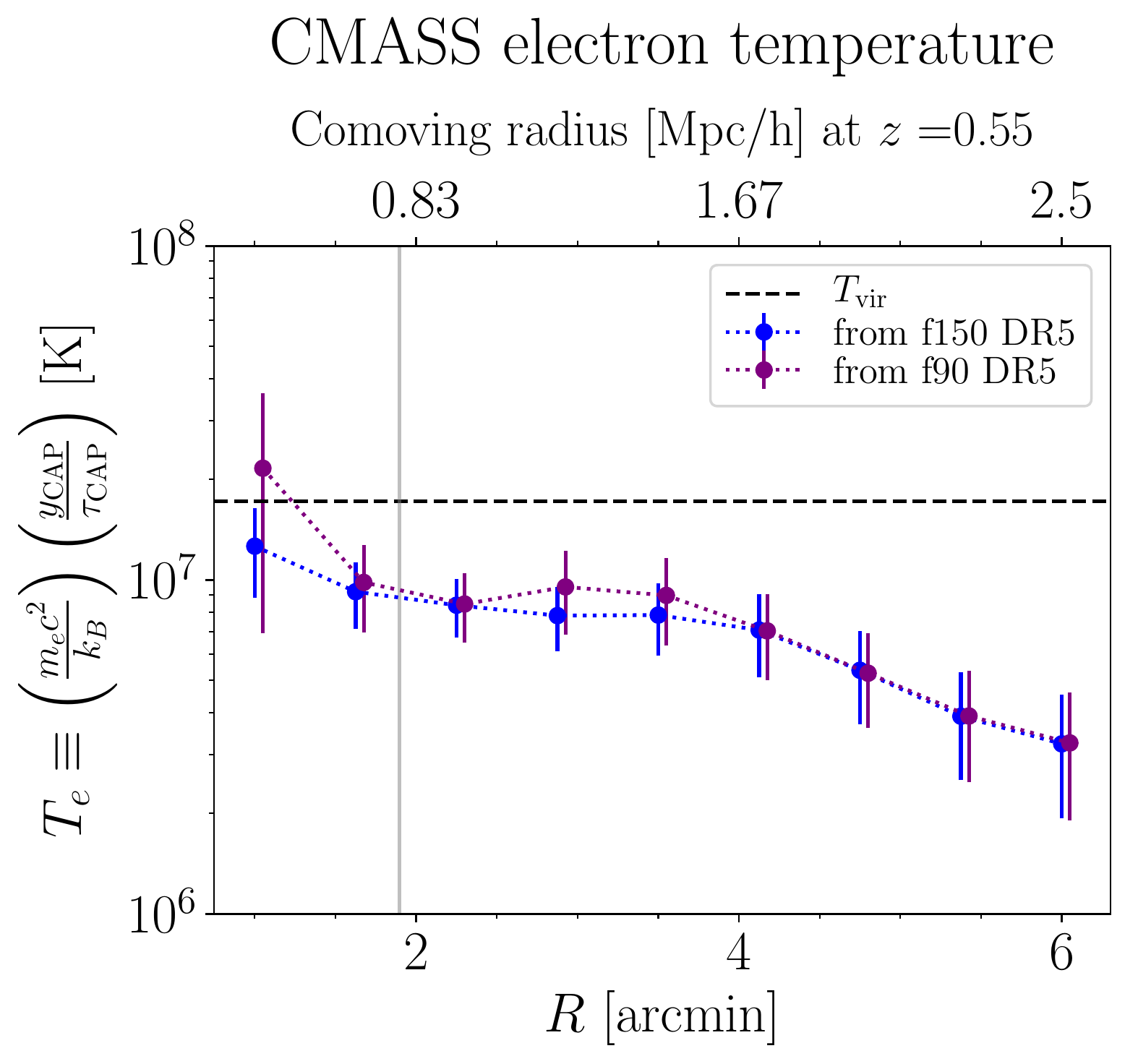}
\caption{
Simplified measurement of the electron temperature around CMASS galaxies. 
For each CAP radius, the electron temperature is simply estimated as the ratio of the tSZ and kSZ measurements (Eq.~\eqref{eq:electron_temperature}).
In comparison, the horizontal dashed line shows the virial temperature estimate for CMASS halos, whose order of magnitude is consistent with the data.
The vertical solid gray line shows the virial radius of the CMASS galaxies ($1.6'$ at $z=0.55$), added in quadrature with the beam standard deviation ($\sigma=\text{FWHM}/\sqrt{8\ln 2}= 1.0'$) of the ILC map with deprojected CIB.
The dotted lines simply connect the data points.
}
\label{fig:electron_temperature_cmass}
\end{figure}

Finally, from the electron pressure $n_e k_B T_e$ information (tSZ data) and the electron number density $n_e$ information (kSZ data), one can estimate the mean electron temperature $T_e$ per CAP filter.
We leave the careful modeling of the electron temperature to \cite{paper2}, and only show a simplified measurement here.
Since $y = \tau \left(\frac{k_B T_e}{m_e c^2}\right)$, we simply estimate the electron temperature as \cite{2020ApJ...889...48L, 2020arXiv200711583L}:
\beq
T_e 
\equiv
\left(\frac{m_e c^2}{k_B}\right)
\left(\frac{y_\text{CAP}}{\tau_\text{CAP}}\right).
\label{eq:electron_temperature}
\eeq
Several caveats are in order.
To form a meaningful ratio, we want $y_\text{CAP}$ and $\tau_\text{CAP}$ to be measured on maps with the same beam. We therefore reconvolved the f90 and f150 maps to the wider beam of the ILC maps with deprojected CIB, from which $y_\text{CAP}$ was measured.
If the CAP filters were simply disk averages, this estimate would be the mean electron temperature, weighted by the electron number density.
Instead, the CAP filters are the difference between the integral in a disk and an adjacent ring of equal area. 
As a result, this estimate is equal to the mean electron temperature only if the temperature is uniform over the whole CAP filter.
Furthermore, being the ratio of two noisy quantities, this estimate is biased high by the noise on the denominator $\tau_\text{CAP}$.
In practice though, we have checked that this is less than a 5\% fractional bias, and is therefore negligible compared to the statistical error.
Nevertheless, it provides a useful order of magnitude for the electron temperature in the CMASS galaxy groups.
To gain more intuition, we compare the measured electron temperature to the expected virial temperature:
\beq
T_\text{vir}
=
\frac{\zeta}{3}
\frac{\mu m_P}{k_B}
\left( \frac{\mathcal{G} M_\text{vir}}{r_\text{vir}} \right),
\eeq
where the parameter $\zeta\sim 1$ depends on the exact density and temperature profile \cite{2010gfe..book.....M}, and $\mu \approx 1.14$ is the mean mass per proton (including electrons and neutrons). 
For the typical assumptions most often adopted in the literature (primordial abundance of helium, and a singular isothermal sphere of gas for which $\zeta=3/2$), this gives \cite{2010gfe..book.....M}:
\beq
T_\text{vir}
=
3.5\times 10^5 K \; \left( \frac{M_\text{vir}}{10^{11} M_\odot} \right)^{2/3},
\eeq
i.e. $T_\text{vir} = 1.7\times10^7 K$ for CMASS and $2.2\times 10^7 K$ for LOWZ.
Since the exact virial temperature depends on the specific shape of the density profile,
we do not expect the measured temperature to match it exactly, but this still provides a rough order of magnitude.
Indeed, the measured temperature, shown in Fig.~\ref{fig:electron_temperature_cmass}, matches this order of magnitude.

\begin{figure}[h!]
\centering
\includegraphics[width=0.49\columnwidth]{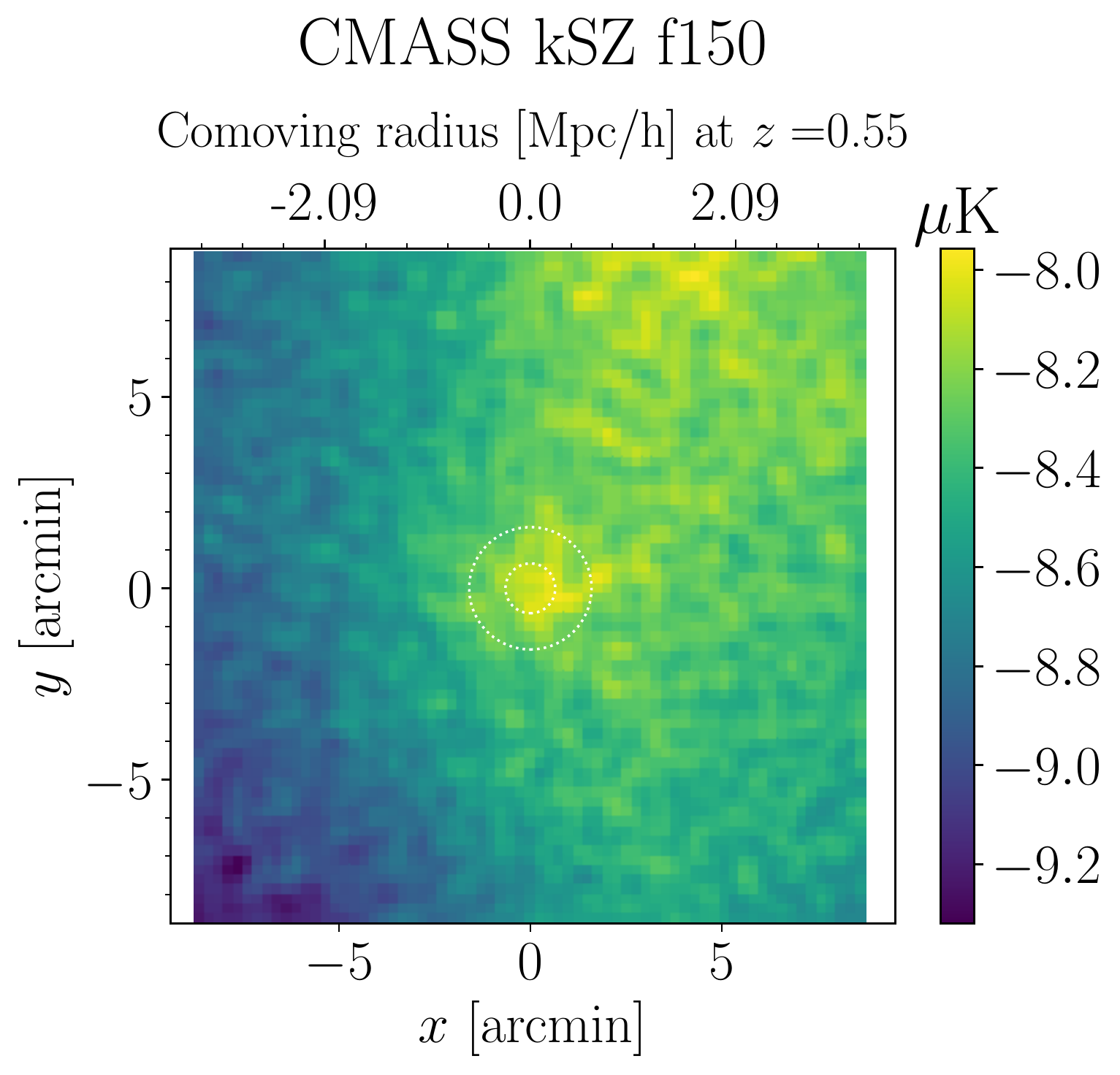}
\includegraphics[width=0.49\columnwidth]{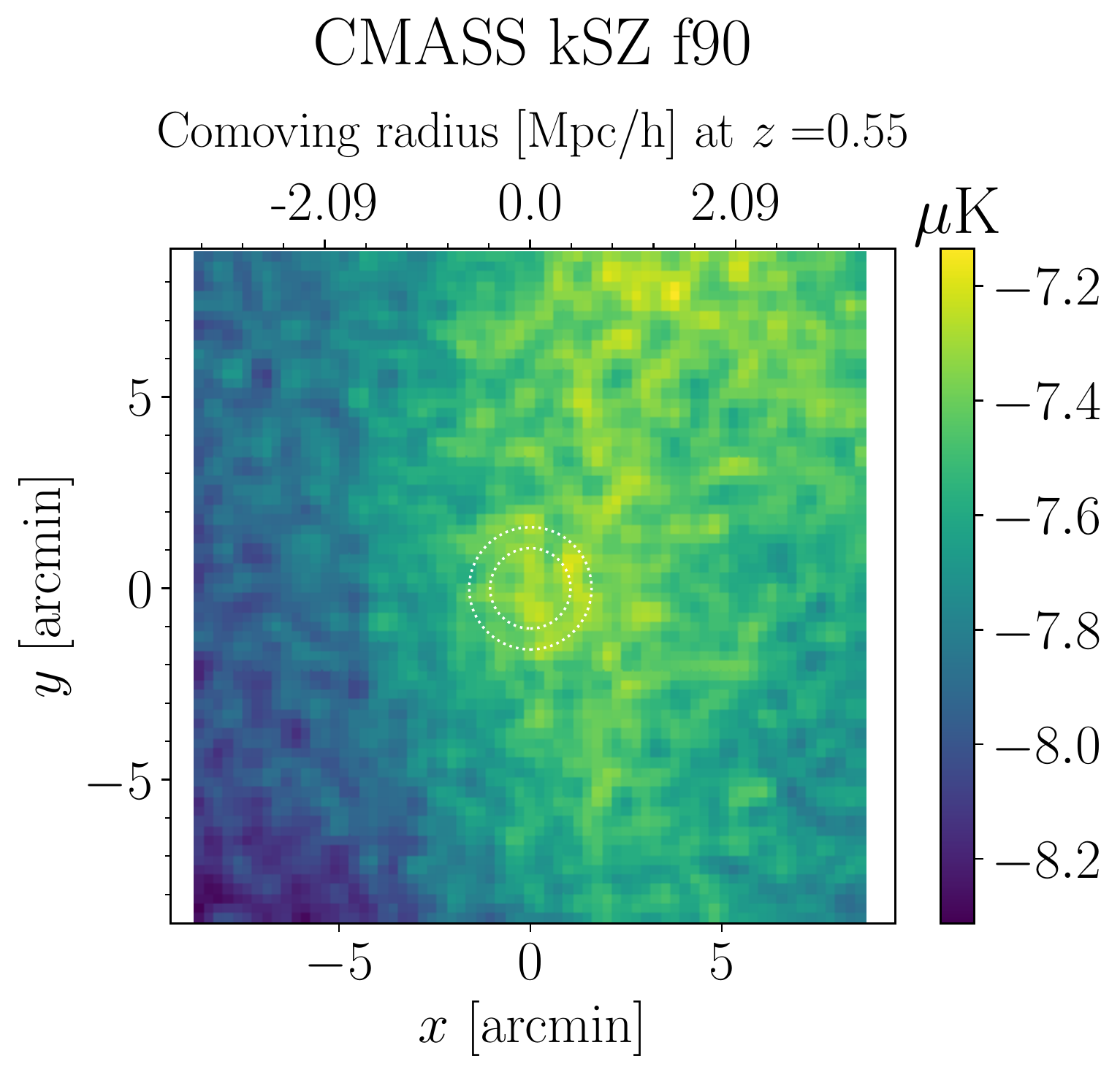}
\includegraphics[width=0.49\columnwidth]{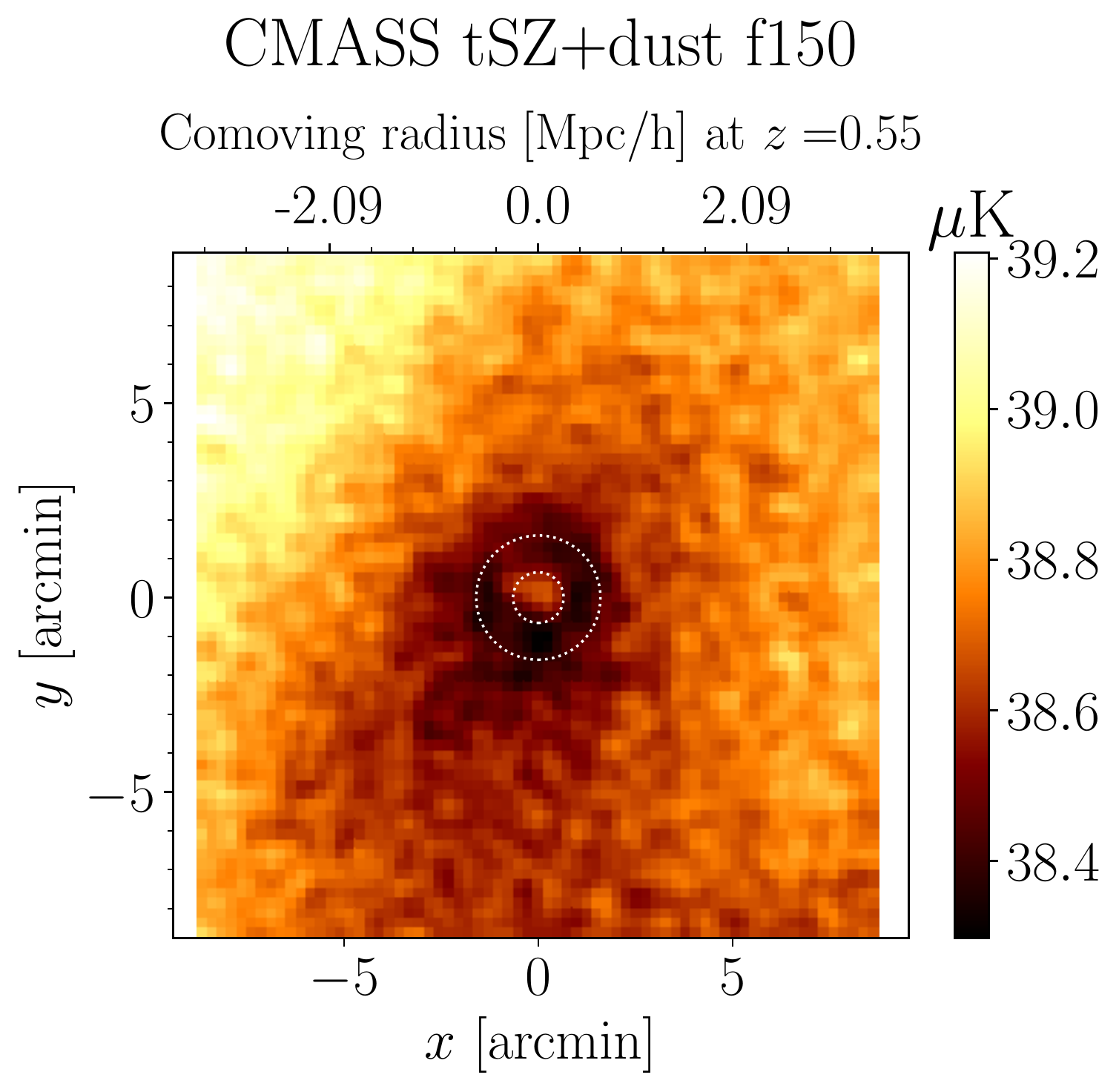}
\includegraphics[width=0.49\columnwidth]{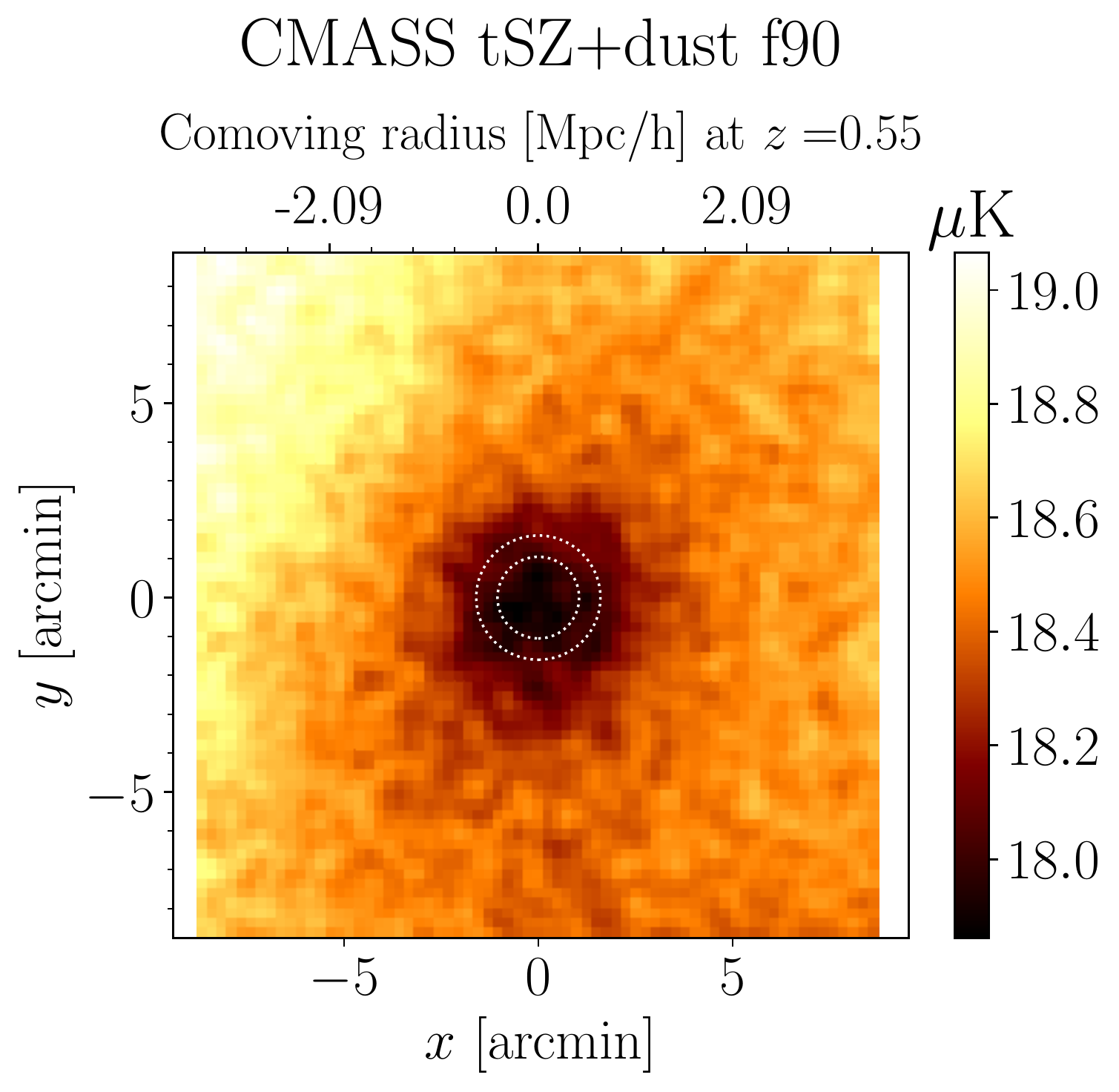}
\includegraphics[width=0.49\columnwidth]{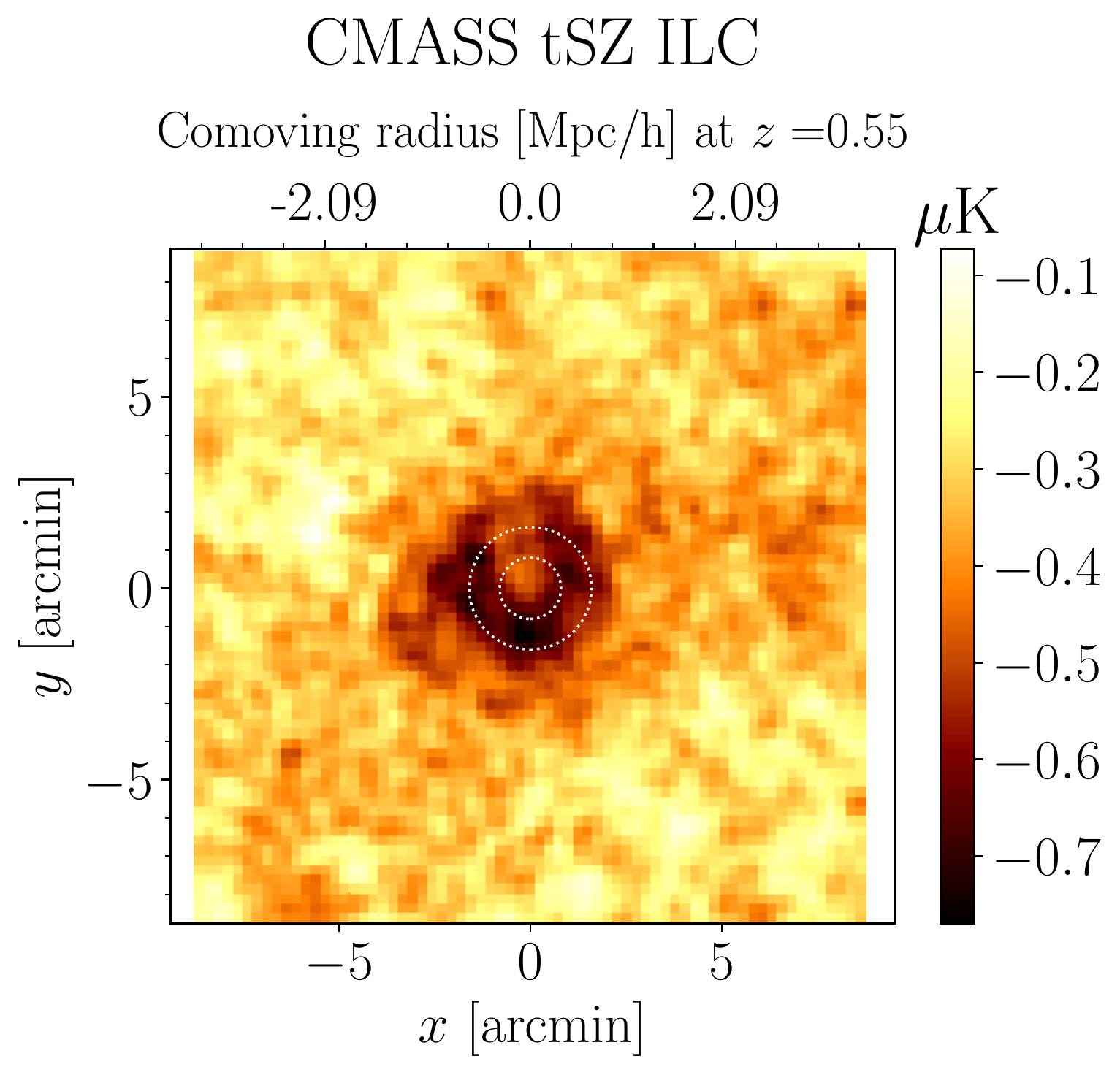}
\includegraphics[width=0.49\columnwidth]{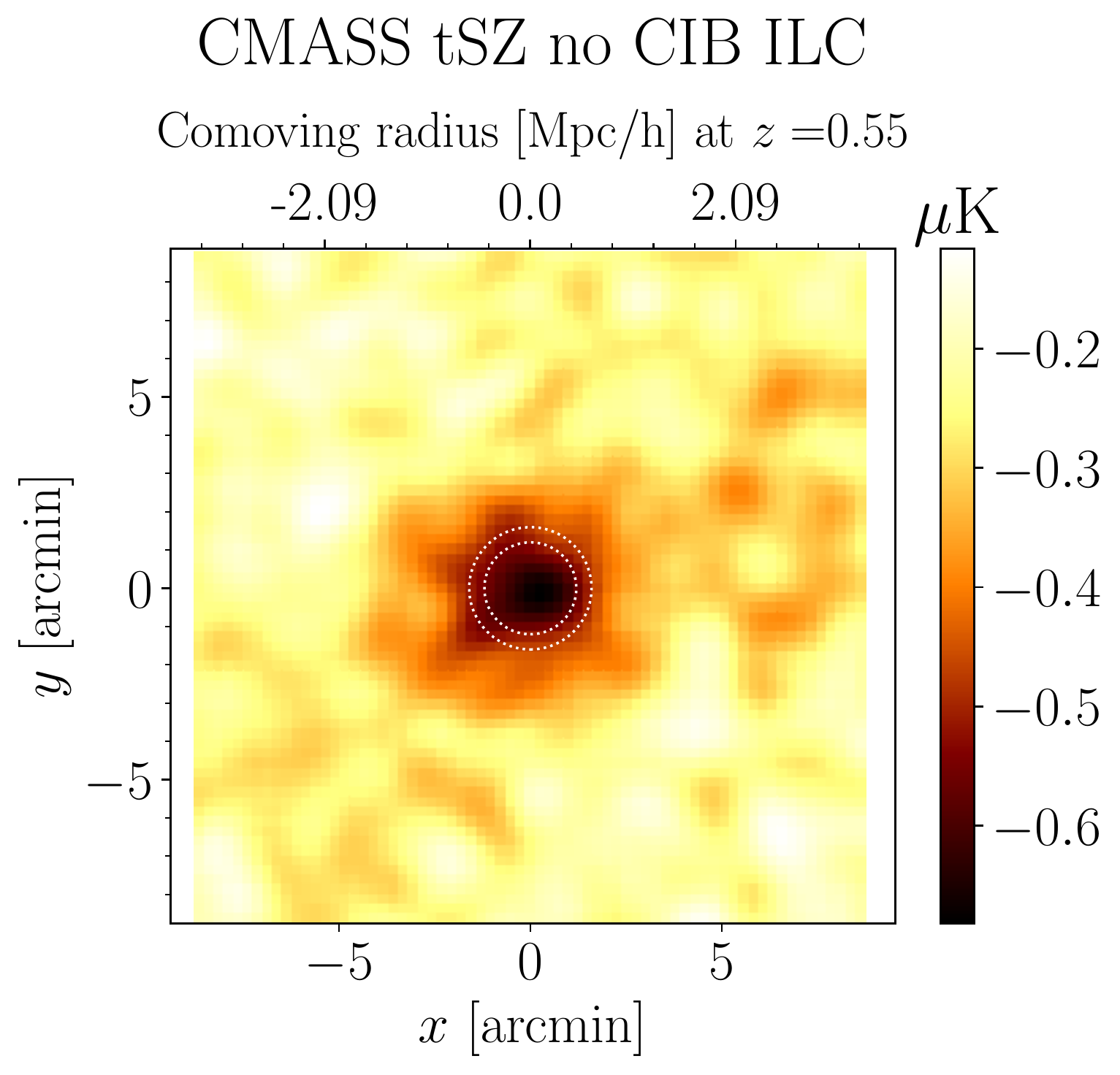}
\caption{
Stacked map cutouts showing the kSZ (top), tSZ+dust and tSZ (middle and bottom): the resolution and sensitivity of ACT allow to image the gas density, pressure, and the dust emission from CMASS objects.
In every cutout, the inner dotted circle has a diameter equal to the beam FWHM, and the outer dotted circle has a radius equal to the Virial radius.
No spatial filtering was applied (other than the beam convolution).
In all cases, the profiles are resolved (wider than the beam) and detectable by eye.
The dust emission fills in the tSZ decrement in f150 and in the tSZ ILC, but is not as visible in f90 and appears absent by eye in the tSZ no CIB ILC.
The dust emission profile is visibly narrower than the gas pressure and density profiles.
}
\label{fig:stacked_maps_cmass}
\end{figure}

To illustrate visually the measurements we have performed, 
we show the stacked 2d map cutouts corresponding to the kSZ, tSZ+dust and tSZ measurements above in Fig.~\ref{fig:stacked_maps_cmass}.
These were obtained by applying Eq.~\eqref{eq:tSZ_est} and \eqref{eq:kSZ_est} to the cutout maps around each CMASS object, as opposed to the CAP filter outputs.
In particular, no spatial filtering (CAP filter or otherwise) was applied. This sacrifices SNR but allows us to show the gas density, pressure and dust profiles without any distortion (apart from the beam convolution).
In \cite{VavagiakisInPrep}, similar 2d cutouts are shown for a different galaxy sample and as a function of luminosity. 
There, the submaps are inverse-variance weighted and normalized to reduce the appearance of large-scale noise.
These images illustrate that the gas density and pressure profiles are resolved:
the inner dotted circle, whose diameter is the beam FWHM, is smaller than the outer dotted circle, whose radius is the Virial radius for the mean CMASS mass and redshift.
They clearly show a dust profile in f150 and the Compton-$y$ ILC map, filling in the tSZ temperature decrement and less extended than the gas pressure profile.
This dust emission is reduced in f90 and apparently removed in the Compton-$y$ ILC deprojecting a fiducial CIB spectral energy density.
We reiterate that these stacked 2d map cutouts are only presented for illustration purposes, and are not used in the quantitative analysis.

\subsubsection{LOWZ}
 
\begin{figure}[h!]
\centering
\includegraphics[width=0.9\columnwidth]{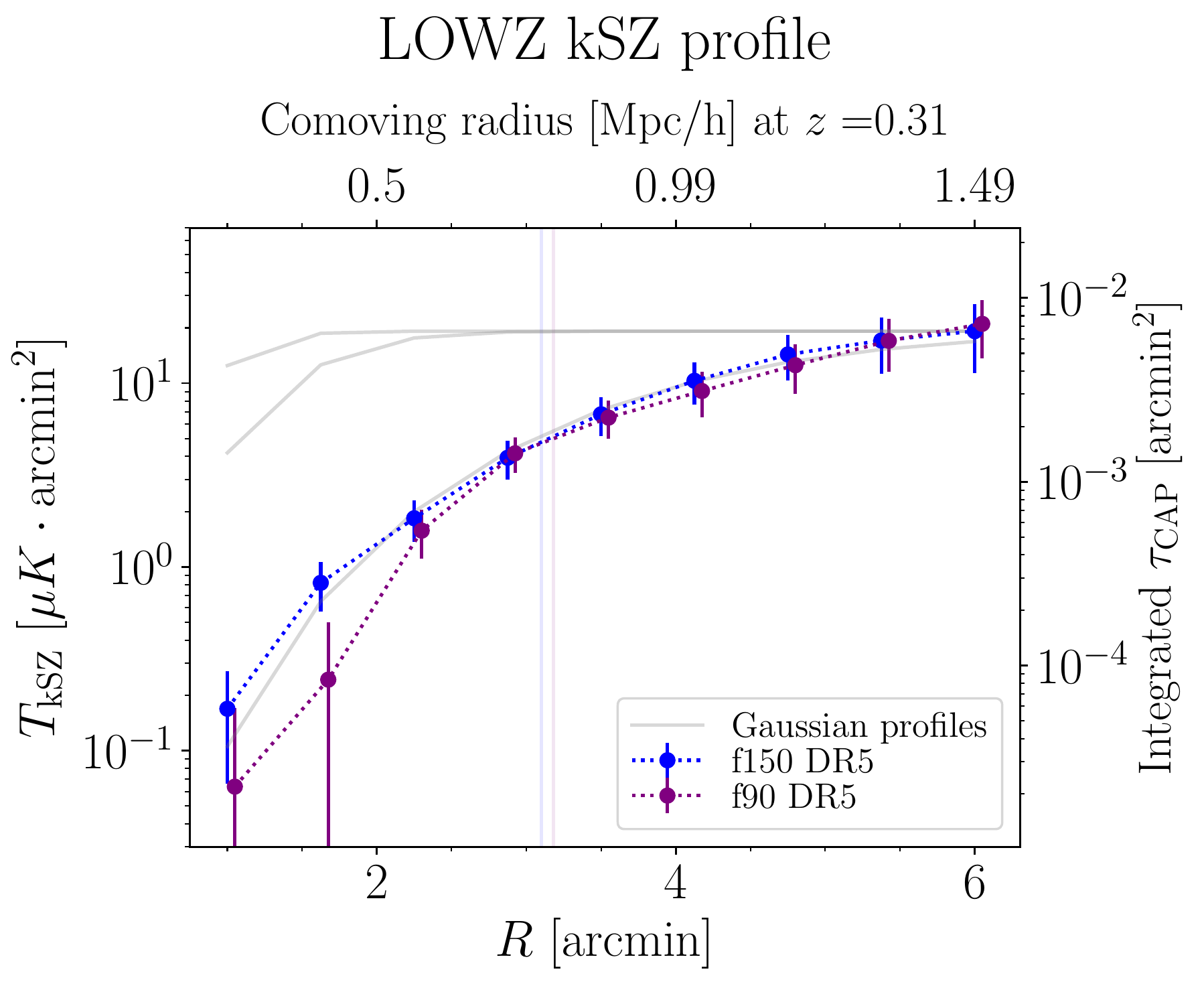}
\caption{
LOWZ kSZ profiles from the coadded maps f150 and f90.
The dotted lines simply connect the points to guide the eye.
The no-kSZ hypothesis is rejected at 2.9$\sigma$ (see Table~\ref{tab:snr}).
The vertical lines show the halo virial radius ($3.1'$ at $z=0.31$) added in quadrature with the beam standard deviations ($\sigma=\text{FWHM}/\sqrt{8\ln 2}= 0.55'$ in f150 and 0.89' in f90).
To guide the eye, the gray lines correspond to Gaussian profiles with $\text{FWHM}=1.3'$ (f150 beam), $\text{FWHM}=2.1'$ (f90 beam) and $\text{FWHM}=6'$ (similar to the measured profile) from left to right. They are normalized to match the largest aperture at 150 GHz.
Null tests are shown in Figs.~\ref{fig:pipe_null_tests_lowz} and \ref{fig:fg_null_tests_ksz_lowz}.
The y-axis on the right converts the measured kSZ signal into the CAP optical depth to Thomson scattering, which counts the number of free electrons within the CAP filter.
The correlation matrix for the different CAP filters and frequencies is identical to Fig.~\ref{fig:kSZ_summary_cmass}.
}
\label{fig:kSZ_summary_lowz}
\end{figure}

Turning to the LOWZ sample, we show the kSZ profiles from f90 and f150 in Fig.~\ref{fig:kSZ_summary_lowz}.
The LOWZ sample is known from clustering \cite{2014MNRAS.441...24A}  to have a more complex halo occupation distribution (HOD) than CMASS, and we do not attempt to model it precisely in \cite{paper2}.
We simply present the measurements here, so that they can be useful for future analyses.
Because we do not model the LOWZ measurements, we do not quote a detection significance (preference of the best fit model over the null hypothesis).
Instead, in Table \ref{tab:snr} we simply quote the significance at which the null hypothesis is rejected, based on $\chi^2_\text{null}$.
As for the CMASS sample, we convert the LOWZ kSZ measurements into integrated optical depth to Thomson scattering in the CAP filter, this time using $v_\text{rms}^\text{true} = 320$ km/s at $z=0.31$, according to linear theory.
The $\tau_\text{CAP}$ values are shown on the y axis of Fig.~\ref{fig:kSZ_summary_lowz}.

\begin{figure}[h!]
\centering
\includegraphics[width=0.9\columnwidth]{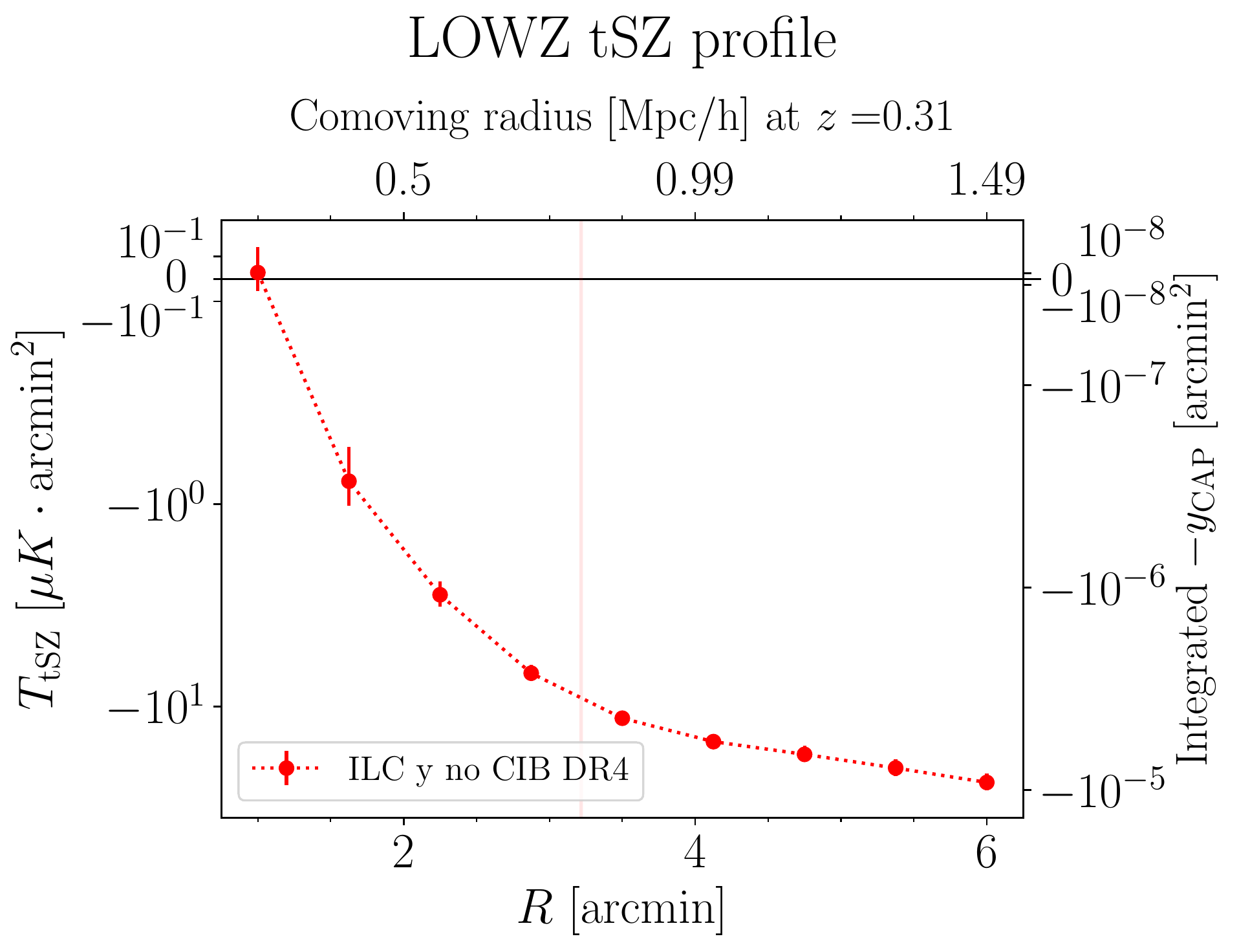}
\caption{
The mean LOWZ tSZ signal in all compensated aperture photometry filters, as defined in Equation \ref{eq:tSZ_est}.
The $y$ map was converted to $\mu$K at 150 GHz with $f_\text{tSZ}(\nu=150 \text{GHz}) T_\text{CMB} = -2.59 \times 10^{6} \mu$K, to allow the reader to compare the tSZ and kSZ signal amplitudes.
The dotted line simply connects the points to guide the eye.
The no-tSZ hypothesis is rejected at 13.9$\sigma$ (see Table~\ref{tab:snr}).
The vertical line shows the halo virial radius ($3.1'$ at $z=0.31$) added in quadrature with the beam standard deviation ($\sigma=\text{FWHM}/\sqrt{8\ln 2}= 1.0'$).
The correlation matrix for the different CAP filters is identical to Fig.~\ref{fig:tSZ_tilec_cmass}.
}
\label{fig:tSZ_tilec_lowz}
\end{figure}

The fiducial tSZ profile from the ILC $y$ map with deprojected dust is shown in Fig.~\ref{fig:tSZ_tilec_lowz}.
 
\begin{figure}[h]
\centering
\includegraphics[width=0.9\columnwidth]{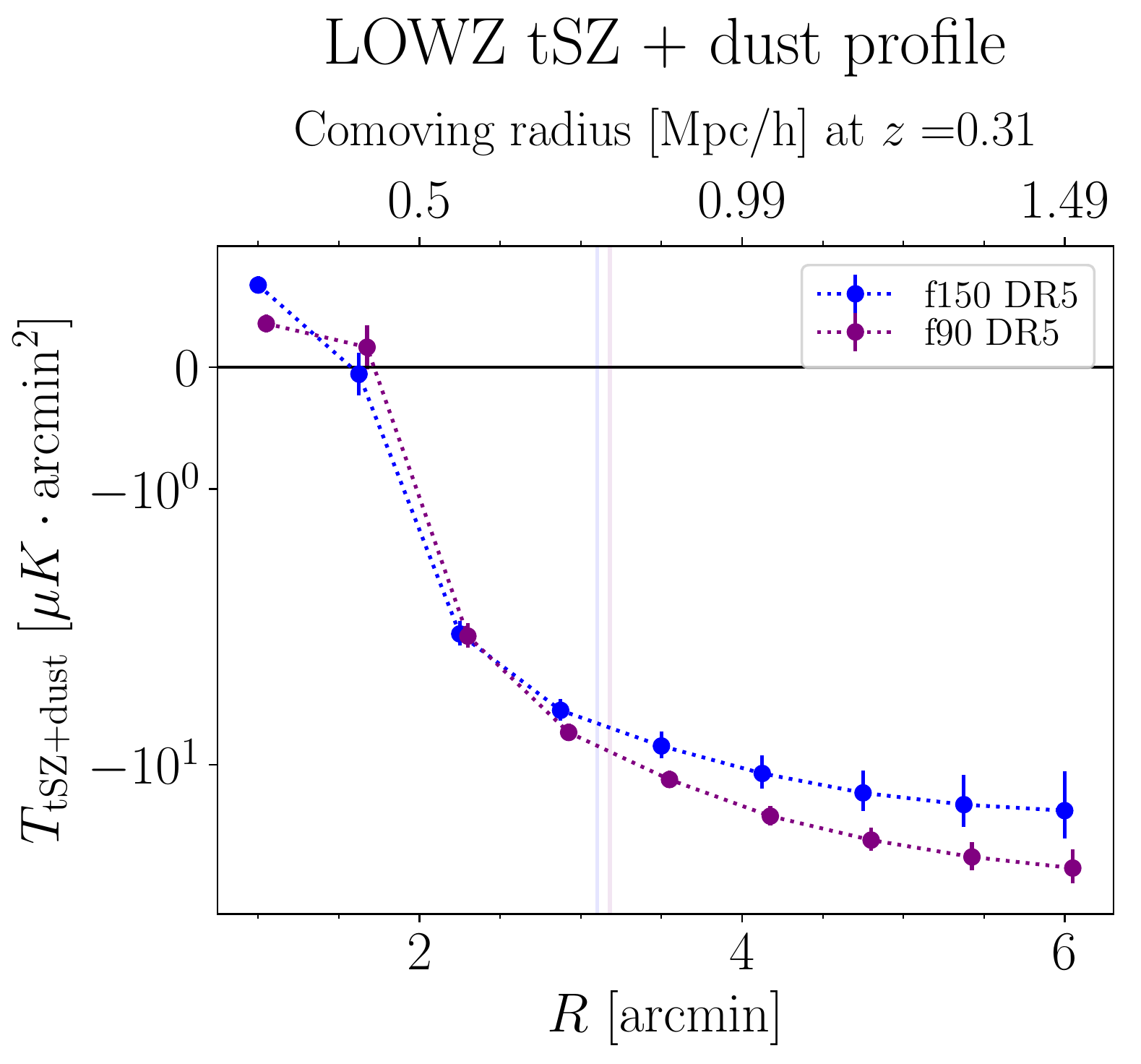}
\caption{ 
The mean LOWZ tSZ + dust signal in all compensated aperture photometry filters, as defined in Equation \ref{eq:tSZ_est}.
The dotted lines simply connect the points to guide the eye.
The no-signal hypothesis is rejected at 16.4$\sigma$ (see Table~\ref{tab:snr}).
The vertical lines show the halo virial radius ($3.1'$ at $z=0.31$) added in quadrature with the beam standard deviations ($\sigma=\text{FWHM}/\sqrt{8\ln 2}= 0.55'$ in f150 and 0.89' in f90). 
The correlation matrix for the different CAP filters and frequencies is identical to Fig.~\ref{fig:kSZ_summary_cmass}.
}
\label{fig:tSZ_coadd_lowz}
\end{figure}

The tSZ + dust measurements from f90 and f150 are shown in Fig.~\ref{fig:tSZ_coadd_lowz}.

\begin{figure}[h]
\centering
\includegraphics[width=0.9\columnwidth]{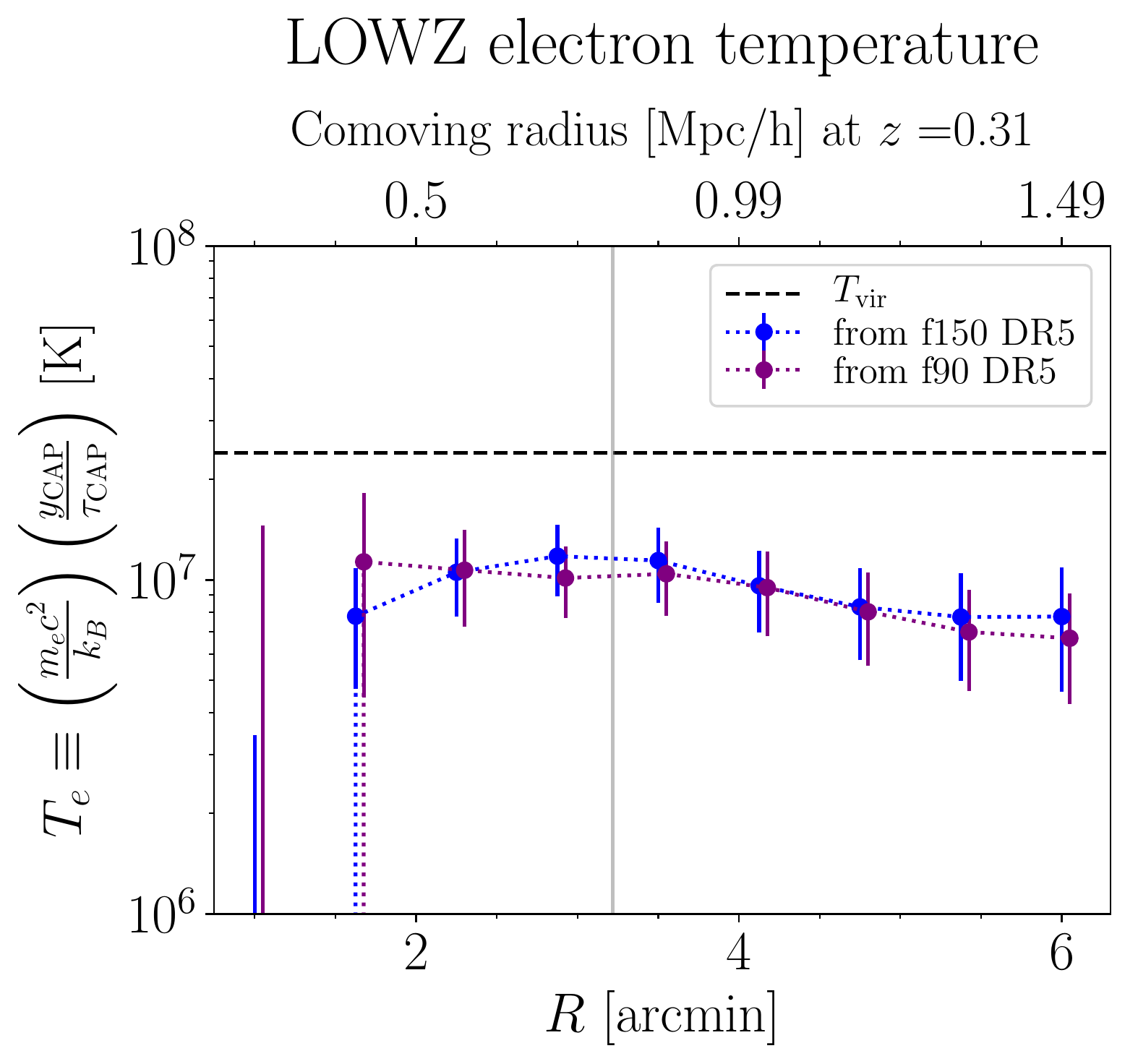}
\caption{
Simplified measurement of the electron temperature around LOWZ galaxies (Eq.~\eqref{eq:electron_temperature}).
In comparison, the horizontal dashed line shows the virial temperature estimate for LOWZ halos, whose order of magnitude is consistent with the data.
The vertical solid gray line shows the virial radius of the LOWZ galaxies ($3.1'$ at $z=0.55$), added in quadrature with the beam standard deviation ($\sigma=\text{FWHM}/\sqrt{8\ln 2}= 1.0'$) of the ILC map with deprojected CIB.
The dotted lines simply connect the data points.
}
\label{fig:electron_temperature_lowz}
\end{figure}

As we did for CMASS, we show a simplified measurement of the electron temperature in Fig.~\ref{fig:electron_temperature_lowz}.
The data is consistent with the order of magnitude of the expected virial temperature.

\begin{figure}[h!]
\centering
\includegraphics[width=0.49\columnwidth]{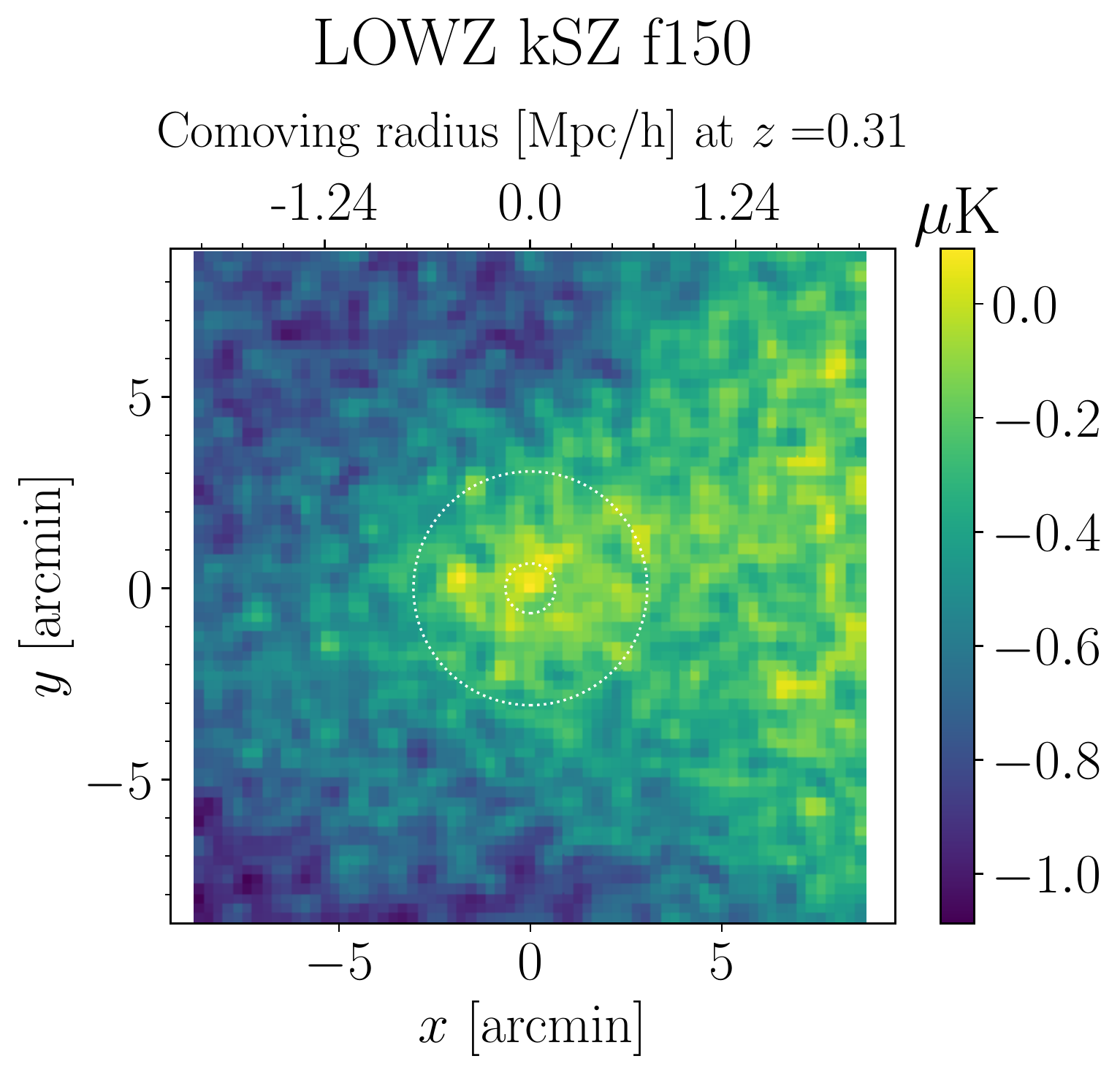}
\includegraphics[width=0.49\columnwidth]{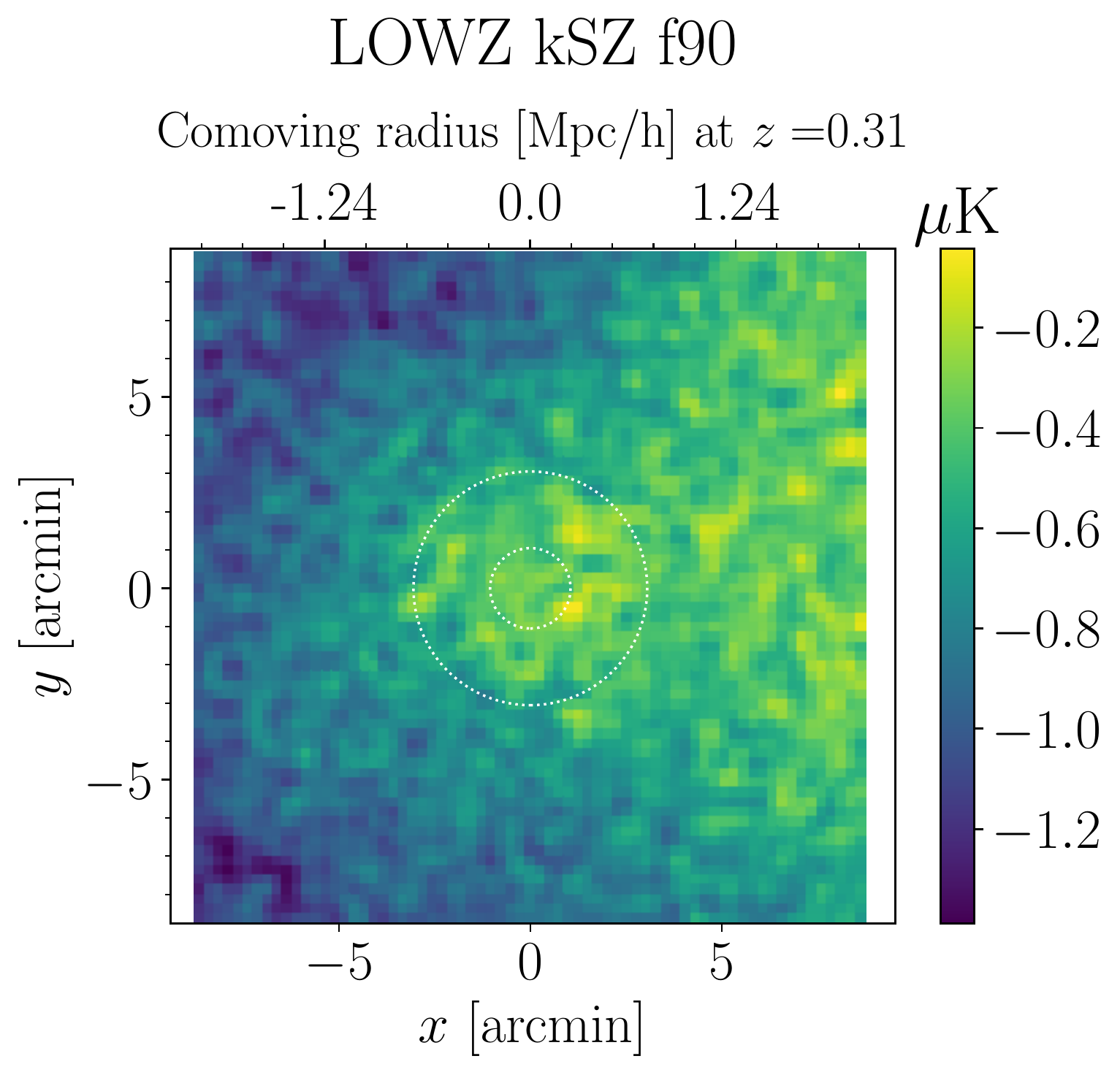}
\includegraphics[width=0.49\columnwidth]{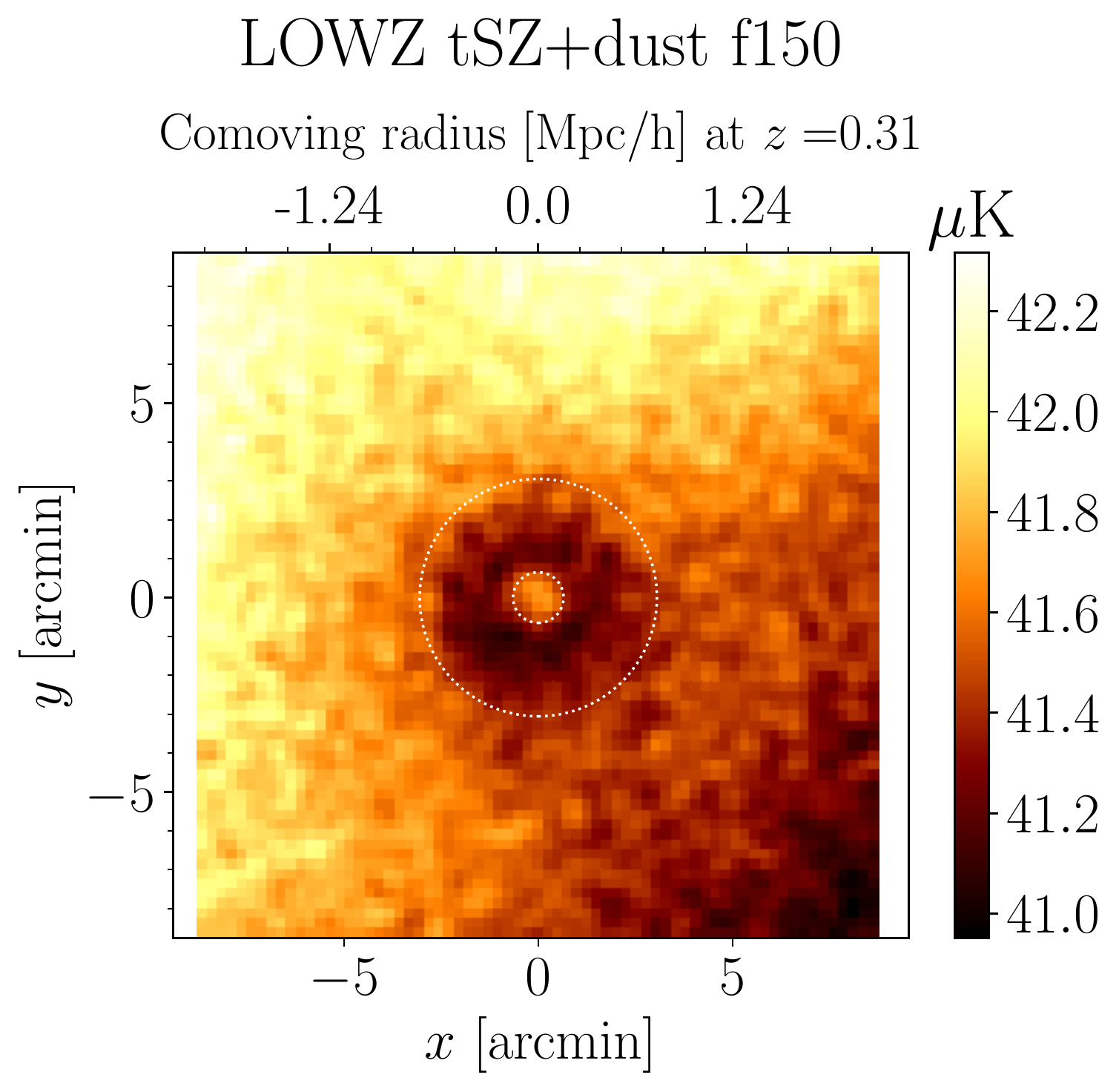}
\includegraphics[width=0.49\columnwidth]{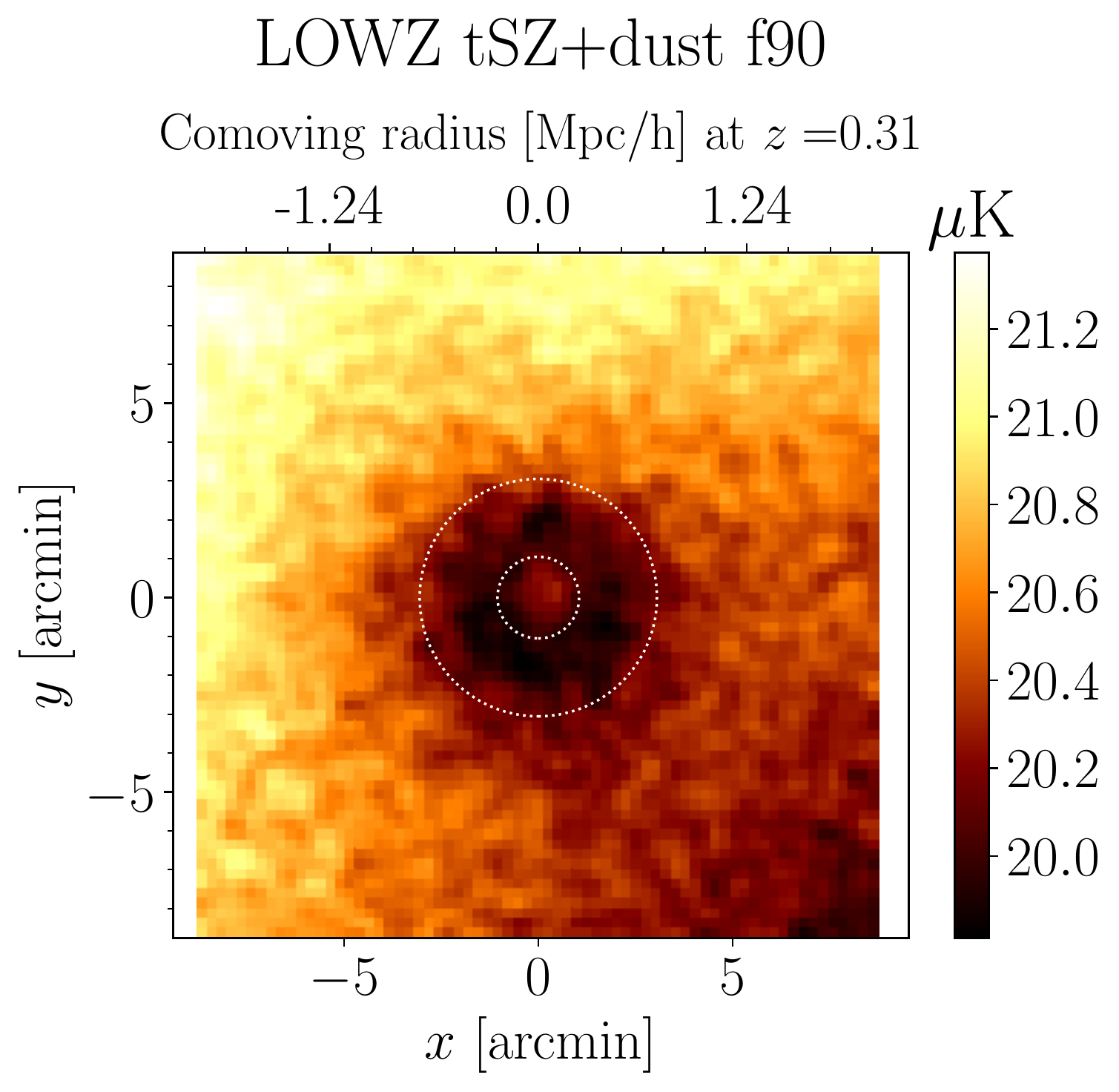}
\includegraphics[width=0.49\columnwidth]{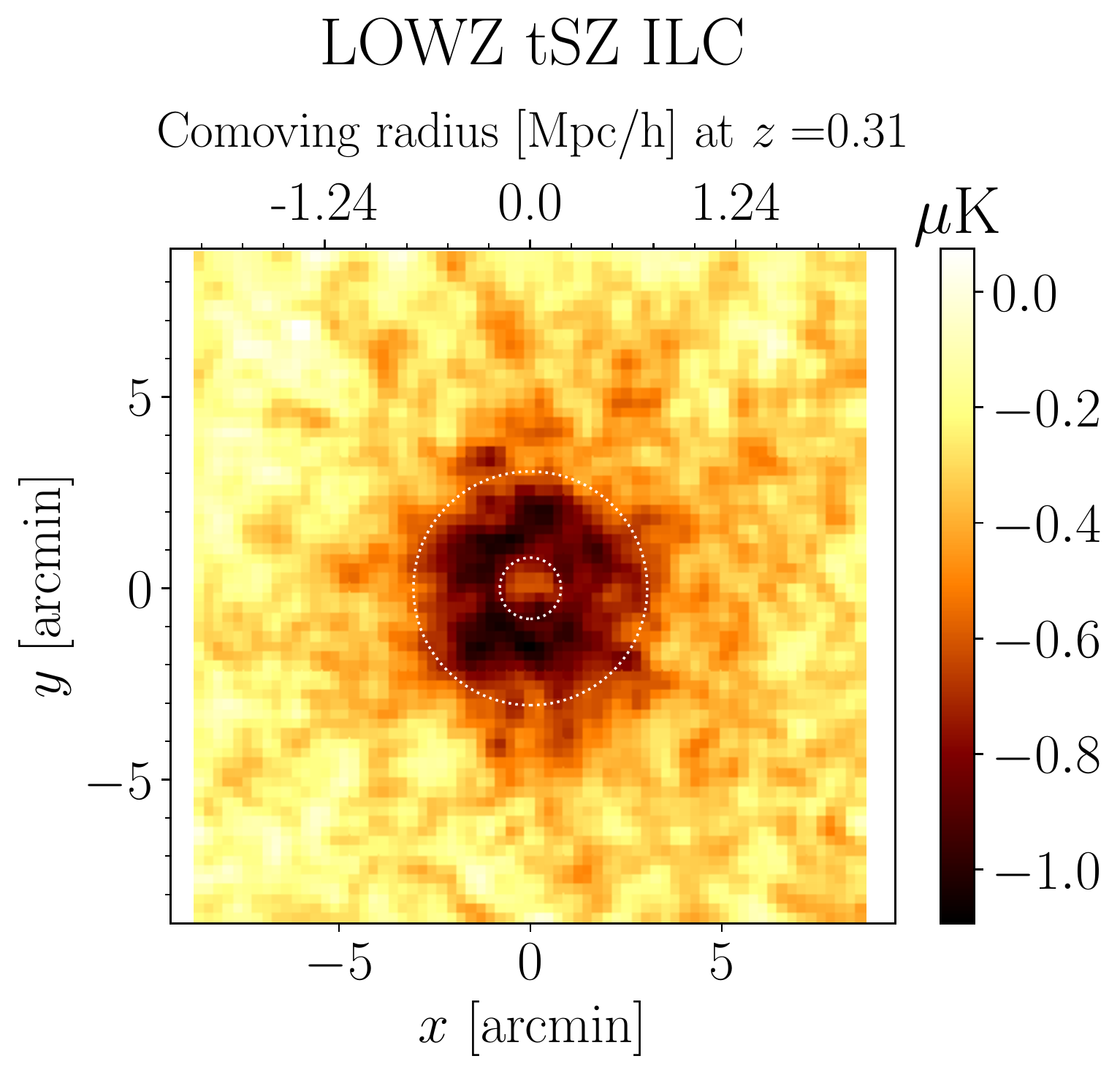}
\includegraphics[width=0.49\columnwidth]{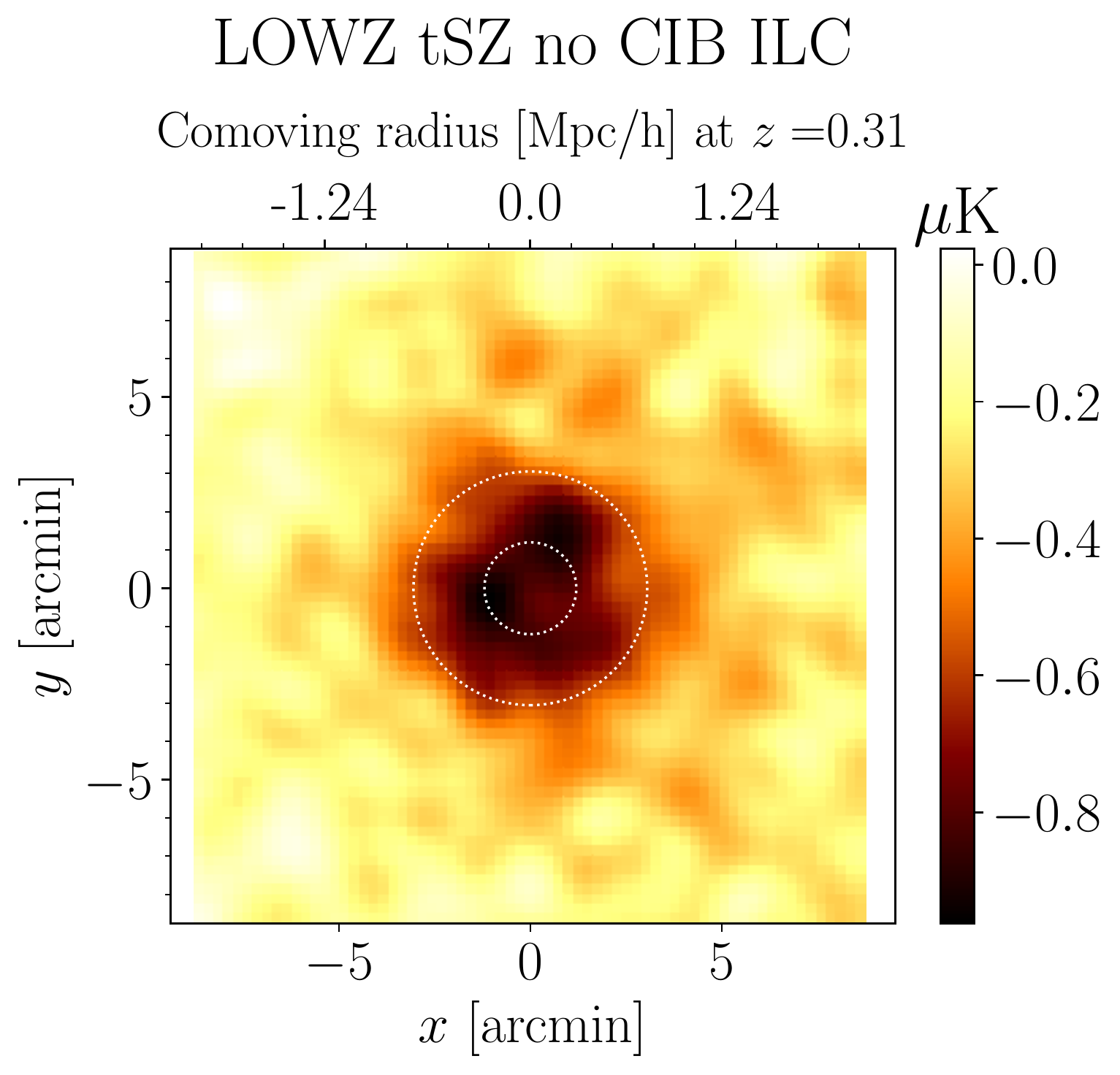}
\caption{
Same as Fig.~\ref{fig:stacked_maps_cmass}, for LOWZ instead of CMASS.
The kSZ SNR is lower than for CMASS, and the dust emission, filling in the tSZ decrement near the center, is visible both in f150 and f90.
}
\label{fig:stacked_maps_lowz}
\end{figure}

Finally, we also show the stacked 2d map cutouts around LOWZ objects in Fig.~\ref{fig:stacked_maps_lowz}.
Again, the gas density and pressure profiles are resolved.
Here, dust emission is clearly visible not only in f150 and the Compton-$y$ ILC, but also in f90, suggesting that the dust emission is brighter. 

We expect a 1.6 times higher dust luminosity for LOWZ than CMASS, due to the 1.6 times more massive host halo.
This effect should be compensated by the 1.5 times higher noise, due to the $1.5^2$ times smaller sample size.
LOWZ galaxies are closer though, with a typical squared luminosity distance smaller than CMASS by a factor 4.6, which translates into a 4.6 higher dust brightness.
One would expect the intrinsic dust luminosity to increase with redshift, due to the higher star formation rate \cite{Liang19}, compensating this effect.
The fact that the dust is more visible in LOWZ than CMASS suggests that this intrinsic evolution between LOWZ and CMASS does not compensate the difference in luminosity distances.
Finally, the tSZ signal scales approximately as $\propto M^{5/3}$, and is independent of redshift, making it $1.6^{5/3}\simeq 2$ times larger for LOWZ than CMASS.
Since the tSZ profile varies on larger scales than the dust profile, it may not be the main limiting factor in our ability to detect the dust.

\section{Discussion and conclusions}
\label{sec:conclusions}

We have measured the gas density using kSZ and pressure using tSZ around the CMASS galaxies in CAP filters of varying sizes, thus tracing the gas profile out to several virial radii from the galaxy group center. 
Our measurement constitutes the highest significance kSZ detection to date and a factor two improvement over our previous one \cite{Schaan:2015uaa}. 
The data shows unequivocally that the gas profile is more extended than the dark matter profile, i.e. that a large fraction of the baryons lies outside of the virial radius.
This conclusion is robust to varying the assumed baryon fraction in CMASS halos.
As a proof of concept, we demonstrated that tSZ and kSZ measurements can be combined to estimate the temperature of the free electron gas around CMASS galaxies.
In a companion paper \cite{paper2}, we explore the physical consequences of our measurements in the context of halo energetics and thermodynamics.
These papers are a stepping stone towards measuring feedback in galaxy formation.

The increase in sample size available with the next generation of surveys will allow us to repeat these measurements as a function of mass, redshift and environment, thus improving our understanding of the complex physics underlying galaxy formation.

These measurements can also be used to calibrate the baryonic effects in weak lensing. Representing roughly 15\% of the total mass, knowledge of the baryon distribution will be essential to correctly interpret the next generation of weak lensing measurements from experiments such as Rubin Observatory, Euclid and Roman Space Telescope. Galaxy-galaxy lensing can be calibrated directly by measuring the kSZ signal around the lens sample.  
In \cite{paper2}, we show that the current measurement is precise enough to pin down the baryon contribution to CMASS galaxy-galaxy lensing measurements and inform the ``lensing is low'' tension on halo scales \cite{2016MNRAS.460.1457S, 2017MNRAS.467.3024L, 2019MNRAS.488.5771L}, by directly measuring the baryon profiles on the relevant scales.
For cosmic shear, some modeling and extrapolation may be required to encompass all of the halos that contribute to the power spectrum on mildly non-linear scales.  Since these are dominated by group-sized halos such as the ones in our sample, we expect kSZ to be useful in calibrating cosmic shear measurements as well, but we defer detailed modeling to future work.

In this paper, we also presented the corresponding measurements for the LOWZ galaxy sample, in addition to CMASS. 
Because the LOWZ catalog has a more complex halo occupation distribution, we leave the interpretation of these measurements to future work.

Once the astrophysical properties of the sample are well characterized, the kSZ signal can also be used to measure the large-scale velocity fields, and reconstruct long-wavelength modes in the matter density with unprecedented precision, providing a new window into the physics of the early Universe \cite{2019PhRvD.100h3508M}, as well as improving our constraints on modified gravity, dark energy \cite{Mueller:2014nsa} and neutrino masses \cite{Mueller:2014dba}.

\acknowledgments

We thank the anonymous referees for their insightful suggestions which improved this article.
We thank Shirley Ho, Eliot Quataert, Chung-Pei Ma, Uro\v{s} Seljak and Martin White for their comments and suggestions on this work.
E.S. is supported by the Chamberlain fellowship at Lawrence Berkeley National Laboratory. 
S.F. is supported by the Physics Division of Lawrence Berkeley National Laboratory.
NB acknowledges support from NSF grant AST-1910021. NB and JCH acknowledge support from the Research and Technology Development fund at the Jet Propulsion Laboratory through the project entitled ``Mapping the Baryonic Majority''.
This research used resources of the National Energy Research Scientific Computing Center, a DOE Office of Science User Facility supported by the Office of Science of the U.S. Department of Energy under Contract No. DE-AC02-05CH11231.
We used the software \texttt{VisIt} \cite{HPV:VisIt} to generate the 3D visualization of the galaxy catalog and reconstructed velocities.

This work was supported by the U.S. National Science Foundation through awards AST-1440226, AST0965625 and AST-0408698 for the ACT project, as well as awards PHY-1214379 and PHY-0855887. Funding was also provided by Princeton University, the University of Pennsylvania, and a Canada Foundation for Innovation (CFI) award to UBC. ACT operates in the Parque Astron\'{o}mico Atacama in northern Chile under the auspices of the Comisi\'{o}n Nacional de Investigaci\'{o}n Cient\'{i}fica y Tecnol\'{o}gica de Chile (CONICYT). Computations were performed on the GPC and \emph{Niagara} supercomputers at the SciNet HPC Consortium. SciNet is funded
by the CFI under the auspices of Compute Canada, the Government of Ontario, the Ontario Research Fund
-- Research Excellence; and the University of Toronto. The development of multichroic detectors and lenses was supported by NASA grants NNX13AE56G and NNX14AB58G.   Colleagues at AstroNorte and RadioSky provide logistical support and keep operations in Chile running smoothly. We also thank the Mishrahi Fund and the Wilkinson Fund for their generous support of the project.
Funding for SDSS-III has been provided by the Alfred P. Sloan Foundation, the Participating Institutions, the National Science Foundation, and the U.S. Department of Energy Office of Science. The SDSS-III web site is http://www.sdss3.org/.
SDSS-III is managed by the Astrophysical Research Consortium for the Participating Institutions of the SDSS-III Collaboration including the University of Arizona, the Brazilian Participation Group, Brookhaven National Laboratory, Carnegie Mellon University, University of Florida, the French Participation Group, the German Participation Group, Harvard University, the Instituto de Astrofisica de Canarias, the Michigan State/Notre Dame/JINA Participation Group, Johns Hopkins University, Lawrence Berkeley National Laboratory, Max Planck Institute for Astrophysics, Max Planck Institute for Extraterrestrial Physics, New Mexico State University, New York University, Ohio State University, Pennsylvania State University, University of Portsmouth, Princeton University, the Spanish Participation Group, University of Tokyo, University of Utah, Vanderbilt University, University of Virginia, University of Washington, and Yale University.
R.D. thanks CONICYT for grant BASAL CATA AFB-170002.
The Flatiron Institute is funded by the Simons Foundation.
EC acknowledges support from the STFC Ernest Rutherford Fellowship ST/M004856/2 and STFC Consolidated Grant ST/S00033X/1, and from the Horizon 2020 ERC Starting Grant (Grant agreement No 849169).
JD is supported through NSF grant AST-1814971.
JPH acknowledges funding for SZ cluster studies from NSF grant number
AST-1615657.
KM acknowledges support from the National Research Foundation of South Africa. %
DH, AM, and NS acknowledge support from NSF grant numbers AST-1513618 and AST-1907657.
MHi acknowledges support from the National Research Foundation of South Africa.

\bibliographystyle{prsty.bst}
\bibliography{ksz}

\appendix

\section{Aperture photometry pipeline}
\label{app:pipeline_description}

The native AdvACT pixel has a typical size of $0.5 '$, not much smaller than the CMASS kSZ and tSZ profile sizes. 
For this reason, properly handling pixelation effects is important.
Here we describe the stacking pipeline we implemented in \texttt{ThumbStack}, 
based on \texttt{pixell}\footnote{\url{https://github.com/simonsobs/pixell}}.

Our stacking pipeline extracts small square cutouts from the ACT map around the position of each galaxy. 
The CAP filters are applied to each cutout separately before being combined together via inverse-variance weighting, with or without the velocity weighting.
The advantage of this approach, compared to stacking the cutouts and finally applying the CAP filters, is that it allows us to adopt a different weighting not only for each galaxy, but also for each aperture filter radius.
This is relevant since the noise in small aperture filters is determined mostly by detector noise, which varies across the AdvACT map.
On the other hand, the noise in large aperture filters comes mostly from the lensed primary CMB, which is uniform across the AdvACT map.
The optimal inverse-variance weight is thus different for small and large apertures.

\begin{figure}[h]
\centering
\includegraphics[width=0.95\columnwidth]{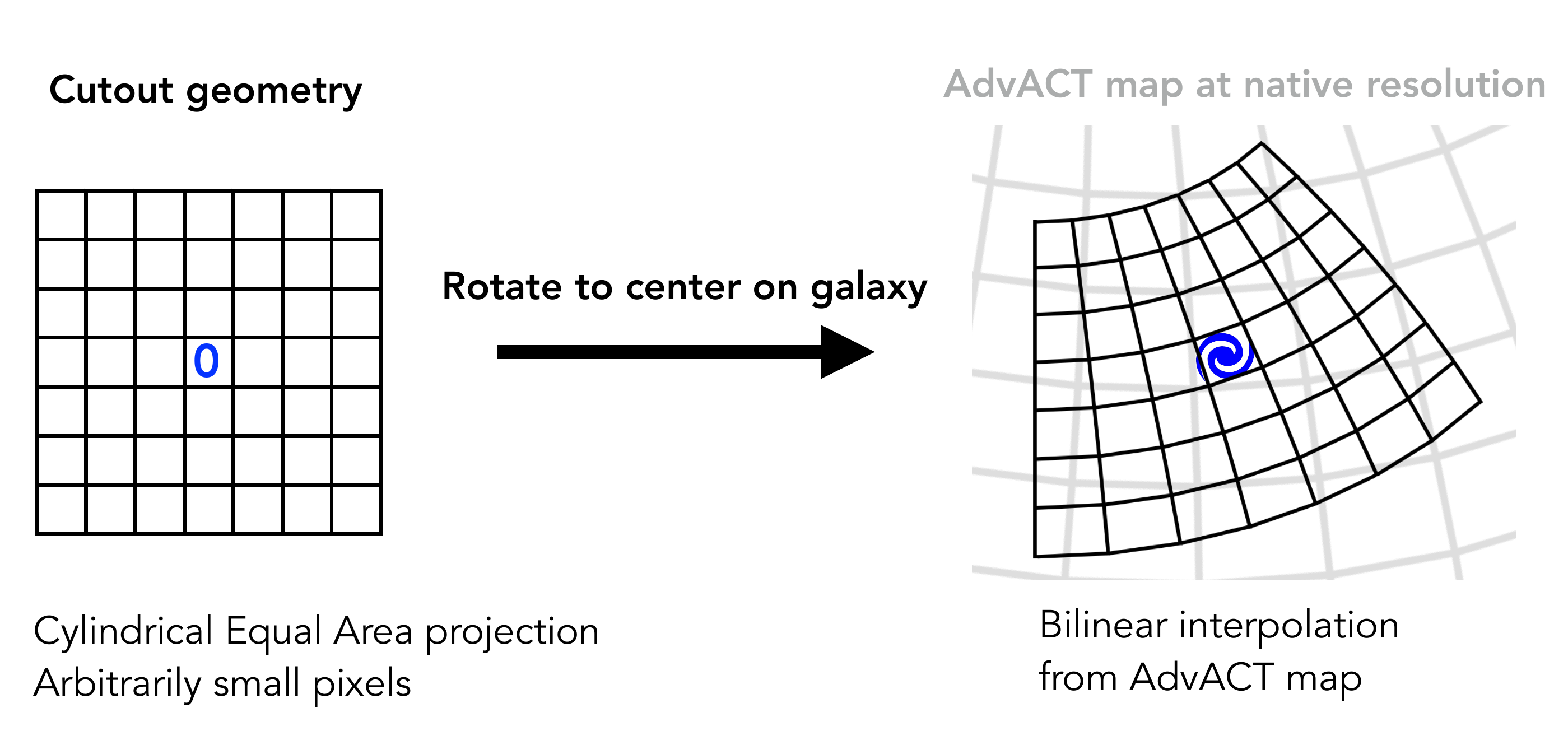}
\caption{
Illustration of our pipeline to extract cutouts from the AdvACT map.
It preserves the flux within pixels to a sufficient accuracy.
It also enables sub-pixel weighting,
meaning that the circular aperture filters can be made arbitrarily smooth
and that the galaxies in the stack can be centered arbitrarily precisely,
rather than placed at the center of the nearest AdvACT pixel.
}
\label{fig:schematic_cutouts}
\end{figure}

The process of extracting cutouts from the AdvACT map is illustrated in Fig.~\ref{fig:schematic_cutouts}.
We first create the desired cutout geometry.
We chose a Cylindrical Equal Area projection, such that each pixel has the same area, simplifying the integration.
This cutout geometry, initially centered around the origin, is then rotated to be centered on the target galaxy, and superimposed with the AdvACT map.
The values of the cutout are then read from the AdvACT map via bilinear interpolation.
This process has the following desirable properties:
\begin{itemize}
\item The bilinear interpolation preserves the flux within pixels exactly for rectangular grids. We have checked that it does so to high accuracy for our realistic curved grid too, as expected since our cutouts are small enough that the flat-sky approximation is adequate. This is typically not the case with higher order spline interpolations.\\
\item The cutout can be defined with arbitrarily high resolution. 
This is equivalent to sub-pixel weighting.
We found $0.25 '$ per pixel to be sufficient (the AdvACT pixel is typically $0.5 '$ on the side).\\
\item As a result, the galaxies can be centered on the cutout grid to arbitrary precision, i.e. to better than the size of the native AdvACT pixel. This means that the measured tSZ and kSZ profiles are not artificially broadened by the AdvACT pixel window function.\\
\item The circular aperture photometry filter can be made arbitrarily circular by increasing the cutout resolution. This amounts to weighting each AdvACT pixel by the exact fraction of its overlap with the aperture filter.\\
\end{itemize}

\section{End-to-end pipeline test \& 2-halo terms for tSZ and kSZ}
\label{app:pipeline_tests_2halo}

\begin{figure}[h!]
\centering
\includegraphics[width=0.45\columnwidth]{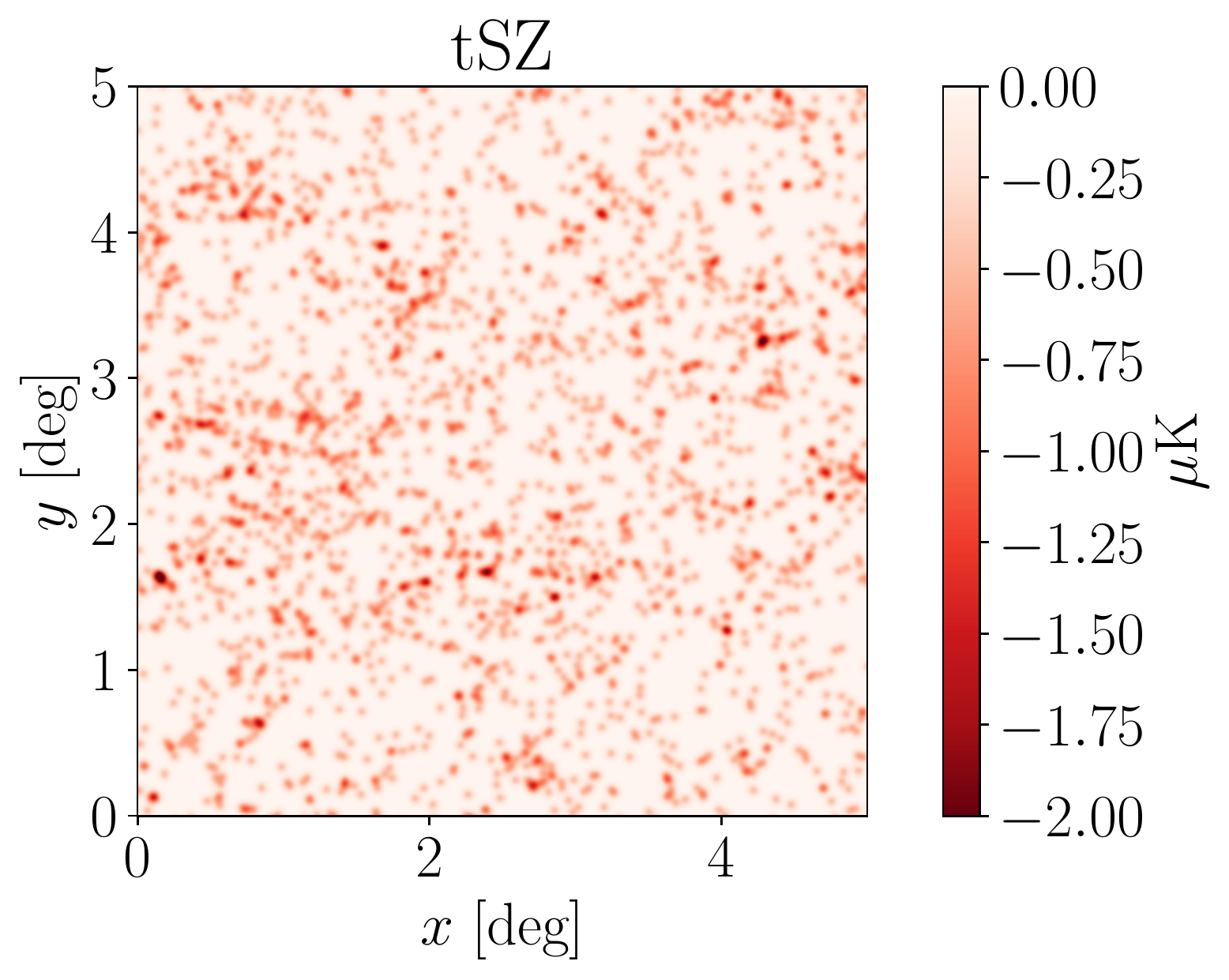}
\includegraphics[width=0.45\columnwidth]{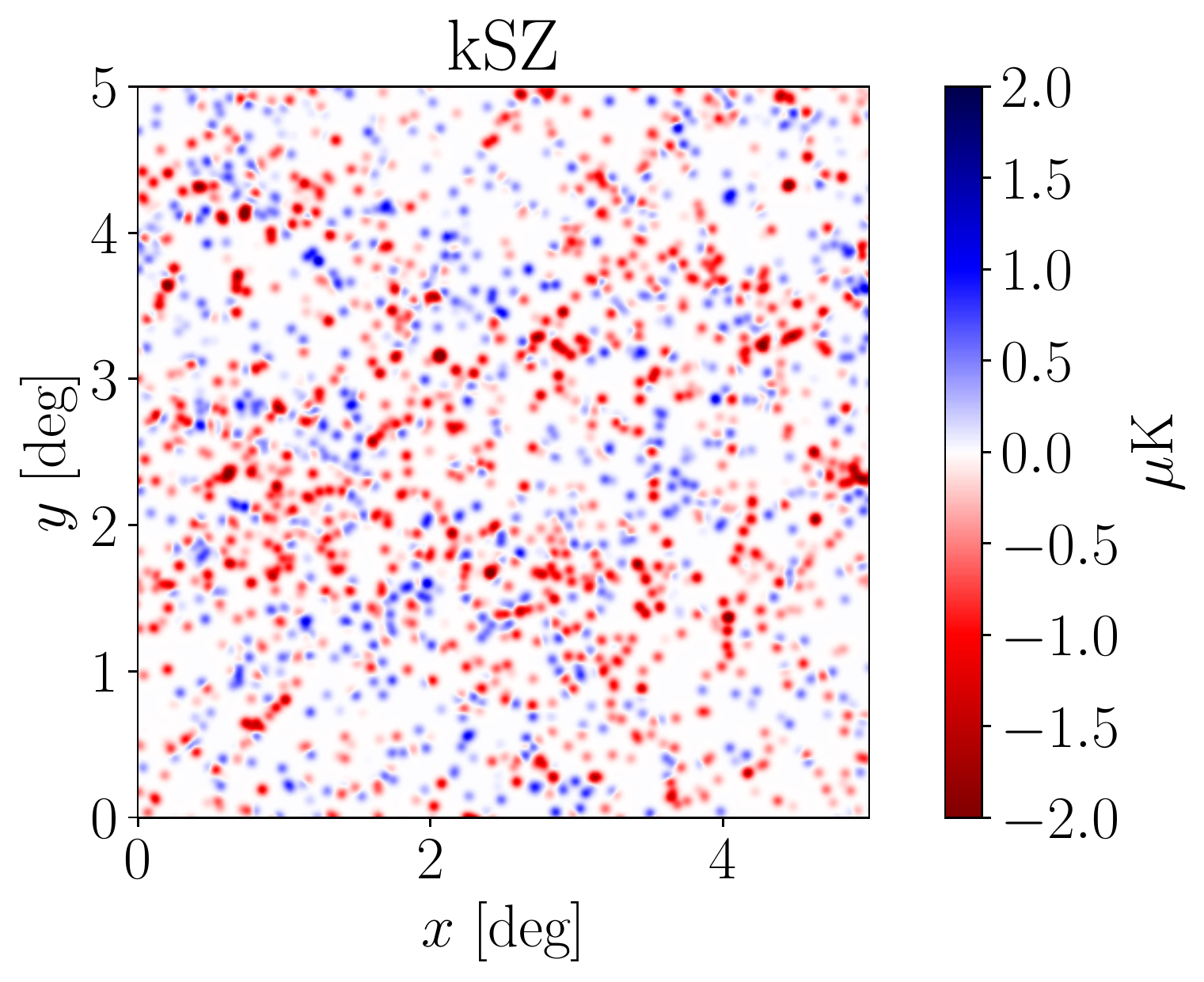}
\includegraphics[width=0.45\columnwidth]{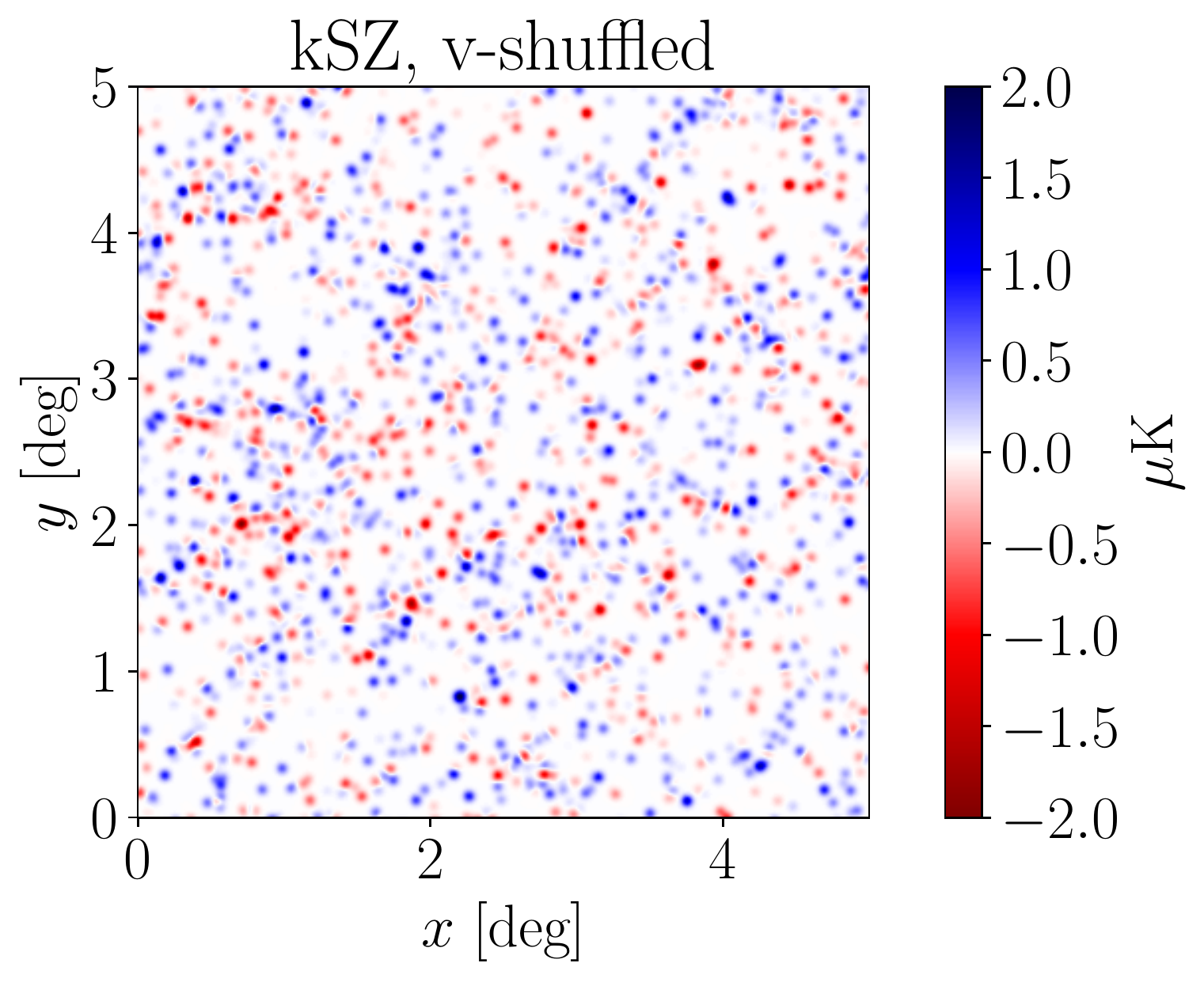}
\caption{
Mocks generated with Gaussian profiles at the true positions of the CMASS galaxies.
Each galaxy is given the same RMS integrated kSZ of 6.3~$\mu$K$\cdot \text{arcmin}^2$ and the same mean integrated tSZ of $-7.0$~$\mu$K$\cdot \text{arcmin}^2$.
The apparent difference in the amplitudes of the galaxy signals between the kSZ and kSZ shuffled cases is entirely due to the clustering of galaxy positions and velocities.
This produces the 2-halo term seen in Fig.~\ref{fig:endtoend_test}.
The kSZ map with shuffled velocities nulls this correlation and 2-halo term, enabling a pipeline test where the 1-halo profile is recovered exactly.
}
\label{fig:signal_mocks}
\end{figure}

To test our pipeline,
we generated mock AdvACT maps (see Fig.~\ref{fig:signal_mocks}) with fiducial tSZ or kSZ signals from halos and no noise (i.e. no CMB, detector noise and other foregrounds).
This allows us to check the accuracy of the pipeline to higher precision than in the real data.

Our mock signal maps have the same geometry and pixelation as the AdvACT maps.
In them, we painted a Gaussian profile with standard deviation $1.5 '$ at the position of each CMASS galaxy,
with the same amplitude for every galaxy (flux normalized to unity for every object).
This Gaussian profile is similar to the actual (beam-convolved) CMASS profiles.
For the kSZ mocks, the signal from each galaxy is multiplied by the reconstructed velocity before painting it on the map.

\begin{figure}[h!]
\centering
\includegraphics[width=0.95\columnwidth]{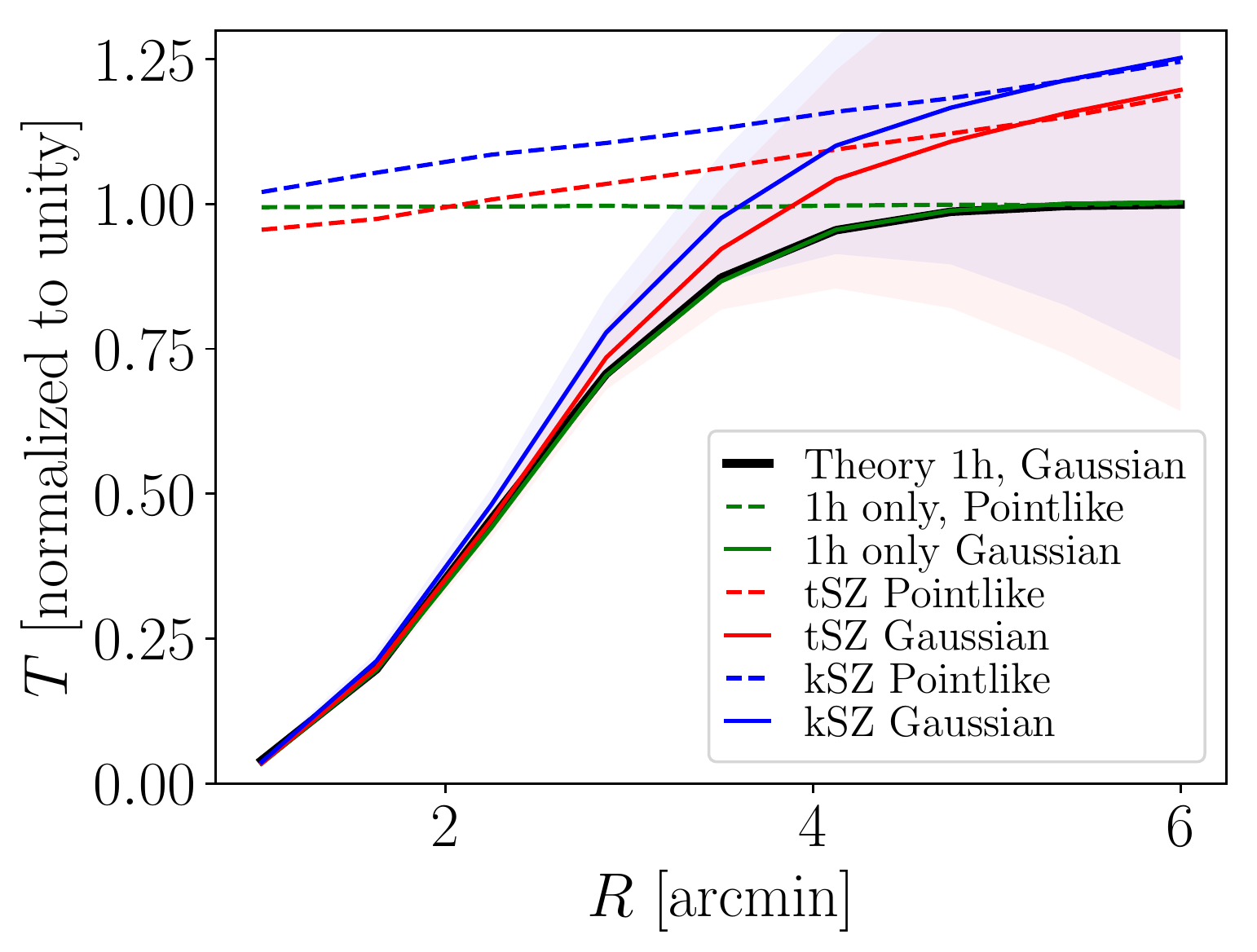}
\caption{
Pipeline test.
We generate mock tSZ and kSZ maps with a Gaussian gas profile, with standard deviation of $1.5 '$ (black line), similar to the measured one.
When using these mock maps, the measured tSZ (red solid) and kSZ (solid blue) profiles differ from the input profile (solid black) due to the 2-halo term from correlated CMASS galaxies.
Comparing to the statistical error in the real data for tSZ (red band) and kSZ (blue band), the 2-halo term only matters at large apertures. 
We account for it in the modeling.
To null the 2-halo term, we shuffle the galaxy velocities before generating the mock kSZ map: the resulting (solid green) profile matches the input perfectly, thus validating the pipeline.
Dashed lines are identical to solid lines, but for pointlike rather than Gaussian gas profiles.
They indicate the effective number of correlated CMASS galaxies around each CMASS target.
When shuffling the velocities (green dashed line), all galaxies are uncorrelated and we get unity at all apertures.
For tSZ (dashed red) and kSZ (dashed blue), the effective number of correlated galaxies rises to $1.2\text{--}1.25$ at the largest apertures.
}
\label{fig:endtoend_test}
\end{figure}

This method reproduces the realistic overlap between nearby galaxies, and the offset between CMASS galaxies and the centers of the closest AdvACT pixels.
More precisely, additional galaxies uniformly distributed around a given CMASS target do not bias the measured CAP filters on average, since the disk and ring of the CAP filter have the same area but opposite sign, and will thus on average cancel.
However, if the additional galaxies are correlated with the CMASS target, then more of them will lie in the disk than the ring, enhancing the signal.
This is simply the 2-halo term in the CMASS$\times$tSZ and CMASS$\times$kSZ cross-correlation function.
Indeed, Fig.~\ref{fig:endtoend_test} shows that the measured tSZ (solid red curve) and kSZ (solid blue curve) are enhanced compared to the input Gaussian profile (solid black curve).
This enhancement is only large ($20\text{--}25\%$) for the largest apertures, as expected, where the statistical error in the real measurement is large. 
The 2-halo term is thus marginally significant here,
although we note that this is only a lower limit to the true 2-halo term: our mock maps only contain the correlated tSZ and kSZ from other CMASS galaxies, not from all the halos in the Universe.
We properly model the tSZ and kSZ 2-halo term in \cite{paper2} accounting for this.
In the future, the 2-halo term will constitute an interesting signal per se, telling us about the free-electron bias.

To make sure that this enhancement is really due to the 2-halo term and not simply a bias in our pipeline, we generated a mock kSZ signal map after shuffling the velocities of the CMASS galaxies.
This removes the correlation between the kSZ signal of adjacent galaxies, thus nulling the 2-halo term.
Indeed, Fig.~\ref{fig:endtoend_test} shows that the signal obtained in this way (solid green curve) matches the input Gaussian profile perfectly, which validates our pipeline.

Finally, to gain more intuition on the effective number of correlated neighbors around a CMASS target, we generated mock signal maps with pointlike profiles for the CMASS galaxies.
Indeed, the CAP filters applied to the target CMASS galaxy will then simply count the excess number of correlated neighbors in the disk compared to the ring.
These are shown in dashed lines in Fig.~\ref{fig:endtoend_test}.
As expected, the mock kSZ maps with shuffled velocities give unity at all apertures, meaning that only the target CMASS galaxy contains correlated signal.
The mock tSZ and kSZ maps give an effective number of neighbors of $0.2\text{--}0.25$.

\section{Null tests}
\label{app:null_tests}

Below, we show the pipeline null tests and foreground tests performed on the CMASS and LOWZ stacked measurements.

Some of the null tests below compare two maps, by performing the stack on a difference map. This is done after reconvolving the map with the narrowest beam to the beam of the other.
For example, the f150 map is reconvolved to the beam of f90, to the ILC beam and to the beam of the ILC map with deprojection, in the corresponding map differences.
Similarly, in the map differences between ILC and ILC with deprojection, the former map is reconvolved to the beam of the latter.
To do so, we use the same beam regularization procedure as outlined in \cite{naessetal20} when reconvolving the coadded f150 maps.
Specifically, at high ell where the beam transfer function is small, the measured values are uncertain. We replace them by the following fitting function, with $v^\star = 0.01$:
\beq
B_\ell = 
\left\{
\bal
& B_\ell^\text{measured} 
\quad\text{if}\quad
\ell \ll \ell^\star\\
& v^\star B_\text{max}^\text{measured} (\ell/\ell^\star)^{2\log(v^\star)}
\quad\text{if}\quad
\ell > \ell^\star\\
\eal
\right.
\eeq
This extrapolating function from \cite{naessetal20} keeps the beam value continuous, as well as its first derivative in the case of a Gaussian beam.
Most importantly, it was chosen to keep the ratios between multiple beams constant (rather than e.g. wildly swinging) in the regime where the beams are
too low to be trustworthy.

\subsection{CMASS}

\begin{figure}[h]
\centering
\includegraphics[width=0.95\columnwidth]{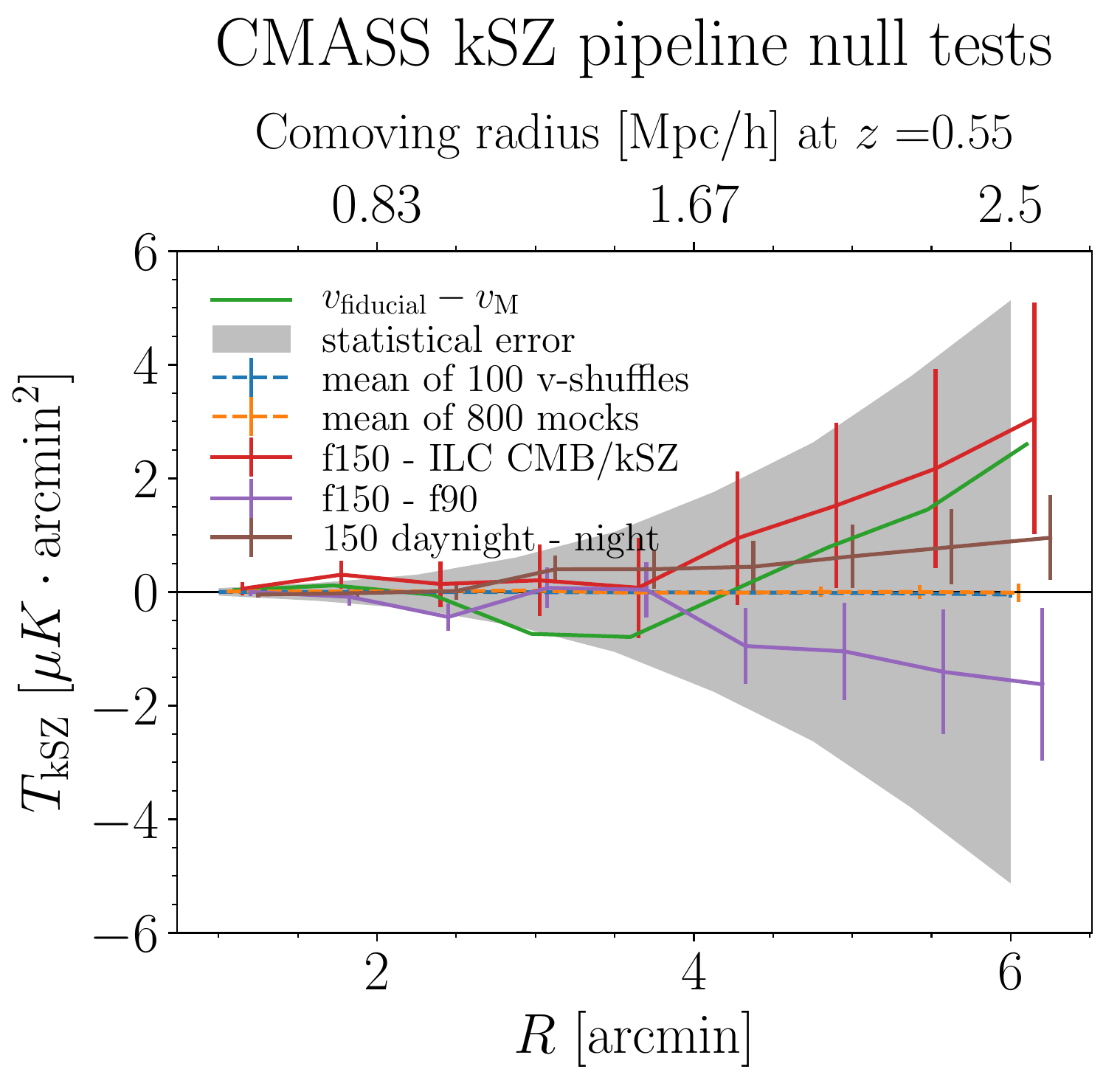}
\caption{
All CMASS kSZ pipeline null tests pass.
The null kSZ signal when shuffling the velocities or stacking on Gaussian random field mock maps are a basic pipeline check.
The stacking on the two difference maps shows that the signal kSZ is signal is stable to changes in the CMB map.
The green curve is the difference of the kSZ signals measured using the two different velocity reconstructions methods, showing that the difference is within the statistical error of the measurement (for f150, grey band).
}
\label{fig:pipe_null_tests_cmass}
\end{figure}

Fig.~\ref{fig:pipe_null_tests_cmass} presents the pipeline null tests for the CMASS kSZ profiles, compared to the statistical uncertainty on the measurement (gray band).
It shows that no kSZ signal is detected when the reconstructed velocities are shuffled, such that each galaxy is attributed the wrong velocity.
It shows that the signal also vanishes when the correct galaxy positions and velocities are used, but the true temperature map is replaced with a Gaussian random field with the same power spectrum.
It shows that the two velocity reconstruction pipelines agree, i.e. that the signal vanishes (within the error bars) when each galaxy is given the difference between the reconstructed velocities from each pipeline.
Finally, we take the differences of the fiducial f150 day+night with f90, with the CMB ILC and with the night-only f150.
This shows that the kSZ measurement is stable with respect to replacing the CMB map.
The signal vanishes when the stack is performed on the difference of f90 and f150, and on the difference of f150 and the ILC CMB/kSZ map (after reconvolving maps to the same beam before differencing).

\begin{figure}[h]
\centering
\includegraphics[width=0.95\columnwidth]{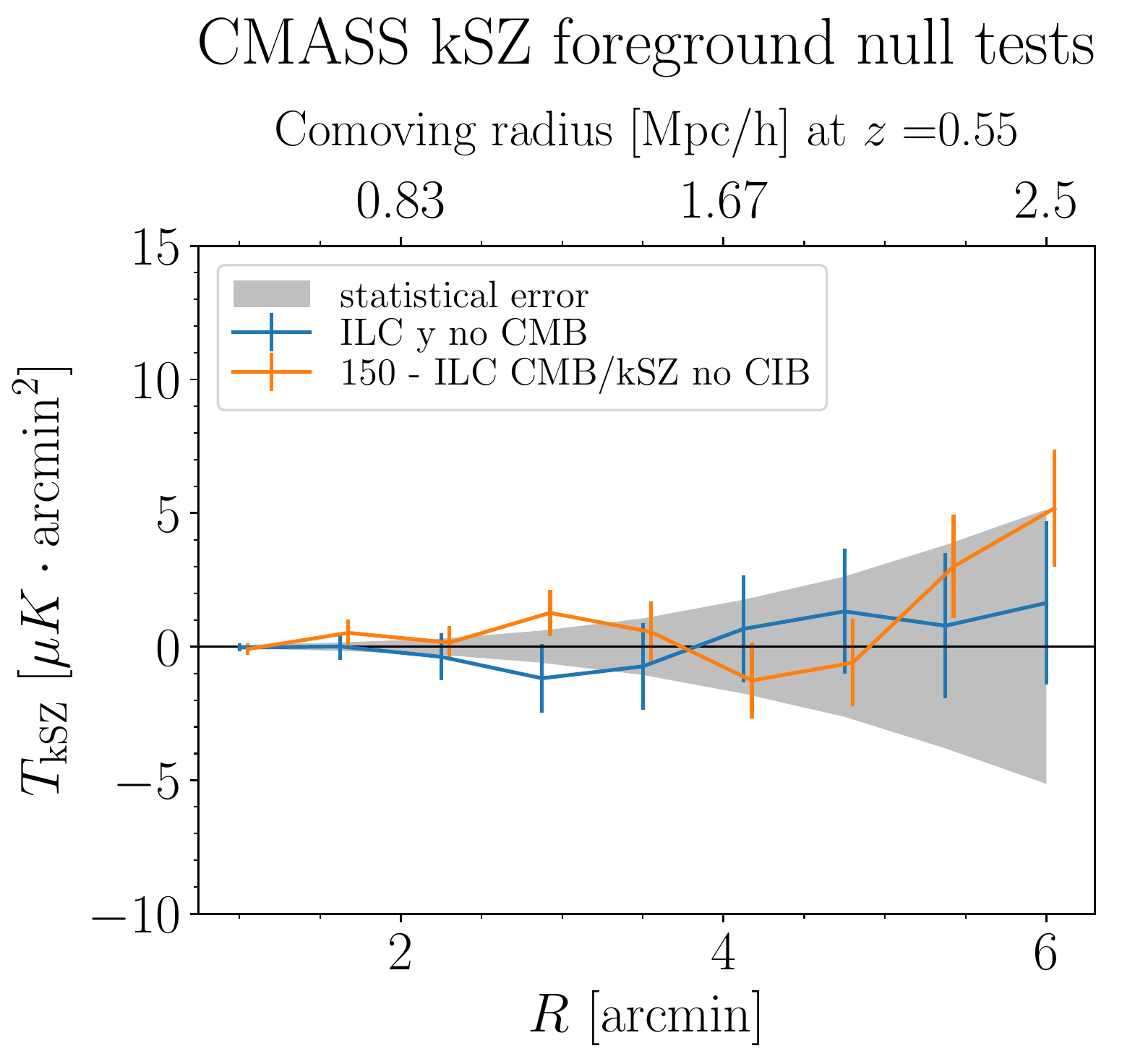}
\caption{
Foreground null tests for CMASS kSZ, showing no contamination from tSZ.
The ILC $y$ map has been converted to temperature units at 150 GHz.
}
\label{fig:fg_null_tests_ksz_cmass}
\end{figure}

We perform foreground null tests in Fig.~\ref{fig:fg_null_tests_ksz_cmass}, checking for a potential tSZ contamination to the kSZ estimator.
To do so, we replace the temperature map with the ILC $y$ map deprojecting CMB. This map has no response to CMB, and therefore no response to kSZ. Any detected signal would come from tSZ (or dust) contamination.
No such signal is seen.
Similarly, we run the stack on the difference between the f150 map and the ILC CMB deprojecting CIB.
This difference map may contain some tSZ and dust. However, it does not bias the kSZ estimator.

\begin{figure}[h]
\centering
\includegraphics[width=0.95\columnwidth]{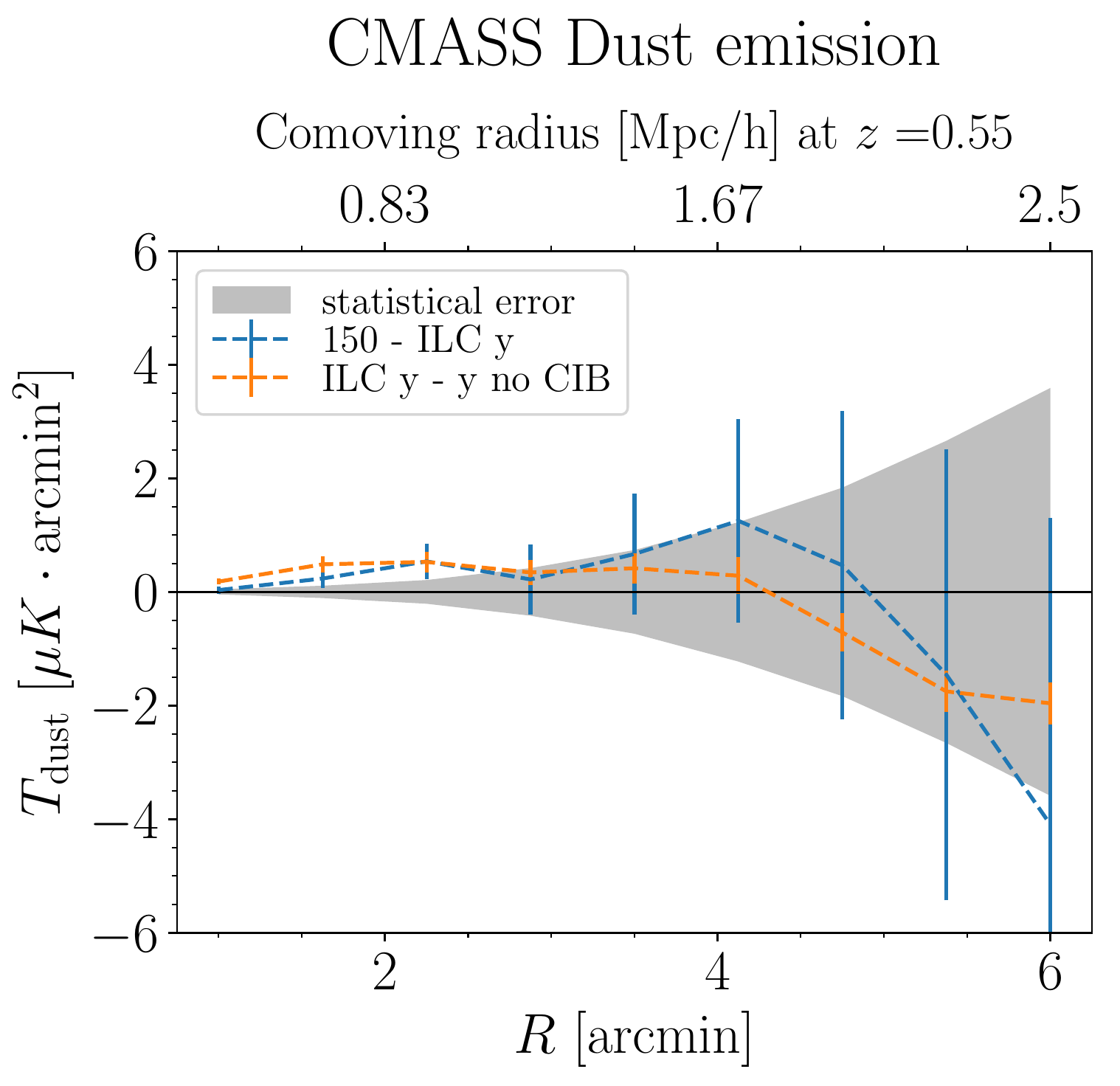}
\caption{
These measurements show the stacked signals from difference maps which null the tSZ signal. Any non-zero residual is therefore an indication of a different dust emission signal in the two maps.
The residuals are non-zero at the smaller radii, suggesting that dust emission is non-negligible in these maps, as expected.
This justifies jointly fitting for tSZ and dust thermal emission in f90 and f150, as we do in \cite{paper2},
or focusing on the ILC $y$ maps with deprojected CIB.
The ILC $y$ maps have been converted to temperature units at 150 GHz.
}
\label{fig:dust_cmass}
\end{figure}

Finally, to get an order of magnitude of the contribution of dust to the tSZ + dust profiles measured from f90 and f150, we use difference maps that null the tSZ signal, in Fig.~\ref{fig:dust_cmass}.
These show that the dust emission is non-negligible, as expected.

\subsection{LOWZ}

\begin{figure}[h]
\centering
\includegraphics[width=0.95\columnwidth]{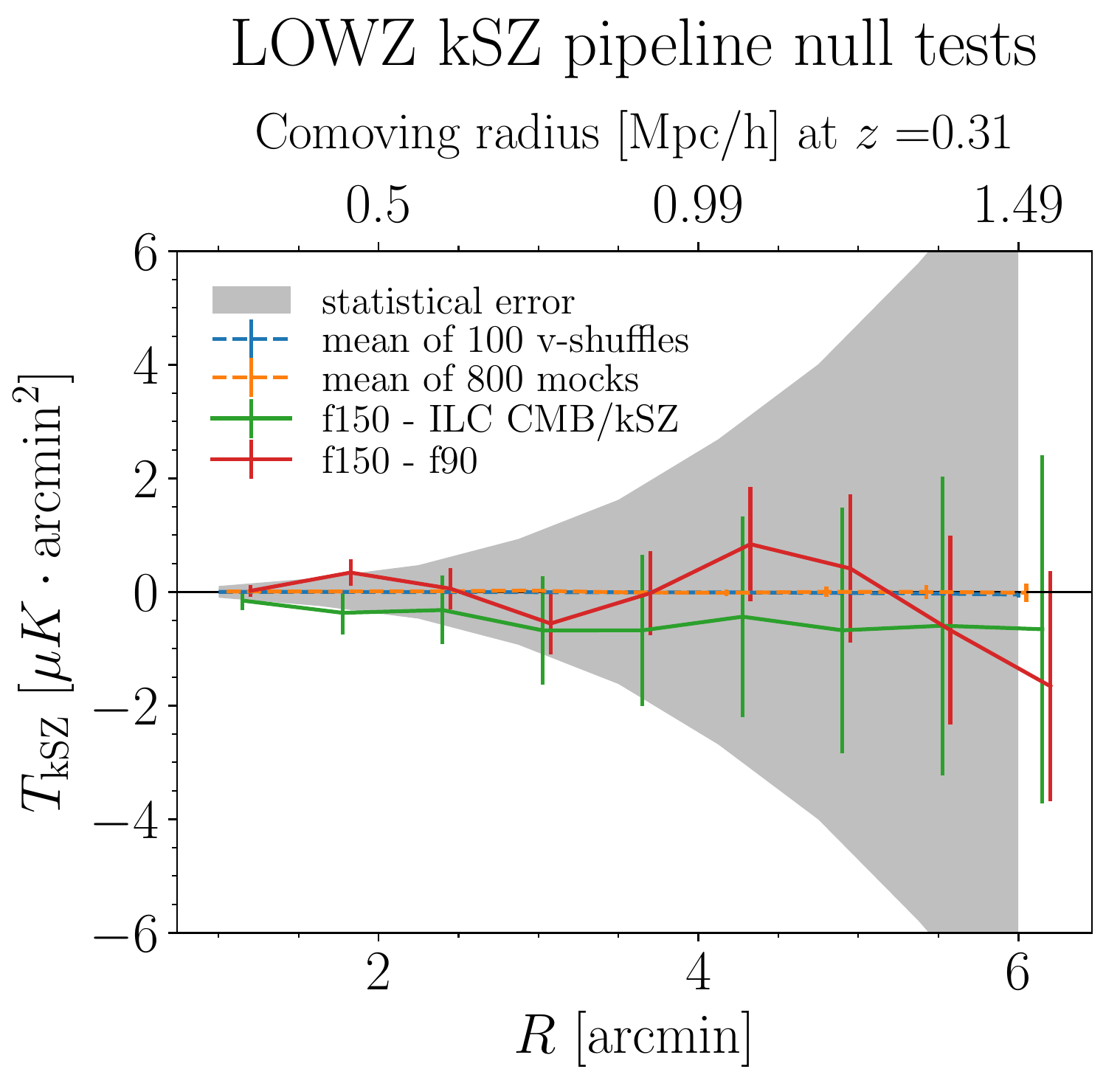}
\caption{
Same kSZ pipeline null tests as Fig.~\ref{fig:pipe_null_tests_cmass} but for LOWZ instead of CMASS.
}
\label{fig:pipe_null_tests_lowz}
\end{figure}

\begin{figure}[h]
\centering
\includegraphics[width=0.95\columnwidth]{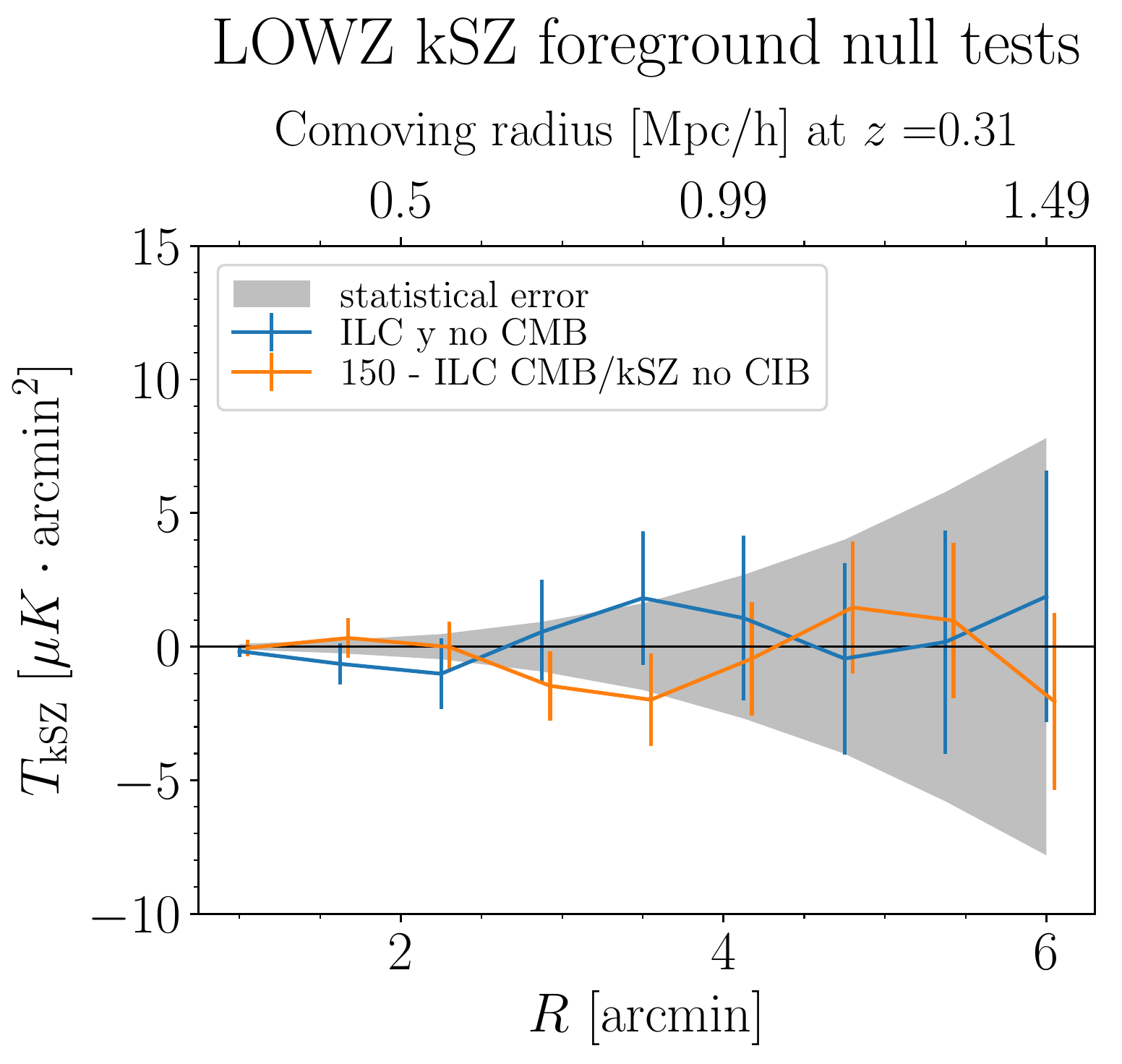}
\caption{
Same kSZ foreground null tests as Fig.~\ref{fig:fg_null_tests_ksz_cmass} but for LOWZ instead of CMASS.
}
\label{fig:fg_null_tests_ksz_lowz}
\end{figure}

\begin{figure}[h]
\centering
\includegraphics[width=0.95\columnwidth]{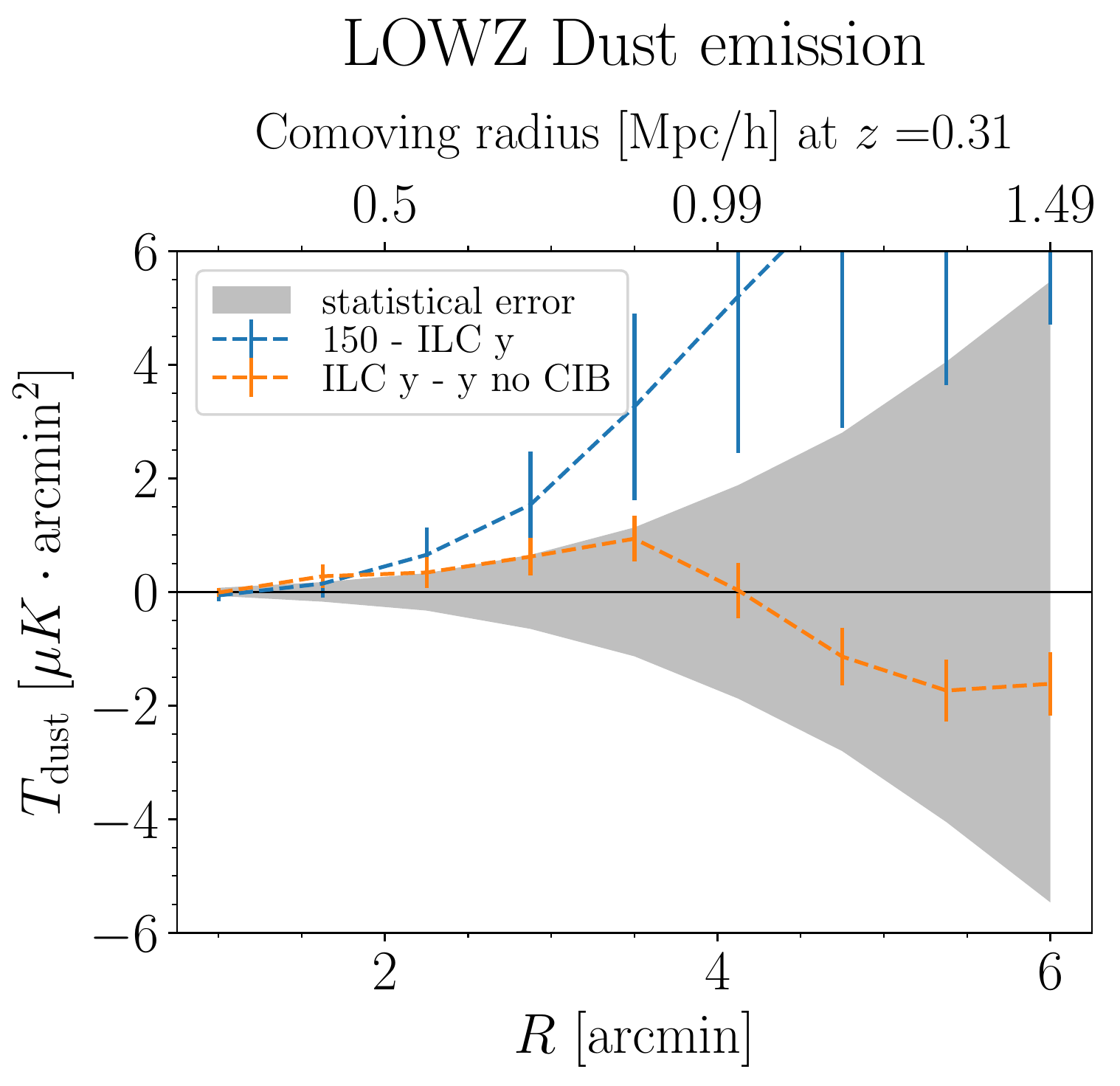}
\caption{
Same as Fig.~\ref{fig:fg_null_tests_ksz_cmass} but for LOWZ instead of CMASS.
Here again, the dust contribution to the measured tSZ+dust profiles is non-negligible, as expected.
}
\label{fig:dust_lowz}
\end{figure}

In Fig.~\ref{fig:pipe_null_tests_lowz}--\ref{fig:dust_lowz}, we perform the same null tests for LOWZ as for CMASS
and find the same conclusions.

\section{Validity of bootstrap for covariance matrices}
\label{app:mock_cov}

Estimating the covariance of the CAP filters using the bootstrap method implicitly assumes that the noise on the CAP filter values is independent from galaxy to galaxy.
This noise comes from detector and atmospheric noise, but also from the lensed primary CMB and all the other foregrounds present in the map.
Because the CMASS galaxies are dense, the CAP filters on different galaxies can be close and even overlap, making their noise correlated.

\begin{figure}[h]
\centering
\includegraphics[width=0.45\columnwidth]{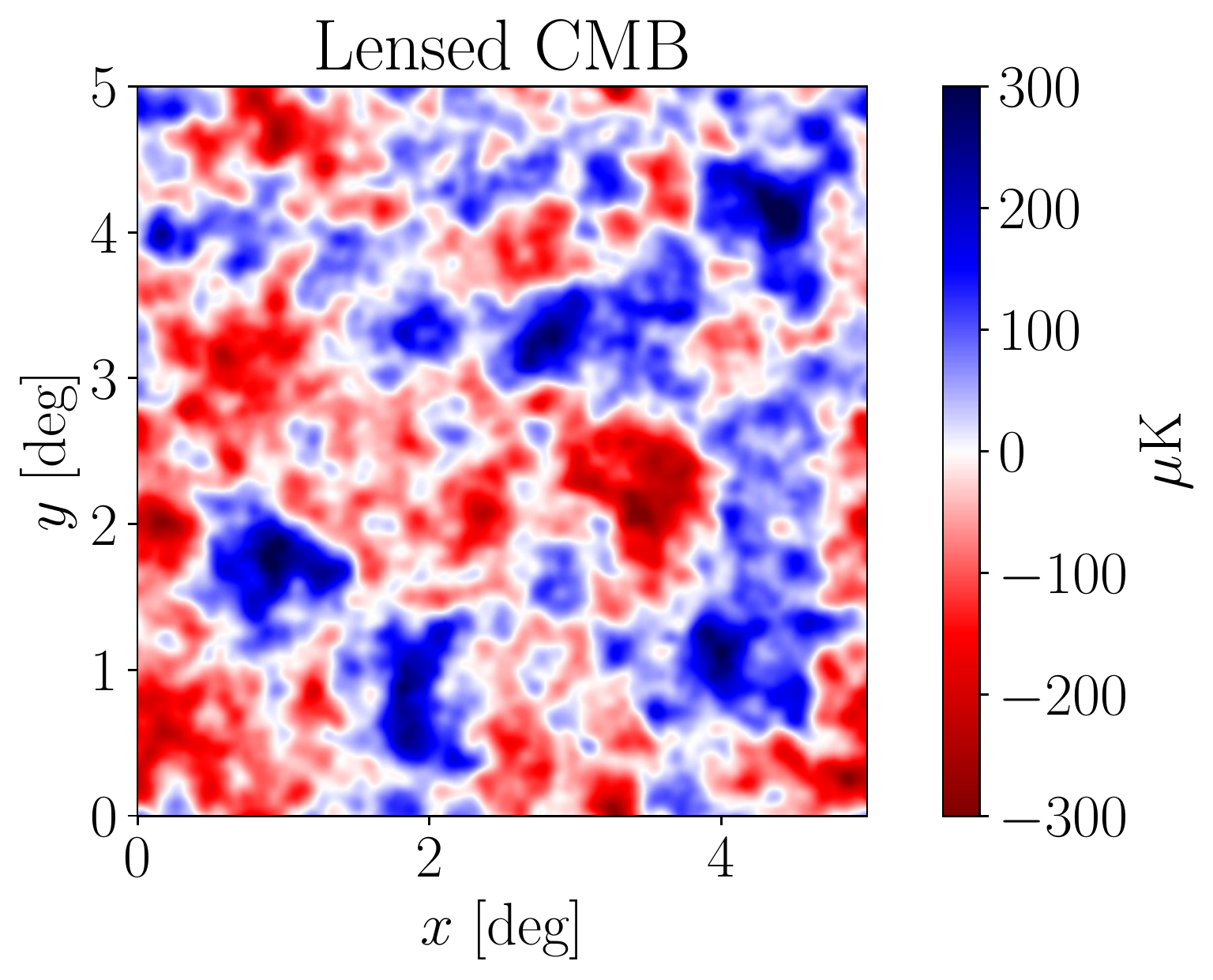}
\includegraphics[width=0.45\columnwidth]{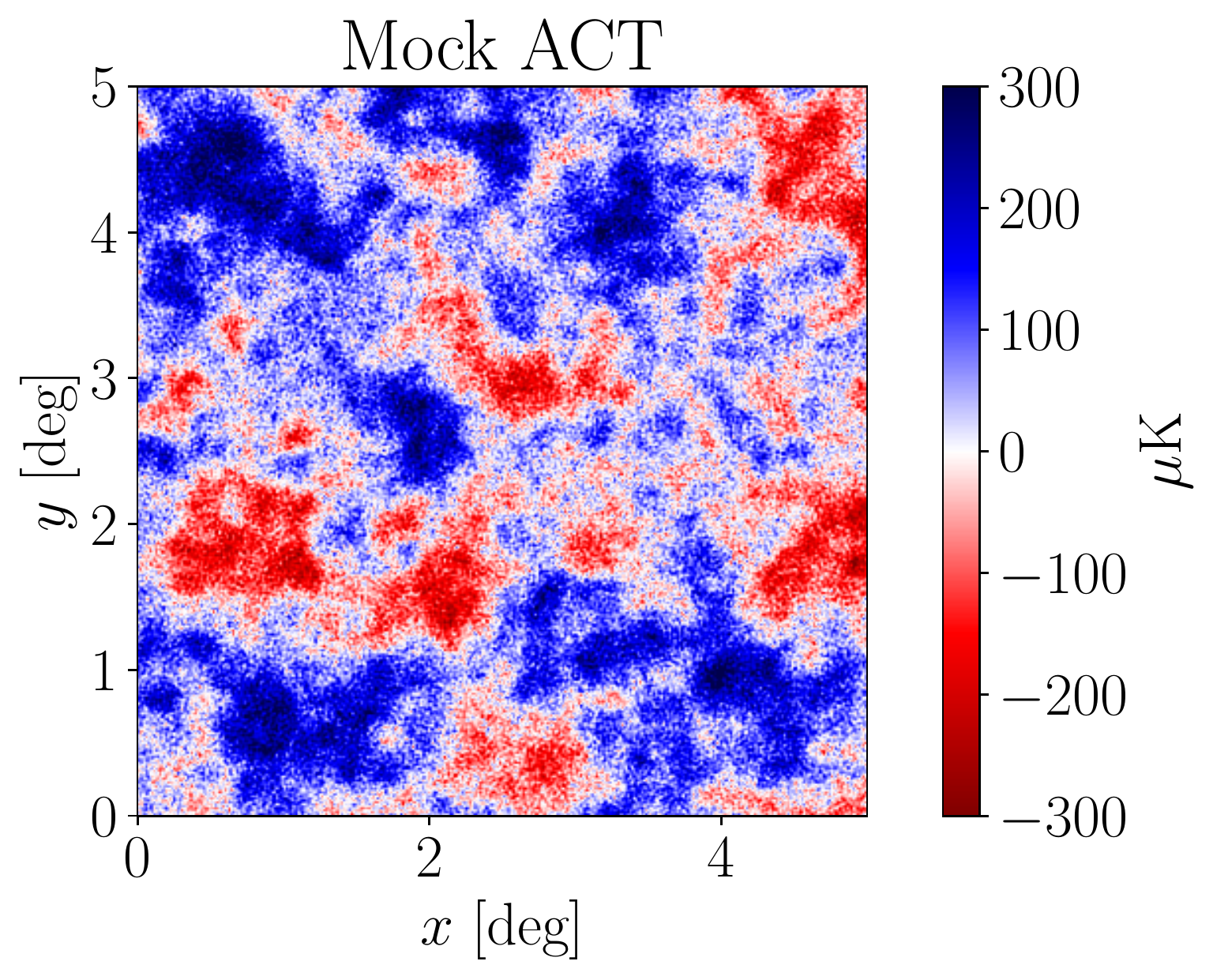}
\caption{
Small cutouts from the 800 Gaussian mocks used to validate the bootstrap covariance matrix.
Left: lensed CMB alone. Right: realistic mocks with power spectrum equal to the AdvACT power, used to test the covariance matrix.
}
\label{fig:noise_mocks}
\end{figure}

To test the impact of this effect, we generate 800 Gaussian mocks of the CMB sky, with the same (average) power spectrum as the AdvACT data (including CMB, foregrounds and noise).
Two cutouts from these mocks are shown in Fig.~\ref{fig:noise_mocks}.

\begin{figure}[h]
\centering
\includegraphics[width=0.95\columnwidth]{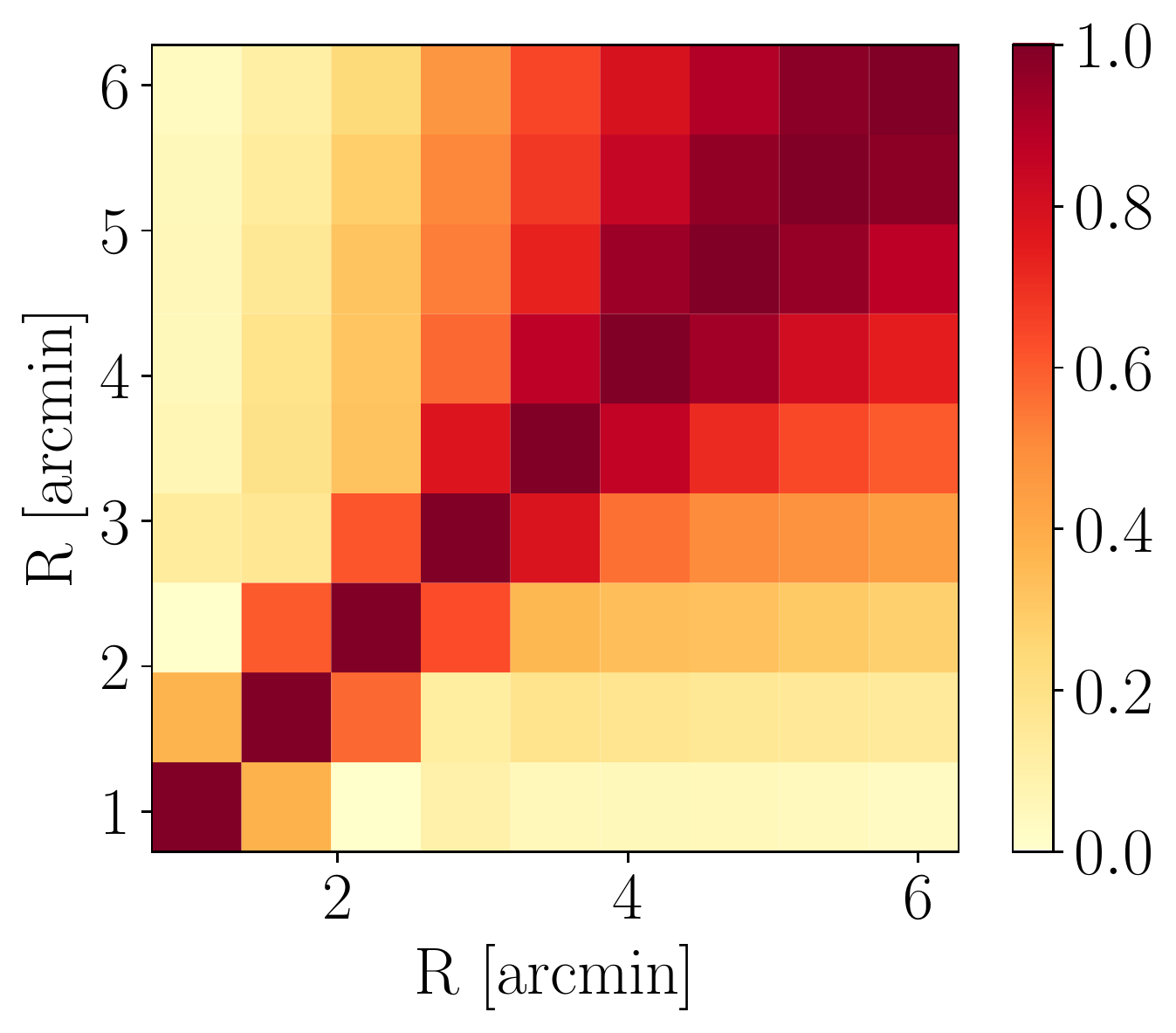}
\includegraphics[width=0.95\columnwidth]{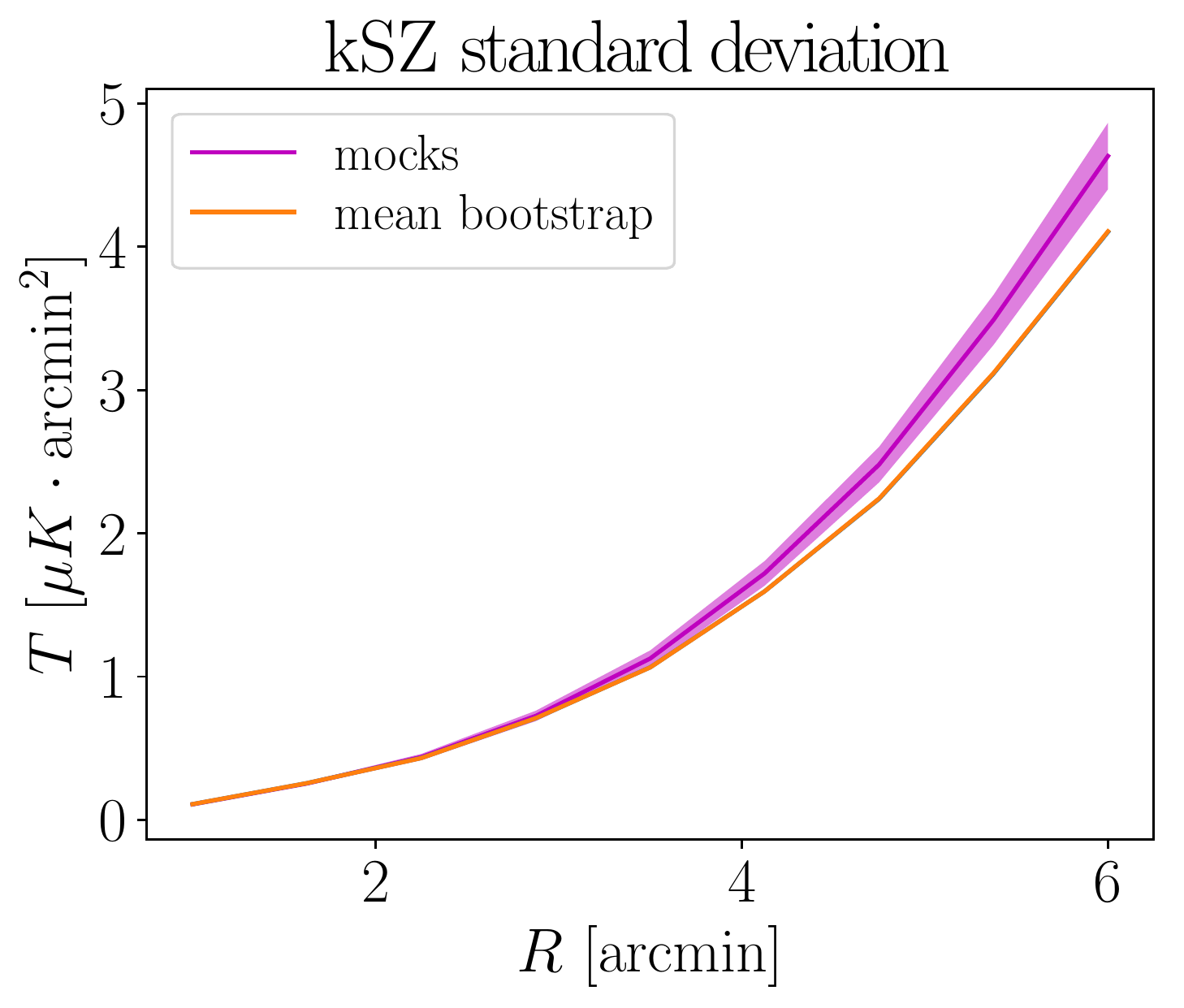}
\caption{
\textbf{Top:} kSZ correlation coefficient matrix.
Above the diagonal: using the scatter across 800 mocks; under the diagonal: averaging bootstrap covariance from each mock.
The visual agreement is good.
\textbf{Bottom:} Comparison of the kSZ error bar (standard deviation) estimated from mocks VS bootstrap.
The bootstrap method underestimates the kSZ covariance by about $10\%$ at large apertures, which contribute almost none of the SNR.
}
\label{fig:cov_ksz_bootstrap_vs_mocks}
\end{figure}

We then perform the same stacking analysis on these mocks as on the real data.
Crucially, we apply the CAP filters at the true galaxy positions.
This ensures that the effect of overlapping filters (which is more important at larger apertures), is correctly taken into account.
In these mock analyses, we then compare the covariance matrix estimated from bootstrap (potentially biased by the filter overlap) to that estimated from the scatter across mocks, which properly includes the effect of filter overlap.
Since these are both estimated from the same mocks, we can quantify the effect of the filter overlap, regardless of any mismatch between the mock maps and the actual AdvACT data.
In particular, the fact that our mock maps have a uniform depth, unlike the actual AdvACT data, is mostly irrelevant.
As shown in Fig.~\ref{fig:cov_ksz_bootstrap_vs_mocks},
we find that bootstrap underestimates the standard deviation by 10\% at large apertures for kSZ, while being more accurate for smaller apertures.
However, these large apertures are also the noisiest, and their weight in the total SNR and fit parameters is negligible.
We conclude that the effect of filter overlap is small for our purposes, but will need to be considered in future analyses.
We have checked that the exact same effect is seen for tSZ.

\section{Consistency with our previous measurement}
\label{app:consistency_schaan+16}

\begin{figure}[h]
\centering
\includegraphics[width=0.95\columnwidth]{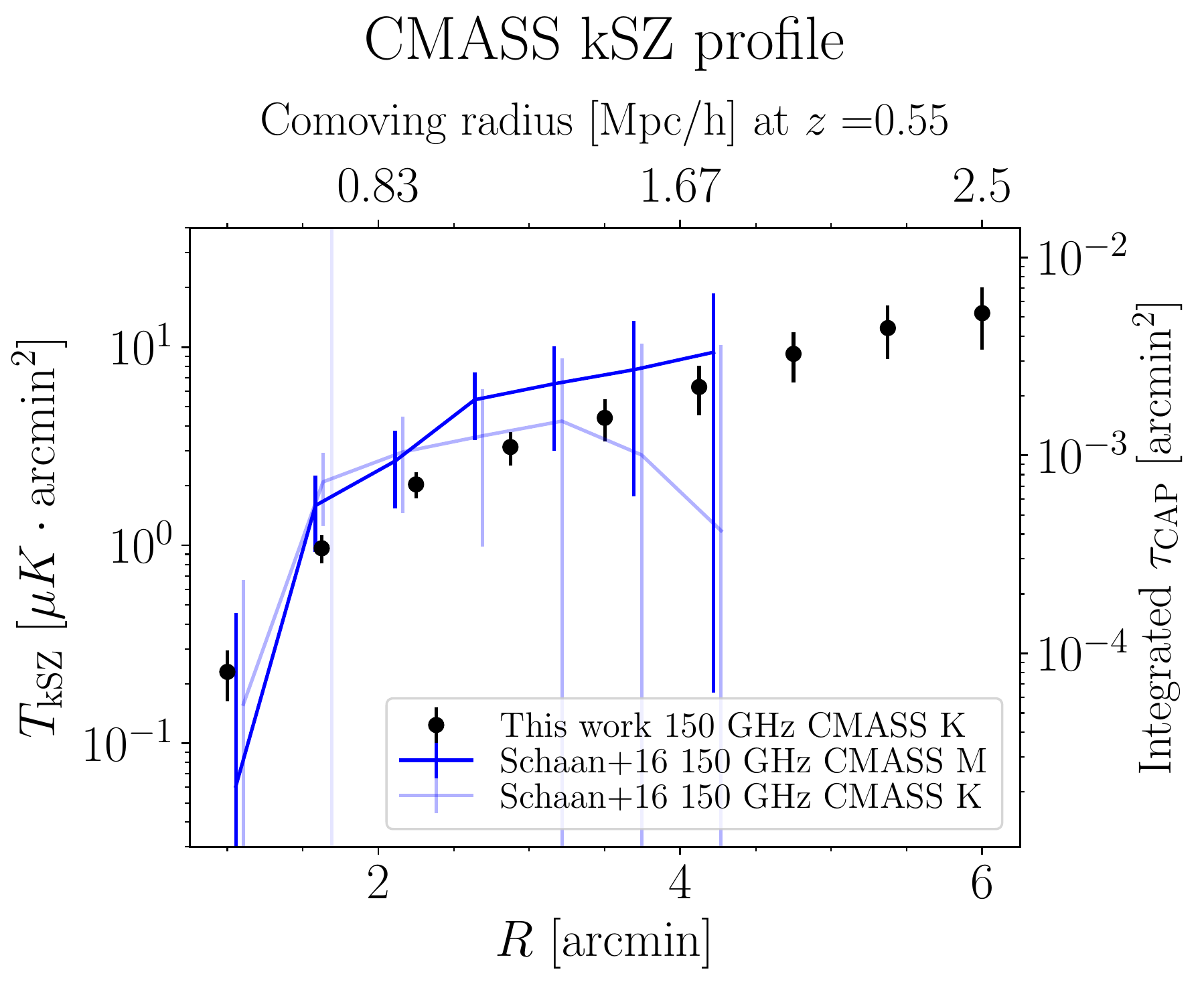}
\caption{
Comparison between the CMASS kSZ profiles as measured in this work and our previous work \cite{Schaan:2015uaa}.
These use the same galaxy samples and velocity reconstruction method.
They differ only by the area and depth of the CMB temperature map, and by a negligible mass weighting in the stack in \cite{Schaan:2015uaa}.
The measurements are consistent, and show the large improvement in sensitivity.
}
\label{fig:consistency_schaan+16}
\end{figure}

We compare the kSZ profiles measured in this work to the ones measured in \cite{Schaan:2015uaa}. These use the same galaxy sample (CMASS) and the same reconstructed velocities (CMASS K and CMASS M), but on a smaller patch of the sky ($\approx 660$ sq. deg), hence fewer galaxies ($\approx 25,500$), and with noisier maps.
Ref.~\cite{Schaan:2015uaa} adopted a mass weighting in the stack, enhancing the contribution from higher mass objects. 
The stacked kSZ profile was shown in terms of the dimensionless $\alpha (\theta_d)$, going from 0 when no baryons are included in the aperture to the fraction of free electrons $f_\text{free}$ (with respect to the total number of electrons) when all the baryons are included in the aperture.
Ignoring the mass weighting, the quantity $\alpha$ in \cite{Schaan:2015uaa} can be converted to our units via:
\beq
\underbrace{T_\text{kSZ}(\theta_d)}
_{\mu\text{K} \cdot \text{arcmin}^2}
=
\underbrace{\frac{1}{r_v}
\frac{v^\text{rec}_\text{rms}}{c}}
_{\text{dim,less}}
\underbrace{T_\text{CMB}}_{\mu \text{K}}
\underbrace{\left(\frac{N_e^\text{vir} \sigma_T}{a^2 \chi^2}\right)}
_{\text{arcmin}^2}
\;
\underbrace{\alpha(\theta_d)}_{\text{dim.less}},
\eeq
where $N_e^\text{vir} \sigma_T / (a^2 \chi^2)$ is the total integrated optical depth to Thomson scattering (in sr or arcmin$^2$). 
The quantity $N_e^\text{vir}$ is the number of electrons from a dark matter plus baryon mass $M_\text{vir}$, assuming cosmological abundance of baryons in the form of fully ionized gas, with a 76\% hydrogen and 24\% helium mass fractions.
Fig.~\ref{fig:consistency_schaan+16} shows that the measurements are consistent, and highlights the large improvement in the error bars.

\section{tSZ bias to kSZ from mass outliers}
\label{app:foreground_outliers}

We have previously argued that kSZ measurements are mostly immune from foreground contamination, due to the cancellation when weighting by velocity. In this appendix we quantify the effectiveness of this cancellation and argue that any potential bias due to imperfections in the cancellation is negligible.

If a few objects in the catalog have a catastrophically wrong mass estimate and are actually massive clusters, their tSZ signal can be much larger than their kSZ signal. 
If the fraction of unidentified massive clusters in the sample increases, the relative bias to the kSZ estimator also increases as
$N_\text{clusters}/N_\text{total}$.
However, because the tSZ signal always has the same sign, the weighting by velocity reduces this bias by a factor
$\langle v \rangle_\text{clusters} / v_\text{RMS typical} = 1/\sqrt{N_\text{clusters}}$.
These effects partially compensate, 
so the overall bias to the kSZ estimator from the tSZ emission of the unidentified clusters is:
\beq
\bal
\text{Relative tSZ bias to kSZ}
&= 
\frac{N_\text{clusters}}{N_\text{total}}
\frac{\langle v \rangle_\text{clusters}}{v_\text{RMS typical}}
\frac{\bar{T}_\text{tSZ cluster}}{\bar{T}_\text{kSZ typical}}\\
&= 
\sqrt{\frac{N_\text{clusters}}{N_\text{total}}}
\frac{1}{\sqrt{N_\text{total}}}
\frac{\bar{T}_\text{tSZ cluster}}{\bar{T}_\text{kSZ typical}}.
\eal
\eeq
In particular, the relative bias grows with the square root of the fraction of clusters in the sample, and decreases with the square root of the total number of objects in the sample.
For example, if $10\%$ of the objects in the $N_\text{total}=10^6$ catalog are actually $10^{14}M_\odot$ clusters, with a tSZ signal $100$ times larger than the typical kSZ signal of the sample, then the overall relative bias to kSZ will be $3\%$.
In our previous measurement \cite{Schaan:2015uaa},
where $N_\text{total}=25,000$, the same fraction of massive clusters would have produced a $20\%$ bias.
This highlights the usefulness of large galaxy catalogs.
However, the statistical precision of the kSZ measurement also scales as $1/\sqrt{N_\text{total}}$,
such that the significance of the tSZ bias to kSZ depends only on the fraction of massive clusters, and not on $N_\text{total}$.

\begin{figure}[h]
\centering
\includegraphics[width=0.95\columnwidth]{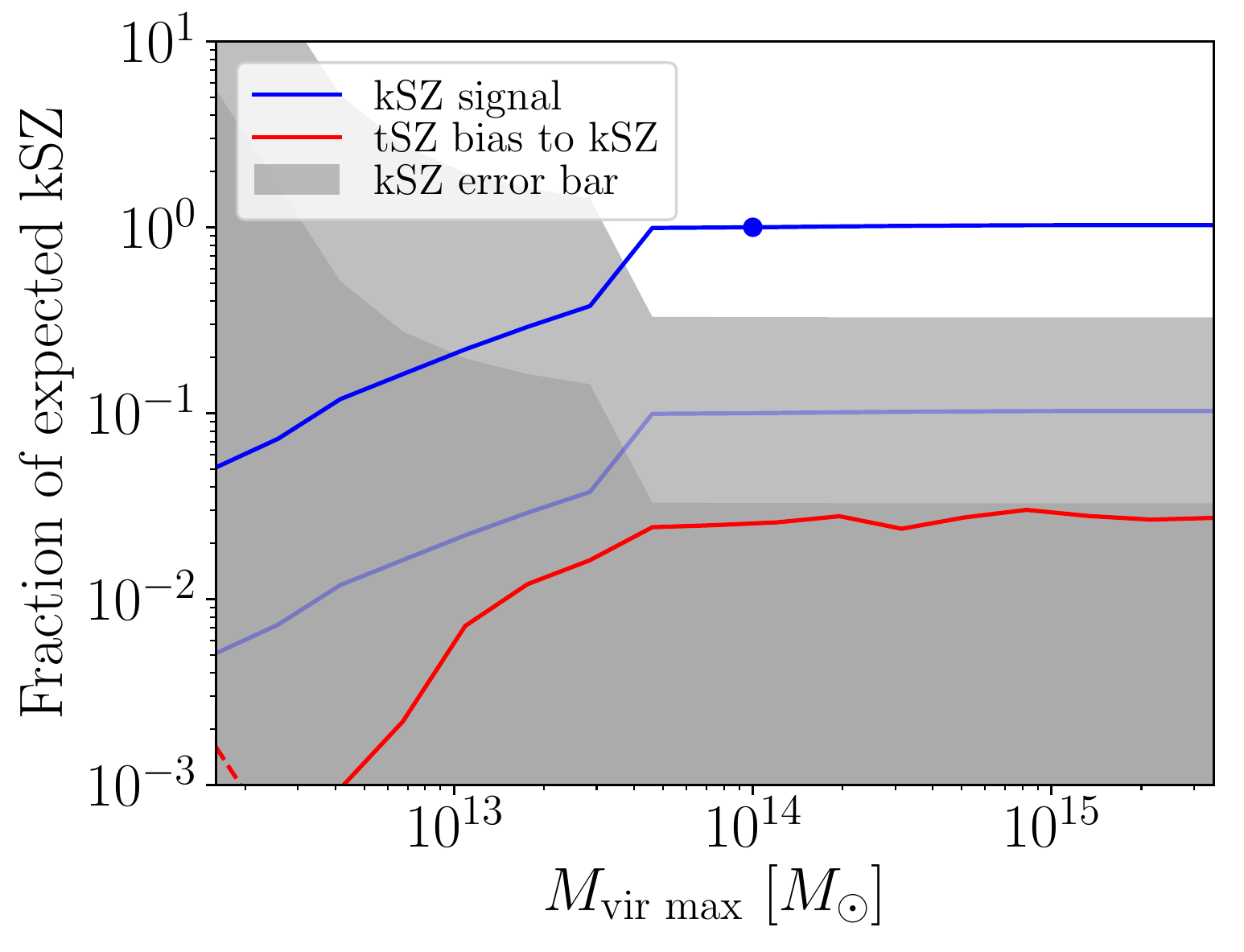}
\caption{
CMASS: the expected fractional bias from tSZ to kSZ (red curve) is shown as a function of the maximum halo mass $M_\text{vir max}$ included in the stack.
This fractional bias is compared to the kSZ signal (dark blue) and one tenth of it (light blue).
It is also compared to the fractional statistical uncertainty (light gray) and one tenth of it (dark gray).
The solid blue point corresponds to the fiducial halo mass cut used in the analysis ($10^{14}M_\odot$), where the expected bias from tSZ is smaller than 10\% of the signal (light blue curve) and than 10\% of the statistical uncertainty (dark gray band).
All of these curves are predictions based on the individual galaxy mass estimates. This plot, and our selection of the maximum mass cut, is blind to the CMB data.
}
\label{fig:tsz_to_ksz_mmax_cmass}
\end{figure}

In practice, we predict the tSZ contamination to kSZ as a function of the maximum halo mass included in the sample.
To do this, we use the individual galaxy stellar mass estimates, which we then convert to halo mass and to tSZ and kSZ signals.
The result is shown in Fig.~\ref{fig:tsz_to_ksz_mmax_cmass}.
Based on this, we select a maximum mass cut of $10^{14} M_\odot$, to ensure that the tSZ contamination is less than $10\%$ of the kSZ signal and than $10\%$ of the statistical uncertainty on the kSZ signal.
This maximum mass cut was selected in a blind way with respect to the kSZ data, since only the information in the galaxy catalog was used (redshifts, masses and velocities).  

\begin{figure}[h]
\centering
\includegraphics[width=0.95\columnwidth]{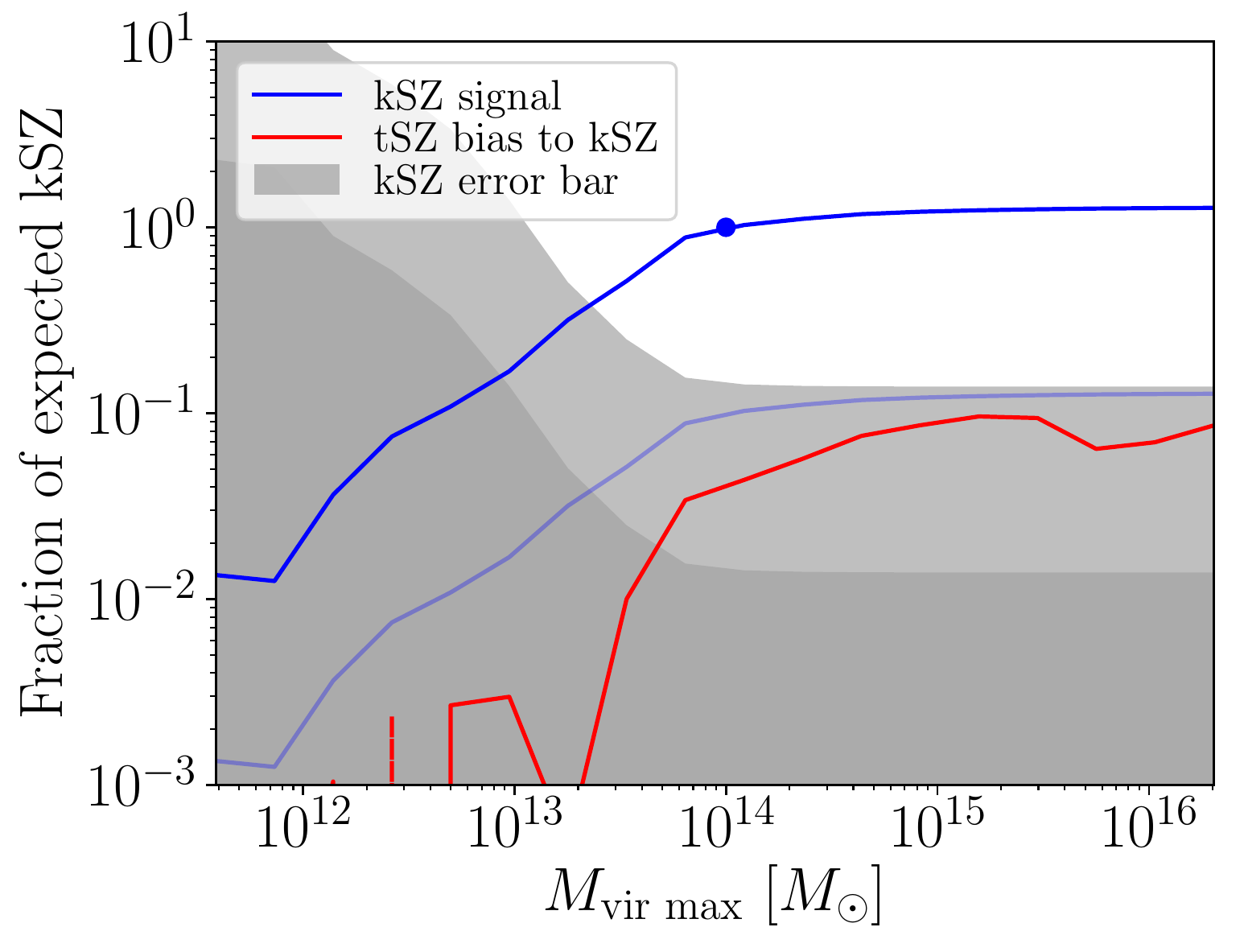}
\caption{
Same as Fig.~\ref{fig:tsz_to_ksz_mmax_cmass}, but for LOWZ instead of CMASS.
}
\label{fig:tsz_to_ksz_mmax_lowz}
\end{figure}

This also shows that a single $10\sigma$ outlier, i.e. a galaxy whose signal is ten times larger than the noise, can cause a $1\%$ bias to the stacked profile (for $N_\text{total}=10^6$ and a noise in temperature equal to $10^3$ times the kSZ signal, as in our case).
We therefore reject outlier galaxies before stacking.
Specifically, we reject objects such that the probability of finding one or more galaxies with such a high absolute temperature value is $5.7\times 10^{-7}$ (i.e. ``$5\sigma$'' threshold).
Because of the large number of galaxies in our sample $N_\text{total}\sim 10^6$, this corresponds to a $7.2\sigma$ cut on the individual temperatures.
In practice, we only find a handful of such outliers, and find no difference in the resulting stacked measurement with or without this outlier rejection.

\section{Host halo mass distribution uncertainties and interpretation of the kSZ signal}
\label{app:mass_distribution}

\begin{figure}[h]
\centering
\includegraphics[width=0.95\columnwidth]{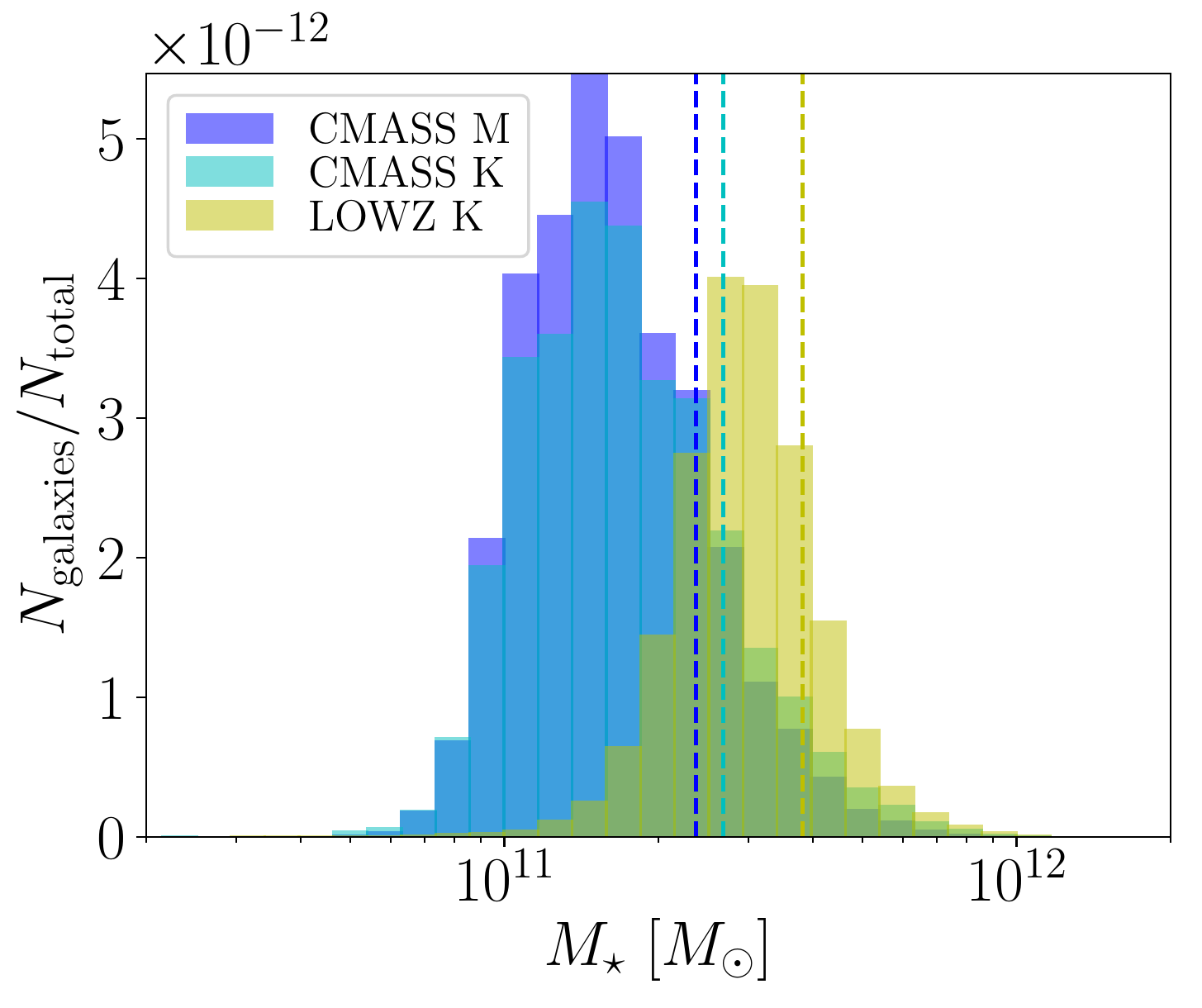}
\caption{
Stellar mass estimates of the LOWZ K (DR10), CMASS K (DR10) and CMASS M (DR12) galaxies
from \cite{2013MNRAS.435.2764M} for CMASS and from the Wisconsin group.
The dashed lines indicate the mean masses for each sample.
}
\label{fig:stellar_masses}
\end{figure}

The virial masses of the host halos of the CMASS and LOWZ galaxies used in this analysis are shown in Fig.~\ref{fig:halo_masses}. 
To estimate these, we start from the stellar mass estimates from \cite{2013MNRAS.435.2764M} for CMASS and from the Wisconsin group\footnote{\url{https://data.sdss.org/sas/dr12/boss/spectro/redux/galaxy/v1_1/}} for LOWZ, shown in Fig.~\ref{fig:stellar_masses}.
We then convert them to halo masses using the stellar-to-halo mass relation of \cite{2018AstL...44....8K}.
Specifically, we use their Eq.~A4, which accounts for the scatter in the stellar-to-halo mass relation.
Uncertainties are large at every step of this process, and the uncertainty in the host halo mass estimates is not well known.

Knowing the host halo masses of the galaxies in the sample is crucial for the interpretation of the kSZ and tSZ signals. 
If a fraction of the objects in the catalog is made of unidentified clusters, these objects will contribute a higher kSZ signal. 
If this is not accounted for, it constitutes a bias in the kSZ modeling. 
This bias scales as the fraction of massive clusters $N_\text{clusters} / N_\text{total}$, but is not reduced by the velocity weighting, since this additional kSZ signal is correlated with the velocities.
The resulting kSZ bias is then simply
\beq
\text{Relative kSZ bias}
=
\frac{N_\text{clusters}}{N_\text{total}} 
\frac{\bar{T}_\text{kSZ cluster}}{\bar{T}_\text{kSZ typical}}.
\eeq
So if $10\%$ of our objects are unidentified $10^{14}M_\odot$ clusters with a 10 times larger kSZ signal, this would produce a kSZ modeling bias of order unity.
This shows how crucial it is to know the HOD of the catalog interest, and in particular the distribution of host halo masses. This modeling is discussed in \cite{paper2}.

\end{document}